\documentclass[11pt,twocolumn,preprintnumbers,amsmath,amssymb,nofootinbib,superscriptaddress,iop,prd,numberedappendix]{openjournal} 
\usepackage{natbib}
\usepackage{newtxtext,newtxmath}
\usepackage{amsmath,amssymb}
\usepackage{graphicx}
\usepackage{dcolumn}
\usepackage{bm}
\usepackage[dvipsnames]{xcolor}
\usepackage{xspace}
\usepackage{lipsum}
\usepackage{multirow}
\usepackage{cancel}
\usepackage{enumitem}

\usepackage{wrapfig}
\usepackage{rotating}
\usepackage{booktabs}

\usepackage{textgreek}
\usepackage[utf8]{inputenc}
\usepackage[english]{babel}

\usepackage{color,colortbl}
\definecolor{linkcolor}{rgb}{0.0,0.3,0.5}

\usepackage{hyperref}
\hypersetup{
    unicode, 
    colorlinks=true,
    linkcolor=linkcolor,
    citecolor=linkcolor,
    filecolor=linkcolor,
    urlcolor=linkcolor,
}
\usepackage{tensind}
\tensordelimiter{?}
\DeclareGraphicsExtensions{.bmp,.png,.jpg,.pdf}
\usepackage{verbatim}
\usepackage[normalem]{ulem}
\usepackage{orcidlink}
\usepackage{soul}

\newcommand{\nv}{\hat{\bf n}}

\newcommand{\wtj}[6]{\left(\begin{array}{ccc} #1 & #2 & #3\\#4 & #5 & #6\end{array} \right)}

\newcommand{\nmt}{{\tt NaMaster}\xspace}
\newcommand{\planck}{{\sl Planck}\xspace}
\newcommand{\txt}{$2\times2$-point\xspace}
\newcommand{\fsb}[3]{\Phi^{#3}_{#2 #1}}
\newcommand{\fsbest}[3]{\hat{\Phi}^{#3}_{#2 #1}}

\definecolor{jaune}{rgb}{0.87953, 0.62395, 0.34134}
\definecolor{bleu}{rgb}{0.3192, 0.3879, 0.63583} 
\definecolor{rouge}{RGB}{157, 67, 75}

\graphicspath{ {./figs/} }

\begin{document}

\title{Constraints from CMB lensing tomography with projected bispectra} 
\author{Lea Harscouet\orcidlink{0009-0008-5331-3728}$^{1,*}$}
\author{David Alonso\orcidlink{0000-0002-4598-9719}$^{1}$}
\author{Andrina Nicola\orcidlink{0000-0003-2792-6252}$^{2, 3}$}
\author{An\v ze Slosar\orcidlink{0000-0002-8713-3695}$^{4}$}
\email{$^*$lea.harscouet@physics.ox.ac.uk}
\affiliation{$^1$Department of Physics, University of Oxford, Denys Wilkinson Building, Keble Road, Oxford OX1 3RH, United Kingdom}
\affiliation{$^2$Jodrell Bank Centre for Astrophysics, Department of Physics and Astronomy, The University of Manchester, Manchester M13 9PL, UK}
\affiliation{$^3$Argelander Institut f\"ur Astronomie, Universit\"at Bonn, Auf dem H\"ugel 71, 53121 Bonn, Germany}
\affiliation{$^4$Brookhaven National Laboratory, Physics Department, Upton, NY 11973, USA}

\begin{abstract}
  We measure the angular power spectrum and bispectrum of the projected overdensity of photometric DESI luminous red galaxies, and its cross-correlation with maps of the Cosmic Microwave Background lensing convergence $\kappa$ from \planck. This analysis is enabled by the use of the ``filtered-squared bispectrum'' approach, introduced in previous work, which we generalise here to the case of cross-correlations between multiple fields. The projected galaxy bispectrum is detected at very high significance (above $30\sigma$ in all redshift bins), and the galaxy-galaxy-convergence bispectrum is detected above $5\sigma$ in the three highest-redshift bins. We find that the bispectrum is reasonably well described over a broad range of scales by a tree-level prediction using the linear galaxy bias measured from the power spectrum. We carry out the first cosmological analysis combining projected power spectra ($gg$ and $g\kappa$) and bispectra ($ggg$ only) under a relatively simple model, and show that the galaxy bispectrum can be used in combination with the power spectrum to place a constraint on the amplitude of matter fluctuations, $\sigma_8$, an on the non-relativistic matter fraction $\Omega_m$. We find that data combinations involving the galaxy bispectrum recover constraints on these parameters that are in good agreement with those found from the traditional ``2$\times$2-point'' combination of galaxy-galaxy and galaxy-convergence power spectra, across all redshift bins.
\end{abstract}

\maketitle

\section{Introduction}\label{sec:intro}
  The analysis of projected probes of the large-scale structure (i.e. proxies of the matter fluctuations integrated along the line of sight and mapped onto the celestial sphere) has been one of the most productive areas of cosmology over the last decade. This is thanks to the wide variety of complementary probes we now have access to, and to the fast growth in the cosmic volume we can cover with them. Examples of these probes are the projected clustering of galaxies \citep{1912.08209,2008.13154,2310.03066}, the gravitational lensing of the Cosmic Microwave Background (CMB lensing) \citep{astro-ph/0601594}, galaxy weak lensing \citep{astro-ph/9912508,1710.03235}, or the Sunyaev-Zel'dovich effect \citep{astro-ph/0208192}. The combination of galaxy clustering and CMB lensing has been particularly fruitful: by measuring the cross-power spectrum between maps of the CMB lensing convergence $\kappa$ and the overdensity of galaxies $\delta_g$ at different redshifts, and combining it with the auto-correlation of these galaxies, we can simultaneously determine the galaxy bias relation and measure the amplitude of matter fluctuations at the redshifts of these galaxies. This technique, often called ``lensing tomography'' or ``\txt analysis''\footnote{This nomenclature is imprecise and may lead to confusion. We only reluctantly subscribe to it due to its now prevalent use by the community. For clarity, in this work, when discussing ``\txt'' analyses, we will refer to the combination of the galaxy-galaxy and the galaxy-convergence power spectra, and not any other pair of cross-correlations.}, has now been used to place constraints on the growth history with percent-level precision spanning the last 12 billion years ($z\lesssim3$) \citep{2105.03421,whiteCosmologicalConstraintsTomographic2022,2306.17748,2309.05659,2407.04607,2410.10808,2503.24385,2506.22416}. Including the CMB lensing auto-correlation\footnote{Often referred to as ``$3\times2$-point'', not to be confused with the same technique when applied to galaxy weak lensing -- even though often more than 3 correlations in total are actually used in both cases.}, one can then extend the growth reconstruction to redshifts above that probed by the highest-redshift galaxy sample used \citep{2409.02109}.

  In spite of this success, further information may still be gained from the joint analysis of galaxies and CMB lensing. As the Universe evolves, the matter overdensities grow via non-linear gravitational collapse, which generates non-Gaussian features in their spatial distribution. In this regime, the two-point correlators used in the standard \txt analysis fail to capture all of the information present in both  $\delta_g$ and $\kappa$. To recover this information, various techniques have been proposed in the literature. On the one hand, field-level analyses constitute the manifestly-optimal approach to extract this information, although they are difficult to deploy in practice, due to their computational complexity, and the need to simulate intricate astrophysical and observational effects present in the data \citep{1203.3639,1909.06396,2108.04825,2304.04785}. The alternative approach is to employ well-educated summary statistics, commonly designed by performing non-linear (and sometimes non-local) operation on the data before compressing it onto a small set of measurements (e.g. some form of correlator) \citep{1211.5213,1710.05162,2204.07646,2206.01709,2206.03877,2206.11005,2310.15250,2403.13985,2409.05695}. Within this category, the bispectrum (i.e. the Fourier- or harmonic-space third-order correlator) stands out for several reasons. First, the bispectrum is the lowest-order non-Gaussian summary statistic, and thus should contain significantly more information than other correlators for mildly non-Gaussian fields. Secondly, it is a theoretically tractable observable, in the sense that predictions for it can be made from first principles (e.g. via perturbation theory), without the need for simulations that must encompass all physical effects present in the data \citep{astro-ph/9808305,astro-ph/0112551,1107.5169,montandonRelativisticMatterBispectrum2023}. 

  Incorporating the bispectrum in cosmological analyses can be a significant challenge, however, in comparison with the standard power-spectrum-based studies. First, optimal bispectrum estimators are often significantly more computationally challenging than their power spectrum versions \citep{scoccimarroFastEstimatorsRedshiftspace2015,2107.06287,2404.07249}. Furthermore, unlike the fairly compact power spectrum, information in the bispectrum measurement is spread over a large number of ``triangle configurations''. Thirdly, it is often difficult to derive robust estimates of the bispectrum uncertainties, and its covariance with the power spectrum \citep{2111.05887}. Simulations may be used in this case, but a large number of them is needed to ensure convergence over the large number of triangle configurations available. Similarly, the mode-coupling matrix -- which accounts for mode loss and coupling induced by the survey geometry -- is also numerically challenging to evaluate due to the sheer number of pairwise combinations of triangle configurations \citep[although the survey window function has been studied for specific bispectrum statistics in e.g.][]{sugiyamaCompleteFFTbasedDecomposition2019}. Finally, bispectrum predictions may be slow to produce from first principles, mostly due again to the large number of configurations. Partly because of these difficulties, the projected bispectrum has not been widely exploited in cosmological large-scale structure analyses. Although the bispectrum has been considered in the context of weak lensing \citep{astro-ph/0004151,astro-ph/0310125,1910.04627}, most existing constraints are based on specific real-space two-point correlations \citep{2110.10141,2201.05227,2309.08601,2309.08602}. One of the few examples present in the literature is the recent measurements of the projected galaxy-galaxy-$\kappa$ bispectrum presented in \cite{2311.04213}.
  
  Recently, \cite{harscouetFastProjectedBispectra2025} (H25 hereafter) introduced the filtered-squared bispectrum estimator (FSB), addressing some of the difficulties mentioned above, and significantly simplifying the analysis of projected bispectra. The FSB is similar in spirit to the skew-spectrum \citep{astro-ph/0105415,0909.1837,1004.1409,1911.05763,2006.12832} (i.e. the cross-correlation between a given field and its square), introducing a filtering operation that allows it to recover different bispectrum configurations. Most importantly, by employing power spectrum techniques, H25 demonstrated that the FSB covariance (including its covariance with the power spectrum) may be accurately estimated using fast analytical methods in a completely data-driven manner, solving one of the main hurdles of bispectrum analysis. In this paper, we exploit the FSB to measure the galaxy-galaxy-galaxy and galaxy-galaxy-$\kappa$ bispectrum in real data, in combination with the standard \txt data vector. We will use these measurements to carry out a first tentative cosmological analysis combining projected power spectra and bispectra, quantifying the ability of the bispectrum to reduce the final parameter uncertainties, and to test the robustness of cosmological constraints, through the consistency between different data combinations.

  The paper is structured as follows. Section \ref{sec:methods} presents the FSB estimator, and the extension we introduce here to the case of multi-field cross-correlations. It also describes the model we will use to interpret our measurements, and our parameter inference pipeline. The galaxy samples and CMB lensing data we use in the analysis are described in Section \ref{sec:data}. Our results, including an assessment of the $ggg$ and $gg\kappa$ bispectrum detections, the complementarity of different data combinations, and the cosmological constraints derived from the $gg$, $g\kappa$ and $ggg$ measurements, are presented in Section \ref{sec:results}. Finally, Section \ref{sec:concl} summarises our results and discusses potential avenues for future exploration. 

\section{Methods}\label{sec:methods}
  \subsection{FSB review}\label{ssec:methods.review}
    In this section we provide a short summary of the Filtered-Squared Bispectrum (FSB), first introduced in H25 -- for a more complete introduction, please refer to section 2.2 of that paper.
    
    \subsubsection{The estimator}\label{sssec:methods.review.estimator}

      Let us consider a scalar field $a(\nv)$ defined on the sphere, and a filter $W_\ell^{L}$ which only has significant support over a compact range of angular scales $\ell \in L$. Note that this choice of filter is not unique, and that we can perform the following steps for a set of filters defined over different multipole ranges -- but for the sake of simplicity we focus for now on one filter $W_\ell^L$. Now let us construct a "filtered-squared" version of $a$, by first filtering it in harmonic space so as to only keep the information coming from the scales selected by the $W_\ell^{L}$ filter: 
      \begin{equation}
        a_{L}(\nv) \equiv {\mathcal S}^{-1}\left[W_\ell^{L}{\mathcal S}[a]_{\ell m}\right]_{\nv},
      \end{equation}
      where we use $\mathcal S, \mathcal S^{-1}$ as shorthands for the spherical harmonic transform and its inverse respectively. We then square the filtered field in real space: $a^2_L (\nv) \equiv(a_L(\nv))^2$.

      The FSB is the angular power spectrum of $(a_L^2)(\nv)$ with the original field $a(\nv)$: in the simplest case (that of full-sky observations, with the power spectrum estimated at each integer $\ell$), this is simply
      \begin{equation}\label{eq:fsb1}
        \fsbest{\ell}{LL}{aaa}\equiv\frac{1}{2\ell+1}\sum_m(a_L^2)_{\ell m}a^*_{\ell m}.
      \end{equation}
      In the presence of a mask $w_a (\nv)$ associated with field $a(\nv)$, and using bandpowers (i.e. bins of $\ell$) labelled by an index $b$, we replace this by the pseudo-$C_\ell$ (PCL) estimator, as implemented in the \nmt code\footnote{\url{https://github.com/LSSTDESC/NaMaster}} \citep{alonsoUnifiedPseudoC_2019}. This version of the FSB estimator accounts for mode-coupling effects introduced when taking the power spectrum of two fields partially defined on the sky. However, the back and forth between real and harmonic space during the early filtering step inevitably introduces some additional mode coupling which is not fully accounted for in the estimator; H25 verified that the estimator remained robust to those residual mode coupling effects, by measuring the (largely negligible) effect caused by masks of varied complexity on the FSB estimator.

\vspace{5mm}
    
    \subsubsection{Relation to the projected bispectrum}\label{sssec:methods.review.reltob}
      We can relate the FSB estimator to the full projected bispectrum of field $a$, $b_{\ell_1 \ell_2 \ell_3}^{aaa}$, by expanding $(a^2_{L})_{\ell m}$ in Eq. \ref{eq:fsb1} in terms of the harmonic coefficients of field $a$. This allows us to write 
      \begin{equation}\label{eq:fsb2}
        \fsbest{\ell}{LL}{aaa} = \sum_{(\ell m)_{123}} {\cal G}_{\ell_1\ell_2\ell_3}^{m_1m_2 m_3} K^{LL\ell}_{\ell_1\ell_2\ell_3} a_{\ell_1m_1} a_{\ell_2m_2} a_{\ell_3 m_3},
      \end{equation}
      where the Gaunt coefficient $\cal G$ ensures that only closed triangle configurations of  $(\ell_1,\ell_2,\ell_3)$ contribute to the estimator, and where we have defined the FSB kernel:
      \begin{equation} \label{eq:fsbkernel}
        K^{L_1L_2\ell}_{\ell_1\ell_2\ell_3} \equiv \frac{W_{\ell_1}^{L_1} W_{\ell_2}^{L_2} W^\ell_{\ell_3}}{2\ell_3+1},
      \end{equation}
      which ensures that the triangles in question are those that have two sides with scales $\ell_1\sim\ell_2\in L$, and one with scale $\ell_3 = \ell$. In the more common case of using bandpowers, $\ell_3 \in b$, the denominator is replaced by the total number of modes in the bandpower, $N_b\equiv\sum_{\ell_3\in b}(2\ell_3+1)$.

      Upon taking the expectation value of Eq. \ref{eq:fsb2}, we recover an expression for the FSB which depends on the projected bispectrum $b_{\ell_1\ell_2\ell_3}$: 
      \begin{align}\label{eq:bsp2FSB}
        \fsb{\ell}{LL}{aaa} = & \sum_{(\ell)_{123}} h^2_{\ell_1\ell_2\ell_3}K^{LL\ell}_{\ell_1\ell_2\ell_3}\,b_{\ell_1\ell_2\ell_3}^{aaa},
      \end{align}
      where
      \begin{equation}
        h^2_{\ell_1\ell_2\ell_3}\equiv\frac{(2\ell_1+1)(2\ell_2+1)(2\ell_3+1)}{4\pi}\wtj{\ell_1}{\ell_2}{\ell_3}{0}{0}{0}^2,
      \end{equation}
      with the squared term a $3j$ Wigner symbol.

      The FSB $\fsb{\ell}{LL}{aaa}$ is therefore an efficient estimator for the projected bispectrum $b_{\ell_1\ell_2\ell_3}^{aaa}$: it is fast to compute, and robust to mode-coupling effects caused by incomplete sky observations. It is also a convenient higher-order statistics for which analytical predictions can be constructed from first principles (as we will see in Section \ref{sssec:methods.th.bth}).

    \subsubsection{FSB covariance}\label{sssec:methods.review.cov}
      Another appealing feature of the FSB is the fact that its covariance matrix can be estimated easily, accurately, and in a model-independent way, using analytical methods. This is possible by reusing various techniques developed in the context of power spectrum covariance estimation. In H25, the FSB covariance estimate is built following these steps:
      \begin{enumerate}
        \item Pushing the FSB reinterpretation of the bispectrum as a power spectrum between $a$ and $a_L^2$, we can use existing techniques \citep{1906.11765} to obtain a first analytical approximation of the covariance, by assuming that both $a$ and $a_L^2$ are Gaussian fields. Although this assumption is in general incorrect, since $a_L^2$ cannot be Gaussian even if $a$ is (which it generally is not), H25 shows it recovers all purely diagonal terms -- both disconnected and connected -- contributing to the covariance. This is not entirely surprising, as the ``Gaussian'' power spectrum covariance is known to dominate the statistical uncertainties even for the highly non-Gaussian fields commonly encountered in large-scale structure \citep{1811.05714}. 
        \item We add to this first estimate two leading-order off-diagonal terms which we derived analytically, coined $N_{222}$ (disconnected term which contributes to the FSB auto-covariance) and $N_{32}$ (partially-connected term which contributes to the FSB-$C_\ell$ cross-covariance).
        \item  The two latter terms are scaled by the inverse of the effective sky fraction $f_\mathrm{sky}$, to account for the impact of mode loss in cut-sky observations.
      \end{enumerate}
      This covariance estimate is reasonably accurate and can be estimated directly from the data, as it only depends on 2-point and 3-point correlators (i.e. the power spectra and FSBs of the field involved). For completeness we repeat here the expressions for $N_{222}$ and $N_{32}$ derived in H25 (before any $f_{\rm sky}$ scaling): 
      \begin{align}\nonumber
        {\rm Cov}_{N_{222}} \left(\fsbest{\ell}{LL}{aaa}, \fsbest{\ell'}{L'L'}{aaa}\right) &  \; = \; \frac{C_\ell^{aa} C_{\ell'}^{aa} W_{\ell'}^L W^{L'}_\ell}{\pi} \quad \times \\ \label{eq:covn222_1} \sum_{\ell_1} (2\ell_1 + & 1) W_{\ell_1}^L W_{\ell_1}^{L'} C_{\ell_1}^{aa} \wtj{\ell}{\ell'}{\ell_1}{0}{0}{0}^2, \\ \label{eq:covn32_1}
        {\rm Cov}_{N_{32}} \left(\fsbest{\ell}{LL}{aaa}, C_{\ell'}^{aa} \right) \; & = \; \frac{4 W^L_{\ell'} C_{\ell'}^{aa} \fsb{\ell}{L\ell}{aaa}}{(2\ell+1)} \,,
      \end{align}
      where $\fsb{\ell}{L\ell}{aaa}$ is a generalisation of the standard FSB in which one of the filtered copies of the field is restricted to the contributions from a single multipole $\ell$.

\vspace{5mm}
  
  \subsection{Multi-field FSB}\label{ssec:methods.multifieldfsb}
    One can easily extend the definition of the FSB to more general cases by modifying some of the analysis choices made, without impacting the efficiency and accuracy of the estimator:
    \begin{itemize}
      \item We could use two different fields $\alpha$ and $\beta$ to build the filtered-squared map, and a third field $\gamma$ to correlate it with. This straightforward modification would allow us to probe the projected bispectrum of multiple fields $b^{\alpha\beta\gamma}_{\ell_1\ell_2\ell_3}$, with obvious applications in the context of cross-correlations. 
      \item We could probe arbitrary triangle configurations beyond the close-to-isosceles ones the FSB is sensitive to. This can be done by replacing the ``squaring'' step by a multiplication of the field filtered on two different scales. The resulting estimator would be directly sensitive to configurations of the form $b_{L_1L_2\ell}$, where $L_1$ and $L_2$ are the two filters being multiplied, and $\ell$ is the multipole at which the power spectrum is calculated.
    \end{itemize}

    One advantage of the FSB with its unique filter definition is that it keeps the dimensionality of the resulting data vector to a reasonable minimum, only selecting triangle configurations of interest (e.g. the squeezed limit of the bispectrum, via the selection of triangles close to isosceles). The main difficulty in the case of the multi-filter estimator is the significantly larger size of the resulting data vector. Unless significant information is encoded in the triangle configurations neglected by the vanilla FSB estimator, this could spoil the efficiency of the estimator for only modest gains in terms of final parameter constraints. We therefore leave the exploration of the multi-filter estimator for future work.

    Instead, in this paper we focus on the multi-field version of the estimator. In particular, we will consider cross-correlating a filtered-squared field $a^2_L (\nv)$ with another field $f(\nv)$, effectively probing their cross-bispectrum $b^{aaf}_{\ell_1\ell_2\ell_3}$. This is the simplest option to generalize the FSB estimator to cross correlations: another option would be to correlate $(a f)_L$ with $a$, but this comes with a few drawbacks. Among these are the fact that real-space multiplication of $a$ and $f$ might discard some information coming from regions of $f$ that do not overlap with $a$; and that $(a f)_L \times a$ exhibits less symmetry than $(a^2)_L \times f$, which would result in a more complicated analytical covariance matrix.
    
    This measurement requires very few changes to the estimator: our original expression for the single-field FSB (Eq. \ref{eq:fsb1}) becomes 
    \begin{equation}\label{eq:fsb_multi}
      \fsbest{\ell}{LL}{aaf} \equiv\frac{1}{2\ell+1}\sum_m(a_L^2)_{\ell m}f^*_{\ell m},
    \end{equation}
    or the corresponding pseudo-$C_\ell$ estimator in the case of cut-sky observations.

    This slight change in the estimator carries over to the covariance: although the analytical Gaussian covariance and $N_{222}$ terms are easily generalised to this multi-field case, the $N_{32}$ term changes more significantly. The multi-field expressions for $N_{222}$ and $N_{32}$ are derived in Appendix \ref{app:multicov}, where we also derive expressions for cross-covariance terms between single-field and multi-field FSBs, as well as the associated $C_\ell$s. For completeness, the most relevant $N_{222}$ and $N_{32}$ terms read:
    \begin{align}\nonumber
      & {\rm Cov}_{N_{222}} \left(\fsbest{\ell}{LL}{aaf}, \fsbest{\ell'}{L'L'}{aaf}\right) = \frac{ W_{\ell'}^L W_{\ell}^{L'} C_{\ell}^{af} C_{\ell'}^{af}}{\pi} \quad \times \\ 
      & \quad\quad\quad\quad \sum_{ \ell_{1}} (2\ell_1 + 1) \begin{pmatrix} \ell & \ell_1 & \ell' \\ 0&0&0 \end{pmatrix}^2  W_{\ell_1}^L W_{\ell_1}^{L'} \, C_{\ell_1}^{aa}, \label{eq:covn32_2} \\ \nonumber
      & {\rm Cov}_{N_{32}} \left(\fsbest{\ell}{LL}{aaf}, C^{af}_{\ell'}\right) = \frac{2 \, W_{\ell'}^L }{(2\ell' +1)} \quad \times \\
      & \quad\quad\quad\quad\quad \left( C_{\ell'}^{aa} \, \fsb{\ell}{L\ell'}{aff} + C_{\ell'}^{af}\, \fsb{\ell}{L\ell'}{aaf}\right) \, .
    \end{align}
    Note that the $N_{32}$ term is split into two contributions (compared to the single term in Eq. \ref{eq:covn32_1}), since some of the symmetries present in the single-field FSB are broken after introducing a different field. This will increase the computational time for this term twofold, which is easily manageable.

    Note that each of the fields involved have, in general, different masks. The resulting mode-coupling will differ from the single-field/single-mask case, but the effect is entirely accounted for by the PCL estimator, which automatically handles the multi-field case for power spectra. Some consideration must be put into the scaling of the $N_{222}$ and $N_{32}$ covariance terms with the effective sky fraction. For instance, if one were to compute the covariance of $\fsbest{\ell}{LL}{aaa}$ with $\fsbest{\ell'}{L'L'}{aaf}$, would they scale these additional terms using $f_\mathrm{sky} \equiv \langle w_a w_a \rangle$ (the natural choice for the single-field FSB), or $f_\mathrm{sky} \equiv \langle w_a w_f \rangle$ (the relevant one for the multi-field case)? Counter-intuitively, the correct approach in this specific example is to use $f_\mathrm{sky} \equiv \langle w_a w_a \rangle$, with no mention of $w_f$. We derive these results in Appendix \ref{app:multicov.fsky}, and empirically confirm the accuracy of these scalings using simulations.
\vspace{10mm}

  \subsection{Maps and power spectrum measurements}\label{ssec:methods.maps}

    \begin{figure}
      \centering
      \includegraphics[width=\linewidth]{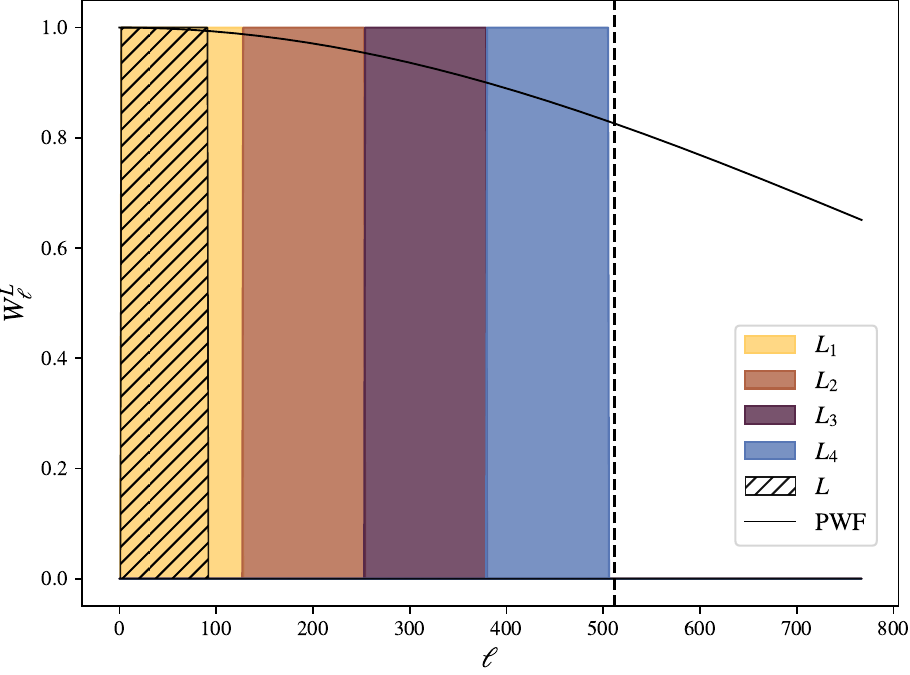}
      \caption{Filters used in this analysis. The colourful set represents the 4 filters used in Section \ref{ssec:results.detection} to detect the FSB on a wide range of scales; the hatched filter is the single filter used for deriving cosmological constraints in Section \ref{ssec:results.constraints}. The latter remains below the conservative $\ell_\mathrm{max}^B$ cut in all redshift bins, and the former are defined below the $2N_\mathrm{side}$ threshold, beyond spherical harmonic transforms may become unreliable. We also show the pixel window function (solid black line) corresponding to the map resolution.}\label{fig:filters}
    \end{figure}
    When analysing the data described in Section \ref{sec:data}, we will construct maps of the galaxy overdensity and CMB convergence fields using the {\tt HEALPix} pixelisation scheme \citep{astro-ph/0409513}, with a resolution parameter $N_{\rm side}=256$, corresponding to a pixel size of $\delta\theta\sim0.23^\circ$.

    We will measure the power spectra and FSBs of these maps as described in the previous sections. In particular, we will measure the galaxy-galaxy and galaxy-convergence power spectra ($C_\ell^{gg}$ and $C_\ell^{g\kappa}$), as well as the ``galaxy$^3$'' and ``galaxy$^2\times$convergence'' FSBs ($\fsb{\ell}{LL}{ggg}$ and $\fsb{\ell}{LL}{gg\kappa}$). We will make measurements of the FSBs using two different sets of filters, shown in Fig. \ref{fig:filters}:
    \begin{itemize}
      \item To determine the detectability of the projected bispectrum, we will measure the FSB in four filters below the $2N_{\rm side}$ threshold. These four consecutive top-hat filters each have width $\Delta\ell=126$, partitioning the multipole range $\ell\in[2,\,506]$. 
      \item To perform a first cosmological analysis of the FSB, we will measure it on a single filter, covering the range $\ell\in[2,92]$. This is to ensure that the bispectrum measurements used in the analysis only include information from angular scales where our tree-level model (see Section \ref{ssec:methods.th}) is reliable.
    \end{itemize}

    The power spectra and FSBs are measured over the full range $\ell<3N_{\rm side}$, although we discard all data above $\ell\geq2N_{\rm side}$, where the {\tt HEALPix} spherical harmonic transform can be unreliable. We will also discard all FSB measurements that break the triangle inequality, where the bispectrum signal vanishes. The spectra we compute in Section \ref{ssec:results.detection} were measured in 24 bandpowers with equal width, $\Delta\ell=21$, covering the range $\ell<2N_{\rm side}=512$.

    Finally, we must also account for the impact of pixelisation on our measurements. The most important effect is the loss of power on scales comparable to that of the pixel size when the map in question is effectively the average of its infinite-resolution version over the pixel area. This loss of power can be captured in terms of the pixel window function (PWF) $\mathscr{b}_\ell$. In practice, only the galaxy overdensity maps used in this analysis are affected by the PWF, as the lensing convergence data is provided natively in the form of its harmonic coefficients, which we then sample at the pixel positions when transforming them into a map. Instead of correcting our measurements for the PWF, we include them in our theoretical model. This can be done by simply modifying the filters entering the FSB kernel (Eq. \ref{eq:fsbkernel}) as e.g.:
    \begin{equation}
      W^L_\ell\rightarrow \mathscr{b}_\ell W^L_\ell.
    \end{equation}
    This must be done for the filters and bandpower window functions corresponding to any field affected by the PWF (i.e. the galaxy overdensity, but not $\kappa$ in our case). Power spectra should be likewise corrected by the PWF where relevant. Note that any pure shot noise contribution (namely the power spectrum shot noise, $N_\ell^{gg}$, and the $ggg$ FSB noise term in Eq. \ref{eq:noiseggg}) should not be corrected for the PWF, as white noise spectra remain constant regardless of the pixelization scheme\footnote{This can be understood in terms of a purely Poisson process, in which the measurements on different pixels remain statistically independent regardless of the pixel size.}.

  \subsection{Theory and modelling}\label{ssec:methods.th}
    This section presents the theoretical framework we will use to make predictions for the power spectrum and FSB measurements we present in Section \ref{sec:results}, as a function of cosmological and galaxy bias parameters.

    \subsubsection{Galaxy clustering and CMB lensing}\label{sssec:methods.th.bias}
    We will study the power spectrum and bispectrum of two large-scale structure tracers projected on the celestial sphere: the CMB lensing convergence $\kappa$, and the angular galaxy overdensity $\delta^{\rm 2D}_g$. These tracers can be connected with the three-dimensional matter and galaxy overdensities ($\delta_m$, $\delta_g$) via line-of-sight integrals of the form
    \begin{align}
      \delta^{\rm 2D}_g(\nv)=\int d\chi\,q_g(\chi)\,\delta_g(\chi\nv,z(\chi)),\\
      \kappa(\nv)=\int d\chi\,q_\kappa(\chi)\,\delta_m(\chi\nv,z(\chi)).
      \label{eq:psp_kernels}
    \end{align}
    Here, $\nv$ is a unit vector pointing along the line of sight, $\chi$ is the radial comoving distance, and the radial kernels for $\kappa$ and $\delta_g^{\rm 2D}$ are
    \begin{align}
      &q_g(\chi)=H(z)\,p(z),\\
      &q_{\kappa} (\chi) = \frac{3}{2}\Omega_m H_0^2 \chi\,(1+z) \frac{\chi_{\rm CMB} - \chi}{\chi_{\rm CMB}},
    \end{align}
    where $H(z)$ is he expansion rate at redshift $z$ (implicitly corresponding to comoving distance $\chi$ in the lightcone) and $H_0$ is the expansion rate today. $\chi_{\rm CMB}$ is the comoving distance to the last-scattering surface, $\Omega_m$ is the fraction of non-relativistic matter, and $p(z)$ is the redshift distribution of the galaxy sample under study. 

    We will use a perturbative expansion in Eulerian space, up to second order, to connect the galaxy and matter overdensities \citep{astro-ph/9302009,0902.0991}. Specifically, in Fourier space this reads
    \begin{align} \label{eq:deltam2order}
      &(\delta_m)_\mathbf{k} = \delta^{(1)}_{\bf k} + \delta^{(2)}_{\bf k}\\ 
      &(\delta_g)_{\bf k} = b_1 \left[\delta^{(1)}_{\bf k} + \delta^{(2)}_{\bf k}\right] + \frac{b_2}2 ((\delta^{(1)})^2)_{\bf k} + b_s (s^2)_{\bf k} \,, \label{eq:deltag2order}
    \end{align}
    were $\delta^{(1)}_{\bf k}$ is the linear-order matter overdensity, which we assume to be Gaussian, $\delta^{(2)}_{\bf k}$ is the contribution to the matter overdensity at second order in standard perturbation theory, and $s^2\equiv s_{ij}s^{ij}$ is the square of the traceless shear field, with
    \begin{equation}
      s_{ij,{\bf k}}\equiv \left(\frac{k_ik_j}{k^2}-\frac{1}{3}\right)\delta^{(1)}_{\bf k}.
    \end{equation}
    The quantities $b_1$, $b_2$, and $b_s$ are the linear, quadratic, and tidal bias parameters. The different terms in Eq. \ref{eq:deltam2order} can be written in terms of the linear overdensity as
    \begin{align}
      \delta^{(2)}_{\bf k} & =\int D{\bf k}_1 D{\bf k}_2\,F_2({\bf k}_1,{\bf k}_2)\,\delta^{(1)}_{{\bf k}_1} \delta^{(1)}_{{\bf k}_2} \,, \\ 
      ((\delta^{(1)})^2)_{\bf k} & = \int D{\bf k}_1 D{\bf k}_2\, Q_2({\bf k}_1,{\bf k}_2) \delta^{(1)}_{{\bf k}_1} \delta^{(1)}_{{\bf k}_2} \,, \\
      (s^2)_{\bf k} & = \int D{\bf k}_1 D{\bf k}_2\,T_2({\bf k}_1,{\bf k}_2)\,\delta^{(1)}_{{\bf k}_1} \delta^{(1)}_{{\bf k}_2} \,,
    \end{align}
    where we use the shorthand notation
    \begin{equation}\nonumber
      \int D{\bf k}_1\,D{\bf k}_2\equiv
      \int\frac{d^3k_1}{(2\pi)^3}\frac{d^3k_2}{(2\pi)^3}\,(2\pi)^3\,\delta^D({\bf k}-{\bf k}_1-{\bf k}_2),
    \end{equation}
    and we have defined the perturbation theory kernel $F_2$, the quadratic kernel $Q_2$, and the tidal kernel $T_2$ as:
    \begin{align}
      F_2({\bf k}_1,{\bf k}_2) & = \frac 57 + \frac12 \mu_{12} \left(\frac{k_1}{k_2} + \frac{k_2}{k_1}\right) + \frac 27 \mu_{12}^2 \,, \\
      Q_2 ({\bf k}_1,{\bf k}_2) & = 1 \,, \\ 
      T_2({\bf k}_1,{\bf k}_2) & = \mu_{12}^2 - \frac13 \, ,
      \label{eq:bsp_kernels}
    \end{align}
    with $\mu_{12} \equiv (\mathbf k_1 \cdot \mathbf k_2) / |\mathbf k_1||\mathbf k_2|$.

    \subsubsection{Power spectrum modelling}\label{sssec:methods.th.pth}
    In a flat universe, the angular power spectrum $C_\ell^{\alpha\beta}$ of two projected quantities $\alpha$ and $\beta$ is related to the power spectrum of the associated 3D fields, $P^{\alpha\beta}(k,z)$ via
    \begin{equation}
      C_\ell^{\alpha\beta}=\int\frac{d\chi}{\chi^2}\,q_\alpha(\chi)\,q_\beta(\chi)\,P^{\alpha\beta}(k_\ell(\chi),z(\chi)),
    \end{equation}
    where $k_\ell(\chi)\equiv (\ell+1/2)/\chi$ is the 3D wavenumber associated with the angular scale $\ell$, and $q_\alpha(\chi)$ is the radial kernel of the $\alpha$ field. This relation is accurate in the Limber approximation \citep{Limber1953, loverdeExtendedLimberApproximation2008}, which holds for the wide radial kernels used in this work.

    Since we will study the galaxy auto-correlation and its cross-correlation with the lensing convergence, we must develop a model for the 3D galaxy-galaxy and galaxy-matter power spectra, $P^{gg}(k,z)$ and $P^{gm}(k,z)$. For simplicity, we will restrict our analysis to large scales, where both spectra are well described at tree level. Using the perturbative bias expansion described in the previous section, the only tree-level contribution is due to the linear galaxy bias term, given by
    \begin{equation}
      P^{gg}(k,z)=b_1^2\,P^{mm}(k,z),
      \hspace{12pt}
      P^{gm}(k,z)=b_1\,P^{mm}(k, z),
    \end{equation}
    where $P^{mm}(k,z)$ is the matter power spectrum. We also account for Redshift-Space Distortions (RSDs) in our modelling of the power spectra by adding a RSD term, linear in the matter density field, to the linear bias expansion.

    We estimate and subtract the contribution from shot noise to the galaxy-galaxy power spectrum as described in \cite{whiteCosmologicalConstraintsTomographic2022}.

    \subsubsection{Bispectrum Modelling}\label{sssec:methods.th.bth}
    As in the case of the power spectrum, the projected bispectrum of any three fields $b^{\alpha \beta \gamma}_{\ell_1 \ell_2 \ell_3}$  can be computed from their 3D bispectrum $B^{\alpha \beta \gamma}(k_1, k_2, k_3, z)$ via the Limber approximation:
    \begin{align} \label{eq:limber} \nonumber
      b_{\ell_1\ell_2\ell_3}^{\alpha \beta \gamma} = \int & \mathrm{d}\chi \frac{q^\alpha(\chi) q^\beta(\chi) q^\gamma(\chi)}{\chi^4} \quad \times \\
      & B^{\alpha \beta \gamma}\left(k_{\ell_1}(\chi), k_{\ell_2}(\chi), k_{\ell_3}(\chi),z(\chi) \right),
    \end{align}
    where $q^\iota$ is the $\iota$ field's radial kernel, and $k_\ell\equiv(\ell+1/2)/\chi$.

  In this work we will focus on the galaxy-galaxy-galaxy and galaxy-galaxy-convergence bispectra ($ggg$ and $gg\kappa$ hereafter), and hence must model the 3D $ggg$ and $ggm$ bispectra. We do not include the convergence-convergence-galaxy ($\kappa \kappa g$) bispectrum in this work, which might require a more complex modelling of the lensing reconstruction noise and is expected to have a lower signal than the $ggg$ and $gg\kappa$ bispectra.
  As in the case of the power spectrum, we will do this at lowest order in perturbation theory (i.e. tree level). Using the perturbative bias expansion introduced in Section \ref{sssec:methods.th.bias}, the corresponding bispectra are given by \citep[see ][]{schmittfullOptimalBispectrumEstimators2015, desjacquesLargeScaleGalaxyBias2018, cagliariBispectrumConstraintsPrimordial2025}: 
  \begin{align} \label{eq:bggg} \nonumber
    B^{ggg} & (k, k', k'') =  \\ \nonumber
    & \quad\quad b_1^3 \, \bigg(2 \, F_2 (\mathbf k, \mathbf k') \,P_k\,P_{k'} + 2 \mathrm{ perms.} \bigg) \\ \nonumber
    & \quad\quad  +  \frac{b_1^2 b_2}2 \, \bigg(2 \, Q_2 (\mathbf k, \mathbf k') \, P_kP_{k'} + 2 \mathrm{ perms.}\bigg) \\
    & \quad\quad  + b_1^2 b_s \, \bigg(2 \, T_2 (\mathbf k, \mathbf k') \, P_kP_{k'} + 2 \mathrm{ perms.}\bigg)\,,\\ \label{eq:bggk} \nonumber 
    B^{ggm} & (k, k', k'') = \\\nonumber
    & \quad\quad b_1^2 \, \bigg(2 \, F_2 (\mathbf k, \mathbf k') \,P_kP_{k'} + 2 \mathrm{ perms.}\bigg) \\ \nonumber
    & \quad\quad +  \frac{b_1 b_2}2 \, \bigg(2 \, Q_2 (\mathbf k, \mathbf k'') \, P_kP_{k''} + 1 \mathrm{ perm.}\bigg) \\
    & \quad\quad + b_1 b_s \, \bigg(2 \, T_2 (\mathbf k, \mathbf k'') \, P_kP_{k''} + 1 \mathrm{ perm.}\bigg)\,.
  \end{align}
  $P_k$ denotes the linear matter power spectrum. The cyclical permutations for the $ggg$ bispectrum are, explicitly, $\{\mathbf k, \mathbf k'\}$, $\{\mathbf k', \mathbf k''\}$, and $\{\mathbf k'', \mathbf k\}$. In turn, the permutations in the $gg\kappa$ case are the same for the linear term, but only $\{\mathbf k', \mathbf k''\}$ and $\{\mathbf k'', \mathbf k\}$ contribute for the quadratic and tidal terms. Once computed, the 3D bispectrum $B^{ggg}(\mathbf k, \mathbf k', \mathbf k'')$ ($B^{ggm}$) can be inserted into Eq. \ref{eq:limber} to recover the projected bispectrum $b^{ggg}_{\ell \ell' \ell''}$ ($b^{gg\kappa}_{\ell \ell' \ell''}$)\footnote{Note that we use the notation $b^{gg\kappa}$ (as opposed to $b^{ggm}$) to highlight the fact that we are using the CMB convergence $\kappa$ as tracer for the matter on the line of sight, $m$ -- this relationship is encoded in the projection kernel $q_\kappa (\chi)$ (Eq. \ref{eq:psp_kernels}).}.

  Due to the stochastic nature of discrete tracers such as galaxies, we must take into account additional contributions to the projected bispectrum due to shot noise. Assuming galaxy tracers can be modelled via Poisson statistics, these terms are given in harmonic space by 
  \begin{align} \label{eq:noiseggg}
    N^{ggg}_{\ell\ell'\ell''} & = \bar{n}^{-2} + \bar{n}^{-1} \left( C_{\ell}^{gg}+C_{\ell'}^{gg}+C_{\ell''}^{gg} \right) \\
    \label{eq:noiseggk} N^{gg\kappa}_{\ell\ell'\ell''} & = \bar{n}^{-1} C_{\ell''}^{g \kappa}
  \end{align}
  where $\bar{n}$ denotes the angular number density of galaxies per steradian in the survey. The derivation of Eq. \ref{eq:noiseggg} can be found in Appendix D of H25; we provide a derivation of Eq. \ref{eq:noiseggk} in Appendix \ref{app:stoch}. The final quantity we convert to an FSB using Eq. \ref{eq:bsp2FSB} is the addition of the projected theoretical bispectrum and of these noise terms. It is worth emphasising that, since these stochastic contributions depend on the signal power spectra ($C_\ell^{gg}$ and $C_\ell^{g\kappa}$), they could potentially help break parameter degeneracies.

    \subsubsection{Magnification}\label{sssec:methods.th.mag}
    Our discussion in the preceding sections has failed to include the impact of lensing magnification in the observed galaxy overdensity. Gravitational lensing modifies the observed positions of galaxies, effectively washing out structure around foreground lenses, and modifies the observed flux via magnification, typically bringing fainter galaxies into the selected sample. The competition between these two effects leads to an additive contribution to the observed galaxy overdensity, given by   \begin{equation}
    \Delta\delta^{\rm 2D,\,mag}_g(\nv)=\int d\chi\,q_\mu(\chi)\,\delta_m(\chi\nv,z(\chi)),
  \end{equation}
  and where the magnification lensing kernel is
  \begin{align}\nonumber
    q_\mu(\chi)\equiv&\frac{3}{2}\Omega_mH_0^2(1+z)\chi\\&\int_{z(\chi))}^\infty dz'\,(5s-2)p(z')\,\frac{\chi(z')-\chi}{\chi},
  \end{align}
  where $s$ is the logarithmic slope of the cumulative number count of sources as a function of magnitude at the magnitude limit.
  
  The presence of magnification modifies the galaxy-galaxy and galaxy-convergence power spectra as
  \begin{align}\label{eq:clgg_mag}
    &C_\ell^{gg} = C_\ell^{gg,0}+C_\ell^{g\mu} + C_\ell^{\mu\mu},\\\label{eq:clgk_mag}
    &C_\ell^{g\kappa} = C_\ell^{g\kappa,0} + C_\ell^{\mu\kappa},
  \end{align}
  where
  \begin{align}
    &C_\ell^{gg,0}=\int\frac{d\chi}{\chi^2}q_g^2(\chi) P^{gg}(k_\ell,z),\\
    &C_\ell^{g\mu}=\int\frac{d\chi}{\chi^2}q_g(\chi)\,q_\mu(\chi) P^{gm}(k_\ell,z),\\
    &C_\ell^{\mu\mu}=\int\frac{d\chi}{\chi^2}q_\mu^2(\chi) P^{mm}(k_\ell,z),\\
    &C_\ell^{g\kappa,0}=\int\frac{d\chi}{\chi^2}q_g(\chi)\,q_\kappa(\chi) P^{gm}(k_\ell,z),\\
    &C_\ell^{\mu\kappa}=\int\frac{d\chi}{\chi^2}q_\mu(\chi)\,q_\kappa(\chi) P^{mm}(k_\ell,z).
  \end{align}
  The magnification bias correction is usually negligible for the $C_\ell^{gg}$ power spectrum, as the additional terms are small compared to the $C_\ell^{gg,0}$ term (due to the small overlap in kernels $q^g$ and $q^\mu$). It can however sometimes be of comparable amplitude in the cross-correlation $C_\ell^{g\kappa}$, as the $q_\kappa$ and $q_\mu$ kernels can have significant overlap -- both represent lensing effects from structures on the line of sight. We will include these contributions when modelling the power spectra, both for $gg$ and $g\kappa$, but we will ignore the contribution from magnification to the projected bispectrum. This is justified since, as in the case of $C_\ell^{gg}$, the contribution from magnification to $\fsb{\ell}{LL}{ggg}$ is likely small. Although this may not be the case for $\fsb{\ell}{LL}{gg\kappa}$ in general, the relatively small signal-to-noise ratio with which we are able to measure this correlation on the scales we will use for cosmological inference makes the contribution from magnification irrelevant.

  \subsection{Inference pipeline}\label{ssec:methods.like}
  
  To quantify the additional information brought about by the projected bispectrum, in Section \ref{ssec:results.constraints} we will use our measurements to derive rudimentary constraints on cosmological parameters. To do so, we will use a Gaussian likelihood of the form 
  \begin{equation} \label{eq:lkl}
    -2 \log p(\mathbf d | \vec \theta) = (\mathbf d - \mathbf t(\vec \theta))^{\mathsf T} \mathsf C ^{-1} (\mathbf d - \mathbf t( \vec \theta)) + {\cal K} \, .
  \end{equation}
  Here, $\mathbf d$ is the data vector containing our combined measurements of the power spectra and bispectra (however we will see in Section \ref{ssec:results.constraints} that the $gg\kappa$ bispectrum will ultimately not be included in the data vector), and $\mathsf C$ is the covariance matrix of those measurements, which is obtained as described in sections \ref{sssec:methods.review.cov}, \ref{ssec:methods.multifieldfsb}, and Appendix \ref{app:multicov}. The theory predictions $\mathbf t (\vec \theta)$ are generated for a set of parameters $\vec \theta$ as described in Section \ref{ssec:methods.th}, and ${\cal K}$ is an irrelevant normalisation constant, independent of $\vec \theta$.

  The parameter set in our study is relatively small, containing only $\sigma_8$, $\Omega_m$, and the linear bias $b_1$. All other cosmological parameters are simply fixed to the best-fit values found by \cite{collaborationPlanck2018Results2020b}. The higher-order bias parameters, $b_2$ and $b_s$ will be determined in terms of $b_1$: in our fiducial analysis, we will make use of the empirical relations between these parameters and $b_1$ found in \cite{lazeyrasPrecisionMeasurementLocal2016} and \cite{lazeyrasLIMDBiasMeasurement2018}. We will refer to these as the ``coevolution relations'' in what follows. These relations have been found to accurately describe the clustering of dark matter haloes as well as galaxies -- although with a much larger scatter in the latter case \citep{barreiraGalaxyBiasForward2021}. Our purpose with these simplifications is to test the constraining power of the projected galaxy bispectrum through a simple cosmological analysis. We leave a more detailed analysis, in which the higher-order bias parameters are determined self-consistently by including their contribution to the one-loop power spectra, and which incorporates external data to place priors on the cosmological parameters we fix here, for future work.

  \begingroup 
    \setlength{\tabcolsep}{10pt} 
    \renewcommand{\arraystretch}{1.5} 
    \setlength\extrarowheight{2pt}
    
    \begin{table}[]
        \centering
        \begin{tabular}{ c c c }  
        \toprule
            Parameter & Fiducial value & Prior \\ \midrule
            {$\Omega_\mathrm{CDM}^\mathrm{fid}$} & 0.2607 & \\
            {$\Omega_b^\mathrm{fid}$} & 0.049 & \\
            {$h^\mathrm{fid}$} & 0.6766 & \\
            {$\sigma_8^\mathrm{fid}$} & 0.8102 & \\
            {$n_s^\mathrm{fid}$} & 0.9665 & \\ \midrule
            
            {$b_1$} & 1 & $\mathcal U \left[ 0,6 \right]$ \\
            {$r_{\sigma_8} = \sigma_8^\mathrm{fid} / \sigma_8 $} & 1 & $\mathcal U \left[ 0, 2 \right]$ \\
            $\Omega_m = \Omega_\mathrm{CDM} + \Omega_b^\mathrm{fid}$ & $\Omega_\mathrm{CDM}^\mathrm{fid} + \Omega_b^\mathrm{fid}$ & $\mathcal U \left[ 0.05, 0.5 \right]$ \\
            \bottomrule 
       \end{tabular}  
       \caption{Fiducial cosmological parameters, corresponding to the values favoured by \planck \citep{collaborationPlanck2018Results2020b}, and parameter priors used in our inference pipeline. For convenience we directly sample $r_{\sigma_8}$, the ratio of $\sigma_8$ to its fiducial value, and instead of sampling $\Omega_m$ directly, we vary $\Omega_\text{CDM}$ while keeping $\Omega_b$ fixed to its fiducial value. We denote a uniform distribution between values $a$ and $b$ using the shorthand $\mathcal U [a, b]$. } 
        \label{tab:fiducialcosmo}
    \end{table}
  \endgroup

  We infer parameters from this likelihood using the Markov Chain Monte-Carlo (MCMC) technique, with the affine-invariant sampler implemented in {\tt emcee} \citep{emcee2013}, and we use the Core Cosmology Library (\texttt{CCL}\footnote{\url{https://github.com/LSSTDESC/CCL}}) \citep{CCL} to carry out most of the theory calculations described below. Generating analytical predictions for the FSB can be computationally costly, given the relatively large numbers of triangle configurations corresponding to a given $\fsb{\ell}{LL}{}$, and the need to perform the convolution in Eq. \ref{eq:bsp2FSB}. Doing so on the fly at every step in an MCMC chain can be prohibitively expensive. We solve this problem by taking advantage of the simple scaling of the FSB and the power spectrum as functions of the model parameters (particularly $b_1$ and $\sigma_8$). Specifically, in the case of the FSB:
  \begin{enumerate}
    \item We generate the perturbation theory, quadratic and tidal terms templates (the terms in round brackets in Eqs. \ref{eq:bggg} and \ref{eq:bggk}) for a fixed cosmology, with parameters given in Table \ref{tab:fiducialcosmo}. To decrease computation time when calculating the projected bispectrum and transforming it into the FSB, we create interpolators for each of these theoretical templates at the level of the projected bispectrum, computing them on a 3D grid of predefined multipoles. These are then used to evaluate the bispectrum at every integer $(\ell_1,\ell_2,\ell_3)$ and transforming it into the FSB via Eq. \ref{eq:bsp2FSB}. All bias parameters are set to 1 when generating these templates. 
    \item We also generate templates for the different terms entering the galaxy-galaxy and galaxy-convergence power spectra, including all contributions from magnification (see Eqs. \ref{eq:clgg_mag} and \ref{eq:clgk_mag}).
    \item To account for the dependence on $\Omega_m$ we simply repeat the previous two steps for $100$ different values of this parameter, spanning the range $\Omega_m\in[0.05, 0.5]$, and generate a cubic spline interpolator for each FSB component as a function of $\Omega_m$. We provide an overview of the final interpolator performance, both in terms of runtime and accuracy, in Appendix \ref{app:interp}. 
    \item To generate predictions for a cosmology with a different value of $\sigma_8$ and $\Omega_m$, we simply rescale each of the templates described above, evaluated at the corresponding value of $\Omega_m$, by appropriate powers of the ratio $r_{\sigma_8}\equiv\sigma_8/\sigma_8^{\rm fid}$, where $\sigma_8$ is the sampled value of this parameter, and $\sigma_8^\mathrm{fid}$ is the fiducial value (given in Table \ref{tab:fiducialcosmo}). The $ggg$ and $gg\kappa$ bispectra both scale with $r_{\sigma_8}^4$, with the exception of the shot noise terms proportional to the power spectrum, which are scaled as $r_{\sigma_8}^2$, and the pure noise term in Eq. \ref{eq:noiseggg} which is independent of $\sigma_8$. In turn, all terms entering the $gg$ and $g\kappa$ power spectra scale with $r_{\sigma_8}^2$. 
    \item Finally, each bispectrum template is scaled by the right factors of $(b_1,b_2,b_s)$, given in Eqs. \ref{eq:bggg} and \ref{eq:bggk}. In the case of the power spectrum, the template $C_\ell^{gg,0}$ is scaled with $b_1^2$, while the templates $C_\ell^{g\kappa,0}$ and $C_\ell^{g\mu}$ are simply multiplied by $b_1$. This also applies to the shot-noise contributions to the bispectrum (Eqs. \ref{eq:noiseggg} and \ref{eq:noiseggk}).
\end{enumerate}
With this approach, each evaluation of the likelihood is relatively fast, allowing us to explore the constraining power of different combinations of power spectra and FSB terms. We reiterate that our model is relatively simple, including terms only at the tree level in both the power spectrum and bispectrum, and using empirical relations to connect the different bias parameters. The analysis presented here is thus only intended to represent a first exploration of the detectability of the projected bispectrum through the FSB approach, its qualitative agreement with theoretical expectations, and its constraining power. As such, the results presented here should be interpreted with a pinch of salt. A more rigorous cosmological analysis of these data is currently in progress and will be presented in future work \citep{2510.17796}. 

To ensure that our theoretical model is valid, we introduce scales cuts in the data vector: we only select scales below $\ell_\mathrm{max} = \chi(\bar z) k_\mathrm{max}$, where $\bar z$ is the effective redshift of the galaxy sample. The value of $k_\mathrm{max}$ depends on the observable: we set $k_\mathrm{max}^P = 0.07\,\mathrm{ Mpc}^{-1}$ for the power spectrum, and a more conservative $k_\mathrm{max}^B = 0.05\, \mathrm{ Mpc}^{-1}$ for the bispectrum. The value of $k_{\rm max}^P$ matches that used by \cite{2407.04607} in the analysis of the LRG sample used here assuming a linear bias relation. Although the tree-level matter bispectrum usually provides accurate fits to simulation results for scales up to $k_{\mathrm{max}}\lesssim 0.1\,h{\rm Mpc}^{-1}$ \citep[see e.g.][]{Takahashi2020}, our $k_\mathrm{max}^B$ cuts are motivated by \cite{ivanovPrecisionAnalysisRedshiftspace2022}, where a scale cut $k_\mathrm{max} = 0.08\,h\mathrm{ Mpc}^{-1}$ is used to model the tree-level bispectrum model in redshift space at $z=0.61$ (which lies between the first and second redshift bins used here, see Section \ref{ssec:data.lrgs}). 
For our fiducial cosmology described in Table \ref{tab:fiducialcosmo}, this cut corresponds to about $k_\mathrm{max}^B = 0.05\,\mathrm{ Mpc}^{-1}$, which in turn gives $\ell_\mathrm{max}^P = 130, \, 166, \, 200, \, 226$ and $\ell_\mathrm{max}^B = 93, \, 119, \, 143, \, 161$ in the four redshift bins studied here. We do not expect these conservative cuts to vary greatly for similar values of $\sigma_8$ or $h$. 
Note that this conservative $k_\mathrm{max}^B$ cut was necessary in \cite{ivanovPrecisionAnalysisRedshiftspace2022} due to the presence of Redshift-Space Distortions (RSDs), which do not affect 2D projected probes as much as their 3D counterparts. It is likely that this scale cut is thus conservative for our purposes, and determining the smallest physical scale up to which projected bispectra may be described by simple perturbation theory models would allow us to improve upon any constraints derived in this analysis.

The scale cut applies to all three legs ($\ell_1$, $\ell_2$ and $\ell_3$) of the bispectrum, and therefore we must make sure that their values satisfy $\ell_i < \ell_\mathrm{max}^B$. Such scale cuts are straightforward to implement thanks to the filtering operation introduced in the FSB estimator. Given that two of the legs ($\ell_1$ and $\ell_2$) can only belong to the scales selected by the filter $L$, one must choose filters that remain below the designated $\ell_{\mathrm{max}}^B$, such that $\mathrm{max} (\{\ell\in L\}) < \ell_\mathrm{max}^B$. If that condition is satisfied, we can then apply the usual scale cuts on the remaining leg $\ell_3$, only keeping values of $\fsb{\ell}{LL}{}$ which satisfy $\ell < \ell_\mathrm{max}^B$. These scale cuts would be difficult to implement on a regular skew-spectrum (i.e. in the absence of filtering before squaring the field): once the projected bispectrum is compressed along one dimension, individual triangle contributions cannot be differentiated. 

The final scale cut applied on the FSB is the minimum $\ell_\mathrm{max}$ set by either $k_\mathrm{max}^B$ or by the triangle inequality: the FSB filter used imposes a limit on the triangle configurations contributing to the estimator -- specifically, there cannot be contributions beyond twice the upper limit of the filter (we only keep values of $\fsb{\ell}{LL}{}$ that satisfy $\ell < 2\,\mathrm{ max} (\{\ell \in L\})$).

As described in Eq. \ref{eq:lkl}, we approximate the likelihood of the data vector $\mathbf d$ to a multivariate normal distribution, described by the data vector covariance $\mathsf C$. This is usually a reasonable approximation at high $\ell$ where the Central Limit Theorem (CLT) applies, but can fail at the largest scales; we therefore apply another fiducial low-$\ell$ cut at $\ell_\mathrm{min} = 20$ for all measured observables -- $\fsb{\ell}{LL}{ggg}$, $\fsb{\ell}{LL}{gg\kappa}$, $C_\ell^{gg}$ and $C_\ell^{g\kappa}$. This should also help mitigate any large-scale inaccuracies introduced by the Limber approximation and neglecting RSD contributions to both FSBs -- although the filter used to define the FSB does include large scales, below $\ell_{\rm min}$, in the bispectrum. As shown in \citet{2510.17796}, however, the larger statistical uncertainties in this statistic makes these effects negligible.
H25 furthermore showed that $\fsb{\ell}{LL}{ggg}$ and $C_\ell^{gg}$ in fact follow a near-Gaussian distribution, even for bandpowers including lower multipoles ($\ell>2$).

We use uninformative top-hat priors for the three free parameters of our model ($\{\sigma_8,\Omega_m,b_1\}$). These are listed in Table \ref{tab:fiducialcosmo}. All other cosmological parameters are fixed to the best-fit \planck values \citep{collaborationPlanck2018Results2020b}, also listed in the table.

  \subsection{Estimating detection significance}\label{ssec:methods.snr}
  To quantify the detection of a potential bispectrum signal in our measurements, we compute the Signal-to-Noise ratio (hereafter SNR), approximately given by 
\begin{equation} \label{eq:snr1}
    \mathrm{SNR} = \sqrt{\chi^2_0 - N_\mathrm{dof}}\,,\hspace{12pt} \chi_0^2\equiv(\mathbf x - \boldsymbol{\mu})^\mathsf{T} \mathsf{C}^{-1} (\mathbf x - \boldsymbol{\mu}),
\end{equation}
where $\mathbf x$ is one FSB measurement, $\boldsymbol{\mu}$ is the expectation value of the estimator for a null detection -- in our case, the noise terms given by either Eq. \ref{eq:noiseggg} or \ref{eq:noiseggk}. The covariance matrix $\mathsf C$ is calculated as described in Section \ref{sssec:methods.review.cov}. $N_\mathrm{dof}$ is the number of degrees of freedom, in this case the number of data points in $\mathbf x$, and represents the expectation value of the previous term in the case of pure noise fluctuations. This method constitutes a good first estimate of the SNR, in that it does not depend on any specific model for the signal we are after. When the SNR is sufficiently high, the argument of the square root is always positive, as $\chi^2_0\gg N_{\rm dof}$. In low-SNR scenarios, noise fluctuations may make the argument of the square root negative. In such cases, we will simply report ${\rm SNR}=0$. We will refer to this first model-independent estimate of the SNR as ``${\rm SNR}_1$'' in what follows.

Another way to compute the SNR would be to fit our measurements to a best-guess template of the expected signal with a free amplitude, and to quantify the significance with which this amplitude differs from zero. As a best-guess template ${\bf T}$ we use the theory predictions described in Section \ref{sssec:methods.th.bth}, using either the fiducial parameter values (Tables \ref{tab:fiducialcosmo} and \ref{tab:lrgs}) in Section \ref{ssec:results.detection} or the best-fit model parameters found in Section \ref{sssec:results.constraints.results}. Scaling this template by a free amplitude $A$ (i.e. ${\bf t}=A\,{\bf T}$, its best-fit and standard deviation can be estimated via generalised least-squares. The result can then be used to estimate the SNR of a given data vector ${\bf x}$:
\begin{equation} \label{eq:snr2}
    \mathrm{SNR} = \frac{A_\mathrm{best}}{\sigma_A} = \frac{\mathbf T^\mathsf{T} \mathsf C^{-1} \mathbf (\mathbf x - \boldsymbol{\mu})}{\sqrt{\mathbf T^\mathsf{T} \mathsf C^{-1} \mathbf T}}\,.
\end{equation} 
Note that here ${\bf T}$ contains only the bispectrum signal (i.e. the FSB version of Eq. \ref{eq:bggg} or \ref{eq:bggk}), with the noise components absorbed in $\boldsymbol{\mu}$, as in Eq. \ref{eq:snr1}. We will refer to this second SNR estimate as ``${\rm SNR}_2$'' in what follows. It is worth noting that this SNR estimate is only as accurate as the model used to construct ${\bf T}$ which, as we discussed in the previous section, may not be the case on small scales ($\ell\geq\ell_{\rm max}^B$).

\section{Data}\label{sec:data}

  \subsection{DESI LRGs}\label{ssec:data.lrgs}

  We study the clustering of photometric luminous red galaxies (LRGs) found in the 9th data release of the DESI Legacy Imaging Survey \citep{1804.08657,2208.08515}. The specific sample used here is described in \cite{zhouDESILuminousRed2023}, and was used by \cite{whiteCosmologicalConstraintsTomographic2022} \citep[and more recently by][]{2407.04607} to extract cosmological constraints from their two-point angular clustering and cross-correlation with CMB lensing data.
  
  The sample is divided into four different redshift bins, covering the redshift range $0.4\lesssim z \lesssim 1$, with number densities in the range $\bar{n}\sim 80$-$160\,{\rm deg}^{-2}$. We use the maps of the LRG overdensity in each redshift bin used in \cite{whiteCosmologicalConstraintsTomographic2022}, and made publicly available\footnote{\url{https://zenodo.org/records/5834378}} together with the analysis mask. We note that changes were made to the sample used in \cite{whiteCosmologicalConstraintsTomographic2022}, as described in \cite{zhouDESILuminousRed2023}, and this revised sample was then used in the cosmological analysis of \cite{2407.04607}. As explained in \cite{zhouDESILuminousRed2023}, the main changes made involve an improvement in the redshift distributions, modifications to the zero-point offsets in the southern sky, and changes in the sky contaminant templates used. This results in a small change in the LRG auto-correlations ($<2\%$). We have verified that none of our FSB measurements (neither $\fsb{\ell}{LL}{ggg}$ nor $\fsb{\ell}{LL}{gg\kappa}$) change significantly within the measurement uncertainties when using the newer sample. Hence, none of the results presented here, regarding the detectability and constraining power of the projected bispectrum, should be affected by these changes.

  We refer to each of the four redshift bins as $z_i$, with $i\in[1,4]$. The shot-noise levels, linear galaxy bias, and estimated number counts slope $s$, were presented in Table 1 of \cite{whiteCosmologicalConstraintsTomographic2022}. We repeat them in Table \ref{tab:lrgs} for convenience. We use the values of $s$ reported in this table to estimate the contribution from magnification bias to the measured power spectra.

  \begingroup 
    \setlength{\tabcolsep}{7pt} 
    \renewcommand{\arraystretch}{1.5} 
    \setlength\extrarowheight{2pt}
    
    \begin{table}[]
    \centering
    \begin{tabular}{cccccccc}
    \toprule
    Bin & $\bar{z}$ & $\bar{n}$ & $10^6N^{gg}$ & $s$ & $b_1$ & $\ell_{\rm max}^P$ & $\ell_{\rm max}^B$ \\
    \midrule
    $z_1$ & 0.47 & 83 & 4.02 & 0.982 & 1.9 & 130 & 93\\
    $z_2$ & 0.63 & 149 & 2.24 & 1.033 & 2.1 & 166 & 119\\
    $z_3$ & 0.79 & 162 & 2.07 & 0.961 & 2.4 & 200 & 143\\
    $z_4$ & 0.92 & 149 & 2.26 & 1.013 & 2.5 & 226 & 161\\
    \bottomrule
    \end{tabular}
    \caption{Properties of the LRG samples used in this analysis. Most of these are taken from Table 1 of \cite{whiteCosmologicalConstraintsTomographic2022}. They include the mean redshift of each sample $\bar{z}$, the mean number of galaxies per square degree $\bar{n}$, the shot-noise amplitude $N^{gg}$, the number counts slope $s$ used to estimate the impact of magnification bias, and the inferred large-scale Eulerian bias $b_1$. We also list the small-scale cuts used in our analysis of the power spectrum $\ell_{\rm max}^P$ and bispectrum $\ell_{\rm max}^B$.}\label{tab:lrgs}
  \end{table}

  \endgroup
  
  \subsection{Planck}\label{ssec:data.planck}
    We use maps of the CMB lensing convergence $\kappa$ constructed from the temperature and polarisation maps of the \planck experiment. In particular, we use the $\kappa$ maps presented in \cite{1807.06210}, reconstructed from the \planck 2018 (PR3) data release maps. In particular we use the minimum variance (MV) $\kappa$ map, which combines information from temperature and polarisation. These data are processed as described in \cite{whiteCosmologicalConstraintsTomographic2022}: the public harmonic coefficients $\kappa_{\ell m}$ are first band-limited and transformed into a {\tt HEALPix} map, and the analysis mask is apodised with a $0.5^\circ$ kernel to reduce the statistical coupling between adjacent multipoles in the estimated power spectra.

    As noted in \cite{2309.05659,2407.04606,2407.04607}, masking, filtering, and specific analysis choices in lensing reconstruction affect the effective normalisation of the reconstructed $\kappa$ map in an inhomogeneous manner. This may be corrected via Monte-Carlo simulations (and hence the associated transfer function is often called the ``MC correction''), but the correction depends on the sky patch under study, and must therefore be recalculated for each cross-correlation. As shown in \cite{2407.04607}, this correction is at the level of $\sim1\%$ of $C_\ell^{g\kappa}$ in the case of the LRG sample used here, and including it leads to a $\sim0.5\sigma$ downwards shift in the value for the amplitude of matter fluctuations recovered by combining information from all redshift bins (corresponding to a $\sim0.3\sigma$ shift in the case of the per-bin constraints we report in Section \ref{ssec:results.constraints}). Since this is commensurate with the impact of other analysis choices, including the treatment of neutrino masses and volume effects \citep[see][]{2407.04607}, we chose not to apply this correction to $C_\ell^{g\kappa}$ in this analysis, for simplicity and in order to facilitate the comparison with the results in \cite{whiteCosmologicalConstraintsTomographic2022}. Furthermore, the impact of this correction on the new measurement of $\fsb{\ell}{LL}{gg\kappa}$ presented here is entirely negligible, given the significantly lower SNR of this measurement compared to $C_\ell^{g\kappa}$ (see Section \ref{ssec:results.detection}).

\section{Results}\label{sec:results}
  In this section, we report the detection of the galaxy bispectrum in all redshift bins introduced in Section \ref{ssec:data.lrgs}, as well as the detection of the galaxy-convergence bispectrum in three out of the four redshift samples. We also discuss the constraining power of the galaxy and galaxy-convergence bispectra, and provide the first (albeit rudimentary) cosmological constraints on the clustering amplitude $\sigma_8$ and the matter density $\Omega_m$ combining power spectra and the projected galaxy bispectrum.

  \subsection{Measurement and detection}\label{ssec:results.detection}

    \begin{figure*}[p]
      \sbox0{\begin{tabular}{@{}c@{}}
      \includegraphics[width=\textheight]{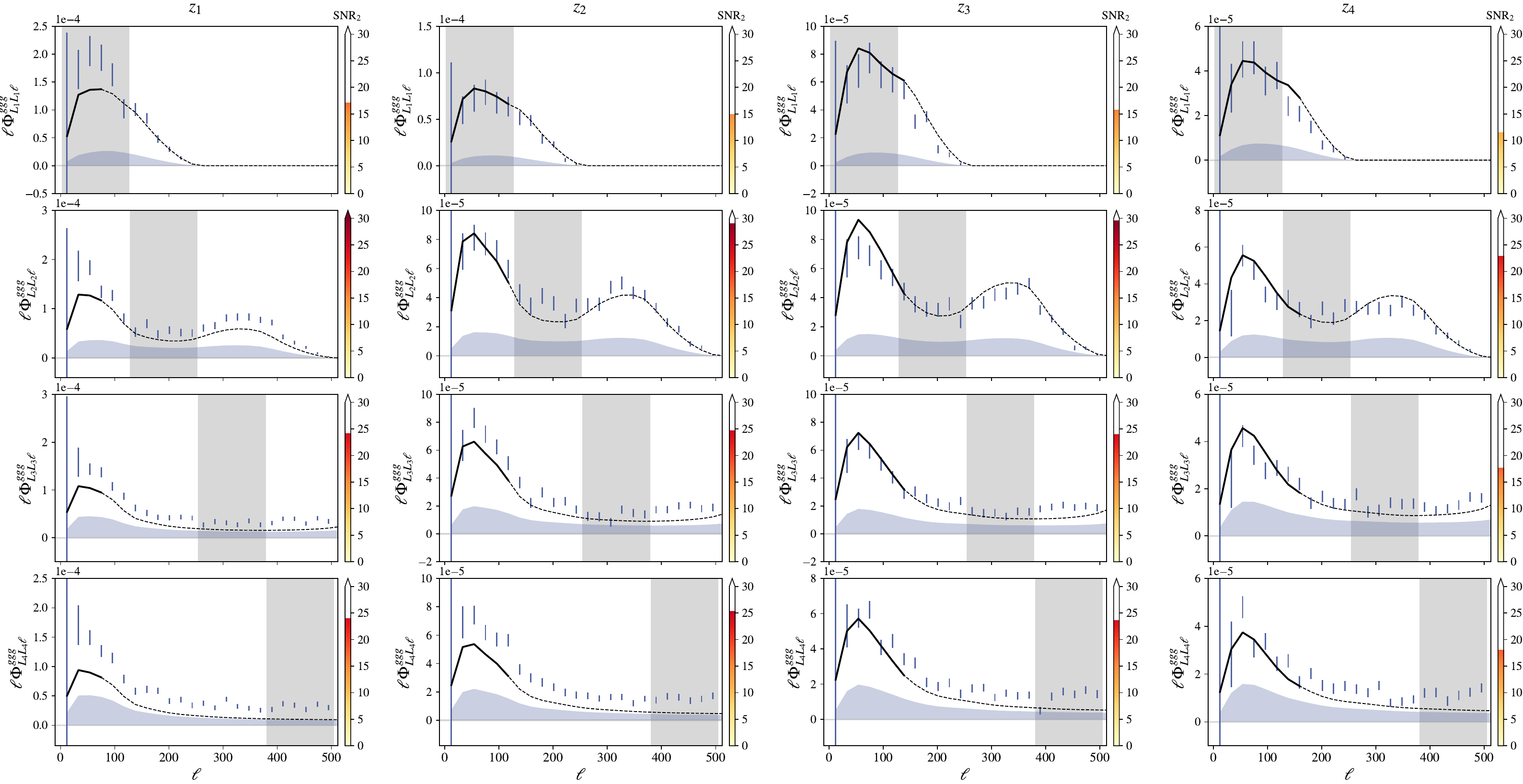}
      \end{tabular}}
      \rotatebox{-90}{\begin{minipage}[c][\textwidth][c]{\wd0}
      \usebox0
      \caption{The $ggg$ FSBs measured in all four redshift bins. Each column corresponds to one redshift bin, and each row to one of the four colourful filters shown in Fig. \ref{fig:filters} -- here represented by the greyed-out band. The FSB measurements are shown in blue, and the black solid line shows the theoretical fit for published values of $b_1$ \citep{whiteCosmologicalConstraintsTomographic2022} and for a fiducial cosmology (Table \ref{tab:fiducialcosmo}). These theory predictions are used as templates in Eq. \ref{eq:snr2} to report the $\text{SNR}_2$ shown on the colour bars on the right of each FSB. Beyond the conservative cuts used in our inference pipeline (see Section \ref{ssec:results.constraints}), we show the predictions as dashed lines to emphasise that the tree-level bispectrum is not necessarily a good theoretical description of the power at those scales. It is worth noting that the FSBs in all but the first filter (i.e. the first row in this plot) also lie in this small-scale regime. The blue shaded area represents the amount of noise (Eq. \ref{eq:noiseggg}) contributing to the overall theory prediction. Note that we plot $\ell \fsb{\ell}{LL}{ggg}$ instead of $\fsb{\ell}{LL}{ggg}$ to improve readability at higher multipoles.} \label{fig:gggdetect}
      \end{minipage}}
    \end{figure*} 

    \begin{figure*}[p]
    \sbox0{\begin{tabular}{@{}c@{}}
    \includegraphics[width=\textheight]{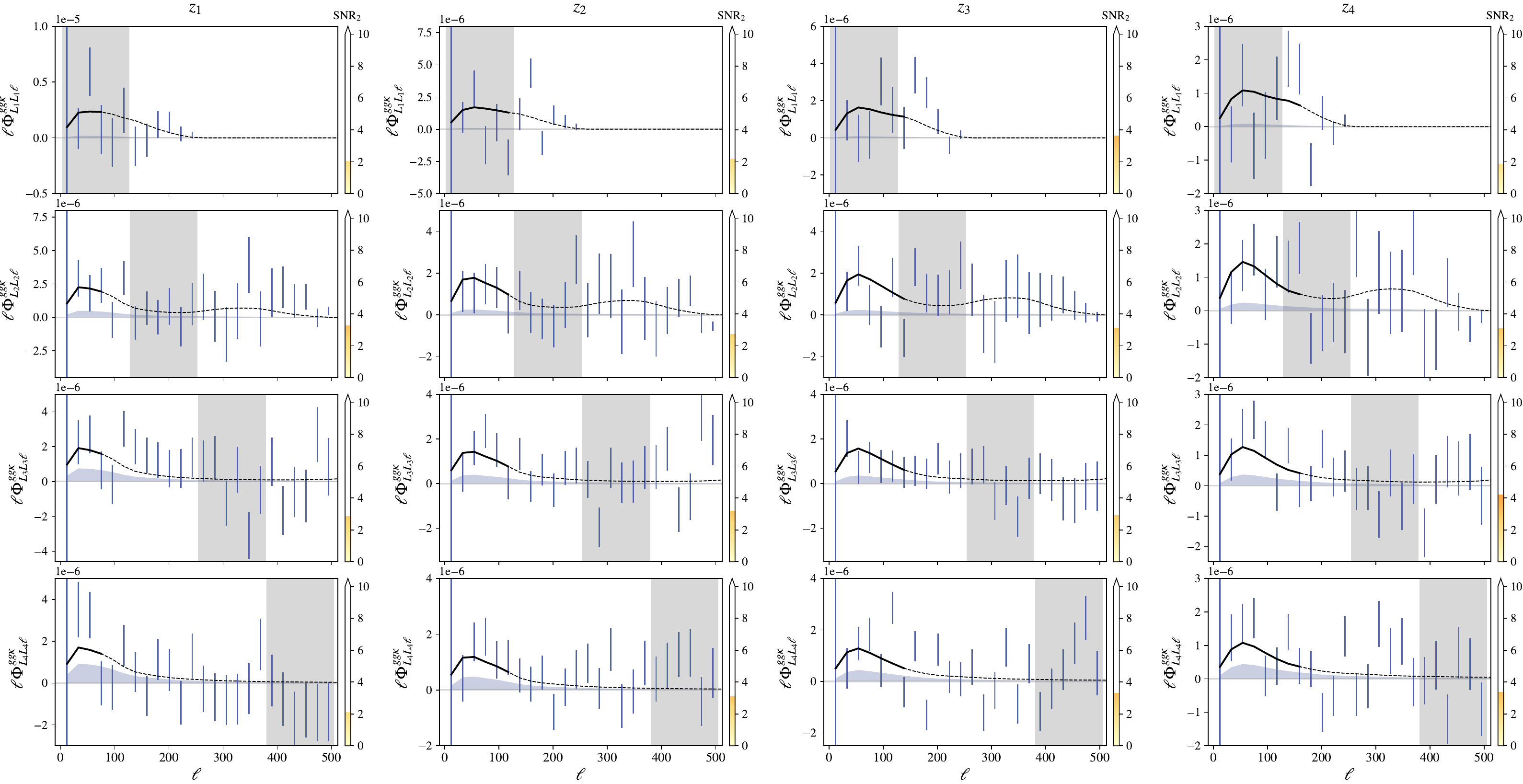}
    \end{tabular}}
    \rotatebox{-90}{\begin{minipage}[c][\textwidth][c]{\wd0}
    \usebox0
    \caption{The $gg\kappa$ FSBs measured in all four redshift bins. Each column corresponds to one redshift bin, and each row to one of the four colourful filters shown in Fig. \ref{fig:filters} -- here represented by the greyed-out band. The FSB measurements are shown in blue, and the black solid line shows the theoretical fit for published values of $b_1$ \citep{whiteCosmologicalConstraintsTomographic2022} and for a fiducial cosmology (Table \ref{tab:fiducialcosmo}). These theory predictions are used as templates in Eq. \ref{eq:snr2} to report the $\text{SNR}_2$ shown on the colour bars on the right of each FSB. Beyond the conservative cuts used in our inference pipeline (see Section \ref{ssec:results.constraints}), we show the predictions as dashed lines to emphasise that the tree-level bispectrum is not necessarily a good theoretical description of the power at those scales. The blue shaded area represents the amount of noise (Eq. \ref{eq:noiseggk}) contributing to the overall theory prediction. Note that we plot $\ell \fsb{\ell}{LL}{gg\kappa}$ instead of $\fsb{\ell}{LL}{gg\kappa}$ to improve readability at higher multipoles.} \label{fig:ggkdetect}
    \end{minipage}}
    \end{figure*}

    We measure the FSBs using the estimator described in Sections \ref{ssec:methods.review} and \ref{ssec:methods.multifieldfsb}, using the filters and bandpowers described in Section \ref{ssec:methods.maps}. We generate basic theoretical predictions (Section \ref{sssec:methods.th.bth}) for these FSBs by rescaling the bispectrum templates according to the method laid out in Section \ref{ssec:methods.like}. These were generated using the values of the linear bias found in \cite{whiteCosmologicalConstraintsTomographic2022} (see Table \ref{tab:lrgs}), calculating the quadratic and tidal bias parameters using the coevolution relations, and setting all cosmological parameters to the best-fit values found by \planck (see Table \ref{tab:fiducialcosmo}). The resulting measurements of $\fsb{\ell}{LL}{ggg}$ and $\fsb{\ell}{LL}{gg\kappa}$, together with these theoretical predictions, are shown in Figs. \ref{fig:gggdetect} and \ref{fig:ggkdetect}, respectively. In each panel, the blue shaded band shows the contribution from shot noise, the vertical gray band marks the filter used to calculate the FSB, and the vertical colour bar shows the SNR with which the signal is detected, which we discuss below.

We expect the tree-level description of the galaxy bispectrum to be reliable when all three legs $(\ell_1, \ell_2, \ell_3)$ remain below the fiducial multipole cuts we describe in Section \ref{ssec:methods.like}. This regime corresponds to only a subset of Figs. \ref{fig:gggdetect} and \ref{fig:ggkdetect}: 
\begin{itemize}
    \item Only the first filter selects scales $(\ell_1, \ell_2)$ below these scale cuts, whereas the three other filters select smaller scales where linear perturbation does not necessarily hold; 
    \item We show the predictions as bold dark lines where $\ell_3$ (which corresponds to the x-axis) falls below the aforementioned scale cuts, and as dashed lines beyond. 
\end{itemize}
Therefore, our theoretical predictions should be most reliable in the first row of each figure, and where theoretical predictions are shown in bold.
    The naive theoretical expectation, with no free parameters, provides a reasonable description of the $ggg$ FSBs within these limits for all except in the first redshift bin, where it systematically underpredicts the signal. As we discuss in Section \ref{sssec:results.constraints.intersections}, this can be accommodated by the combination of a higher value of the linear bias and a lower $\sigma_8$ \citep[the latter in qualitative agreement with the lower clustering amplitude seemingly preferred by this sample in][]{whiteCosmologicalConstraintsTomographic2022}. Interestingly, the theoretical predictions are surprisingly accurate even beyond these scale cuts: in the first and second row of Figs. \ref{fig:gggdetect} and \ref{fig:ggkdetect}, the tree-level bispectrum still provides a somewhat reasonable fit to the measurements over an extended multipole range.
    It would be interesting to determine the range of scales over which the projected bispectrum can be safely predicted. Since the projected galaxy overdensity is necessarily more Gaussian than its 3D counterpart, it may be possible to extend the analysis to smaller angular scales, thereby recovering more cosmological information. We leave this study for future work. For now we will remain cautious and implement conservative scale cuts for the analysis in Section \ref{ssec:results.constraints}. However, to quantify the significance with which the bispectrum signal is detected in these measurements, independently of our ability to accurately model it, we will use the full range of scales $\ell<2N_{\rm side}$. Visually, the detection of the non-Gaussian contributions to the FSB can be seen by comparing our measurements with the blue band showing the expected contribution from shot noise.

    The SNR with which the $ggg$ and $gg\kappa$ FSBs are detected in each redshift bin, combining the signal from all FSB filters, are reported in Table \ref{tab:snr}. The combined SNR is computed following the two methods described in Section \ref{ssec:methods.snr}, using as input the concatenated data vector containing the FSB measurements in all filters.  We use the theoretical predictions shown in Figs. \ref{fig:gggdetect} and \ref{fig:ggkdetect} (or rather their pure signal contributions) to build the theoretical template used to estimate ${\rm SNR}_2$. It is worth noting that, since the theoretical template is not accurate on small scales (this is clearly visible in the figures as the FSB filter moves to higher $\ell$s and higher filters -- as seen also in H25), the approach used to estimate ${\rm SNR}_2$ may become unreliable. Thus, we also provide the estimates of ${\rm SNR}_1$, which do not rely on any theoretical template (but instead become unreliable for low-SNR data).

    \begingroup 
    \setlength{\tabcolsep}{10pt} 
    \renewcommand{\arraystretch}{1.5} 
    \setlength\extrarowheight{2pt}
        \begin{table}[]
        \centering
        \begin{tabular}{cccc}
        \toprule
        Bin & FSB type & $\mathrm{SNR}_1$ & $\mathrm{SNR}_2$ \\
        \midrule
        \multirow{2}{*}{$z_1$} & $ggg$ & 45.06 & 38.37 \\
        & $gg\kappa$ & 2.39 & 3.53 \\ \midrule
        \multirow{2}{*}{$z_2$} & $ggg$ & 43.82 & 39.27 \\
        & $gg\kappa$ & 6.06 & 4.60 \\ \midrule
        \multirow{2}{*}{$z_3$} & $ggg$ & 43.65 & 40.08 \\
        & $gg\kappa$ & 7.45 & 5.44 \\ \midrule
        \multirow{2}{*}{$z_4$} & $ggg$ & 33.58 & 30.70 \\
        & $gg\kappa$ & 6.50 & 5.04 \\ 
        \bottomrule
        \end{tabular}
        \caption{Combined SNRs corresponding to the FSBs shown in Figs. \ref{fig:gggdetect} and \ref{fig:ggkdetect}.}
        \label{tab:snr}
        \end{table}
    \endgroup
    
      \begin{figure*}
        \centering 
        \includegraphics[width=0.7\textwidth]{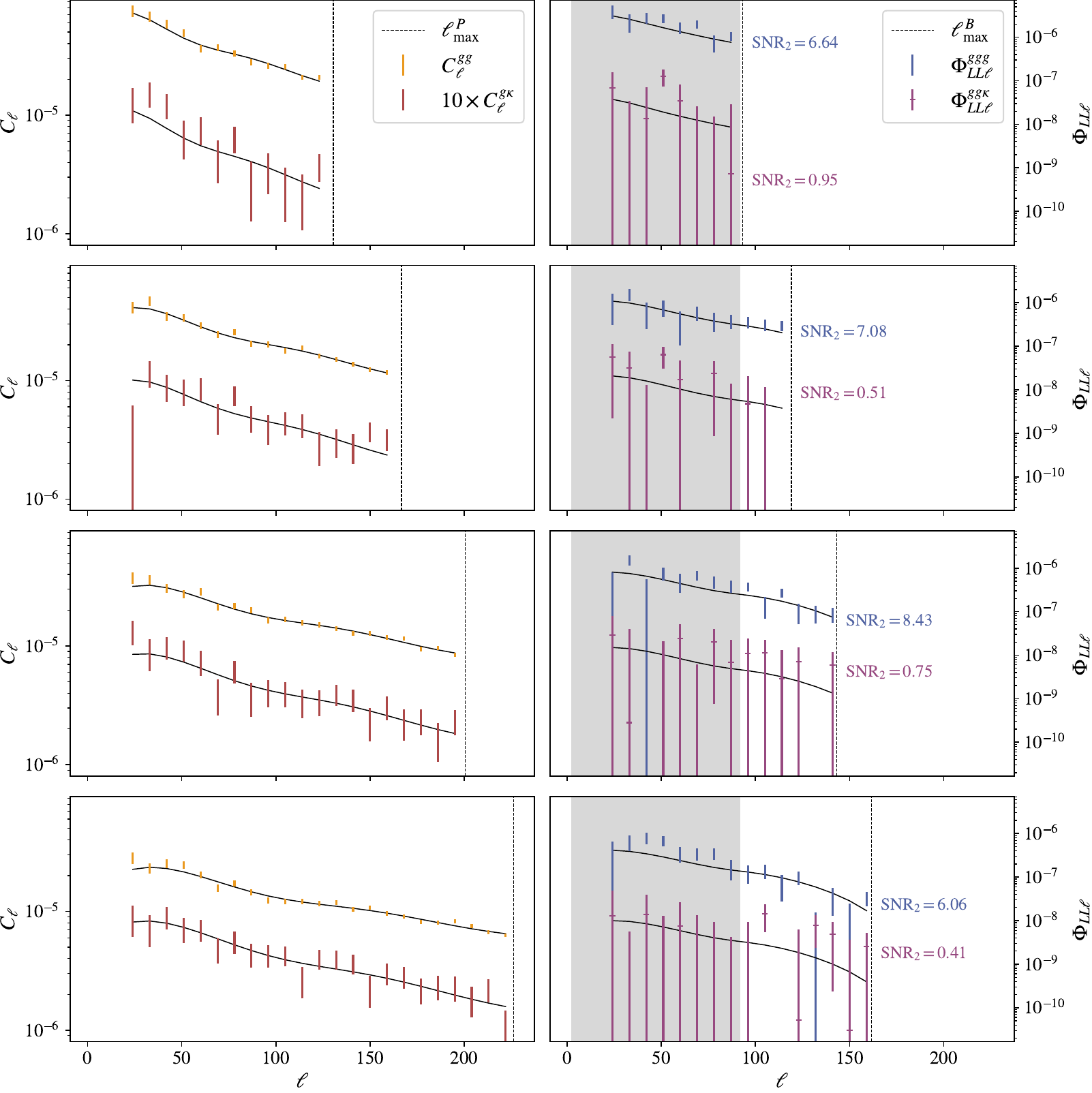}
        \caption{A visual representation of the data vectors used to derive constraints on $\sigma_8$ and $\Omega_m$, with power spectra $C_\ell^{gg}$ (yellow data points) and $C_\ell^{g\kappa}$ (red, multiplied by a factor of $10$ for readability) measurements on the left, and FSBs $\fsb{\ell}{LL}{ggg}$ (blue) and $\fsb{\ell}{LL}{gg\kappa}$ (purple) on the right. Each row corresponds to a different redshift bin, starting from lowest redshift at the top. The multipole cuts evolve accordingly (the number of data points included in the analysis thus increases with redshift). The new filtered range (for a comparison with previous filters, see Fig. \ref{fig:filters}) is shown in grey, and fits within the most restrictive $\ell_\mathrm{max}^B$ cut (lowest redshift bin). The black lines show theory predictions for the best-fit values obtained from constraining the models with the full data vector, $\fsb{\ell}{LL}{ggg} + C_\ell^{gg} + C_\ell^{g\kappa}$. These best-fit values are reported in Table \ref{tab:cosmo}, and the corresponding predictions are used as templates to report $\text{SNR}_2$ (Eq. \ref{eq:snr2}) indicated next to each FSB.} \label{fig:datavectors}
      \end{figure*}

    The detection of the $ggg$ FSB is indisputable, with a combined $\mathrm{SNR} > 30$ in all redshift bins. We find that both estimates, ${\rm SNR}_1$ and ${\rm SNR}_2$ are in qualitative agreement, with ${\rm SNR}_1$ providing a consistently larger estimate (by $\sim10$-$20\%$). This is likely due to the fact that the model used to create the theoretical template ${\bf T}$ used to calculate ${\rm SNR}_2$ (see Eq. \ref{eq:snr2}) under-predicts the signal on small scales, leading to a lower inferred signal amplitude.
 
    In turn, the SNR of the $gg\kappa$ FSBs is significantly lower. Nevertheless, the signal is clearly detected, with ${\rm SNR}>5$ in the three last redshift bins, and only tentatively in the first bin (${\rm SNR}\sim3$). Judging from Fig. \ref{fig:ggkdetect}, much of the signal seems to be concentrated in the higher-$\ell$ filters, which will not be used in the cosmological analysis presented in the next section. Thus, although detected, we do not expect $\fsb{\ell}{LL}{gg\kappa}$ to contribute significantly to the final cosmological constraints.

  \subsection{Constraining power}\label{ssec:results.constraints}
    \subsubsection{FSB measurements}\label{sssec:results.constraints.datavectors}
      As described in Section \ref{ssec:methods.like}, we implement conservative scale cuts to the data to ensure our theoretical predictions are reliable. This means we must also make sure the FSB filters chosen are such that all triangle configurations contributing to the estimator satisfy these cuts. As described in Section \ref{ssec:methods.maps}, to ensure this, we use new measurements of the FSB, using a single filter lying below the most conservative $\ell_\mathrm{max}^B$ cut (corresponding to the lowest redshift bin $z_1$, shown in Fig. \ref{fig:filters}), and reduce the bandpower width to $\Delta\ell = 9$. Our data vector, for a given redshift bin, will therefore consist of a single $\fsb{\ell}{LL}{ggg}$ and $\fsb{\ell}{LL}{gg\kappa}$ for this filter, in addition to the power spectrum measurements $C_\ell^{gg}$ and $C_\ell^{g\kappa}$. These measurements are shown in Fig. \ref{fig:datavectors} for the four redshift bins, truncated at their respective $\ell_\mathrm{max}$ cuts. Unsurprisingly, as discussed in the previous section, the SNR of the $\fsb{\ell}{LL}{gg\kappa}$ measurements within these scales is very low (below 1 in all cases). We will therefore discard the $gg\kappa$ measurements in our analysis, keeping only the $ggg$ FSBs, which are detected above $6\sigma$ in all cases. Increasing the SNR should allow us to make the most of the constraining power of the $gg\kappa$ FSBs. This could be achieved by either exploiting all scales within the studied multipole range (allowing for the filter to be defined up to $\ell_\mathrm{max}^B(z)$ in each bin, instead of using the same, large-scale filter for all) or pushing our model to smaller scales (thus allowing us to relax the conservative scale cuts used so far in this study). Using more sensitive CMB lensing data, from ground-based experiments such as the Atacama Cosmology Telescope \citep{2304.05203}, could also improve the SNR of these measurements. We will study these potential avenues for optimisation in future work.

      We also add in Fig. \ref{fig:datavectors} the best-fit theoretical predictions given by the constraints obtained in Section \ref{sssec:results.constraints.results} for the combined probes $\fsb{\ell}{LL}{ggg}$, $C_\ell^{gg}$, and $C_\ell^{g\kappa}$. As we will quantify in Section \ref{sssec:results.constraints.results}, these provide an excellent fit to our measurements over all the scales studied.

    \subsubsection{Probe complementarity}\label{sssec:results.constraints.intersections}
      \begin{figure}
      \centering
      \includegraphics[width=.85\linewidth]{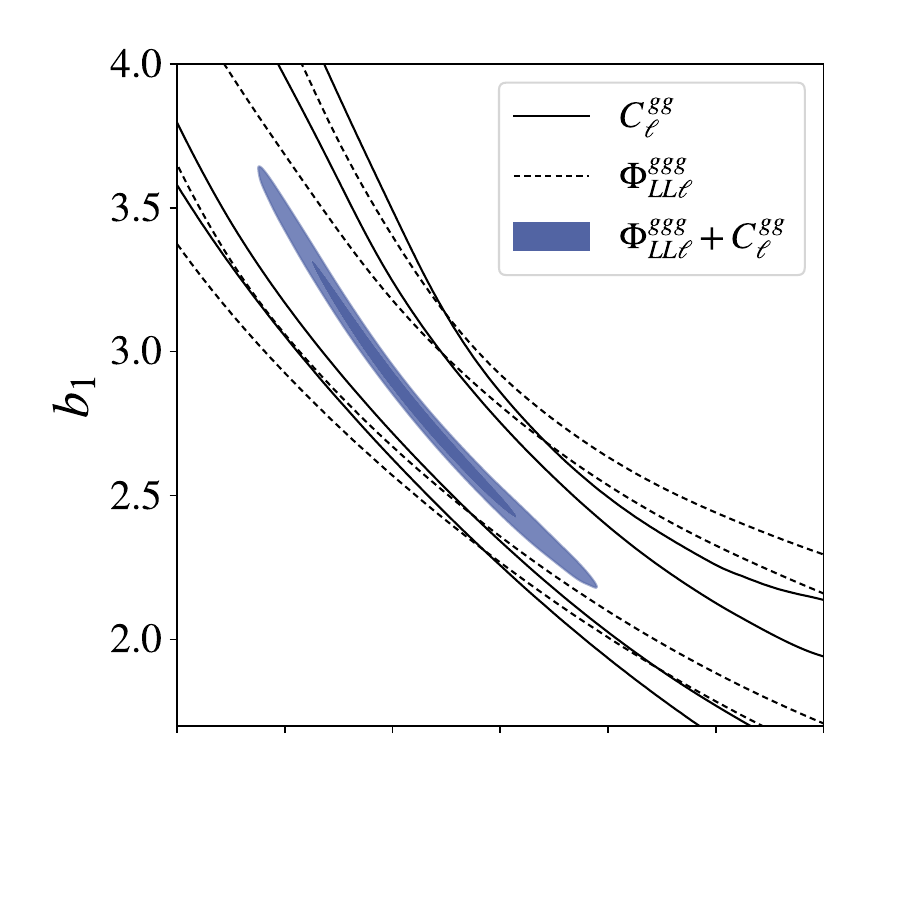}\vspace{-12.5mm}
      \includegraphics[width=.85\linewidth]{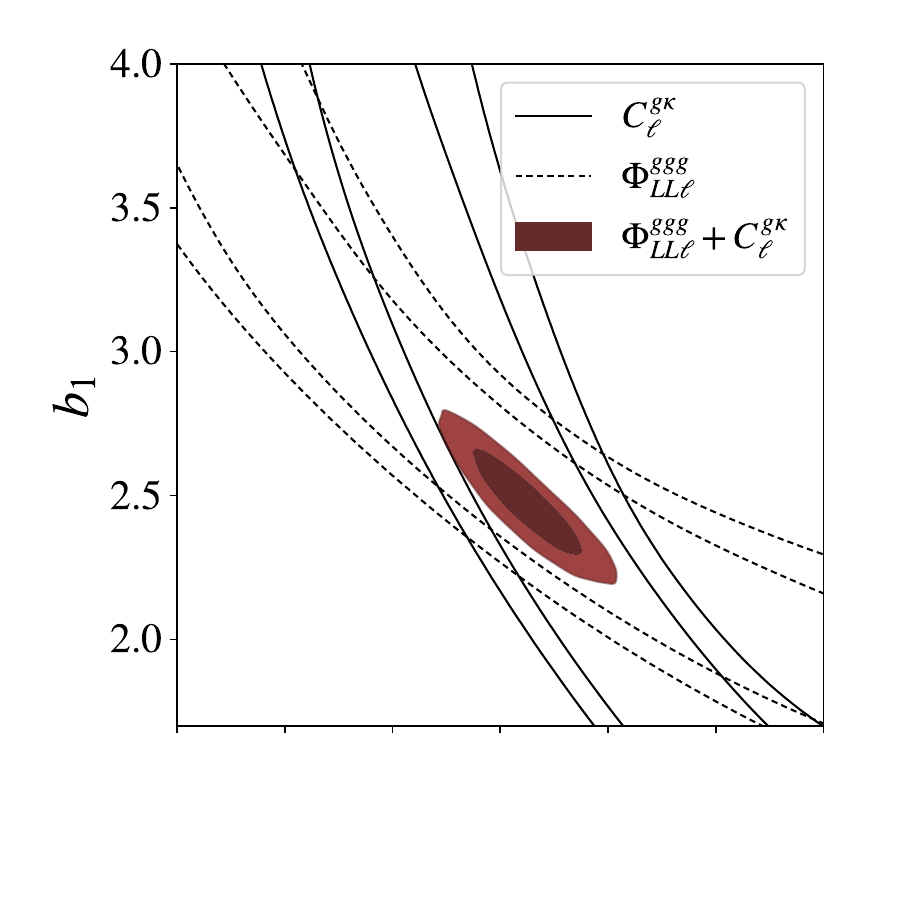}\vspace{-12.5mm}
      \includegraphics[width=.85\linewidth]{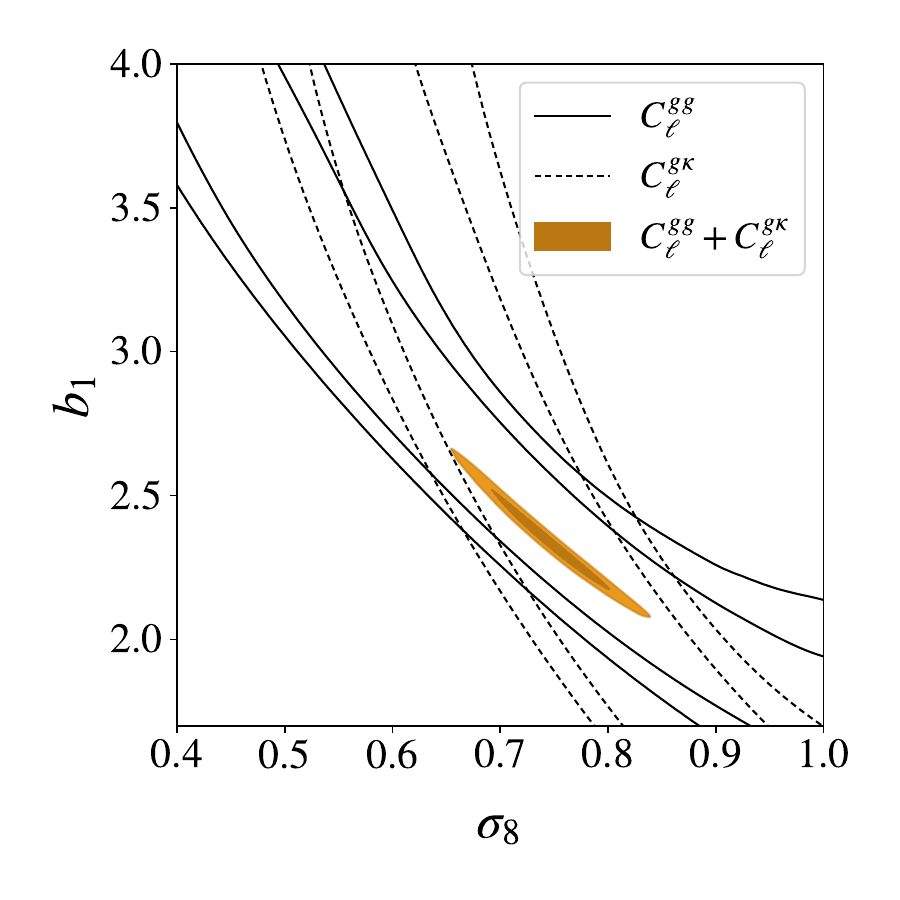}
      \caption{Constraints on $\sigma_8$ and $b_1$ from three data combinations: $\fsb{\ell}{LL}{ggg} + C_\ell^{gg}$ (blue contours in top panel), $\fsb{\ell}{LL}{ggg} + C_\ell^{g\kappa}$ (red contours, middle panel), and $C_\ell^{gg} + C_\ell^{g\kappa}$ (yellow contours, bottom panel). The black contours show the degeneracy directions of individual power spectrum or FSB measurements, with the filled contours showing the combined constraints. The constraints shown were obtained for redshift bin $z_3$.} \label{fig:intersections}
      \end{figure}
      As a pedagogical detour, before presenting a cosmological analysis of the data vector presented in the last section, let us study the ability of the galaxy bispectrum to break parameter degeneracies and improve cosmological constraints. To do so, let us use a simpler theoretical model in which we keep $\Omega_m$ fixed to its fiducial value (see Table \ref{tab:fiducialcosmo}), while varying both $\sigma_8$ and $b_1$, the two parameters that affect the overall amplitude of the clustering statistics (both $C_\ell$s and FSBs). For simplicity, we will also focus on the third redshift bin, $z_3$, alone, where the $ggg$ FSB has the highest SNR.

      We can exploit the different dependence on $b_1$ and $\sigma_8$ of the power spectra and bispectra. For simplicity, and without loss of generality, let us ignore the contributions from magnification bias, as well as the bispectrum contributions dependent on the quadratic and tidal bias parameters. At the tree level, the amplitude of the different power spectra and FSBs considered here depends on $\sigma_8$ and $b_1$ as follows:
      \begin{align}
        C_\ell^{gg}\propto\sigma_8^2b_1^2, \hspace{12pt} & C_\ell^{g\kappa}\propto\sigma_8^2b_1,\\
        \fsb{\ell}{LL}{ggg} \propto\sigma_8^4b_1^\alpha,\hspace{12pt}
        & \fsb{\ell}{LL}{gg\kappa} \propto\sigma_8^4b_1^\beta,
      \end{align}
      with $\alpha=3$ and $\beta=2$ ignoring contributions from $b_2$ and $b_s$. The complementary scaling with $b_1$ and $\sigma_8$ of these spectra makes it possible to constrain both parameters by combining any two of them\footnote{Except for the combination of $C_\ell^{g\kappa}$ and $\fsb{\ell}{LL}{gg\kappa}$, which constrain only the combination $\sigma_8^2b_1$ in both cases. In this sense, the loss of the $gg\kappa$ FSB due to its poor SNR is not a major setback for this analysis.}. In practice, however, the scaling of the FSB with $b_1$ is more complicated due to the non-linear bias contributions at tree level that are the same order of magnitude as the terms containing only $b_1$. In the next section we will assume co-evolution relations, which express $b_2$ and $b_s$ as a function of $b_1$. In this case, we find that at the characteristic scale $\ell\sim75$ the effective values of $\alpha$ and $\beta$ in each of the redshift bins are $\alpha\simeq4.5$, $\beta\simeq3.1$, although in general these values are scale-dependent and sample-dependent. Given that co-evolution equations are good, but imperfect approximations, it is necessary to marginalize over all three bias parameters, perhaps with an informative prior on bias relations.

      \begin{figure*}
        \centering
        \includegraphics[width=0.4\linewidth]{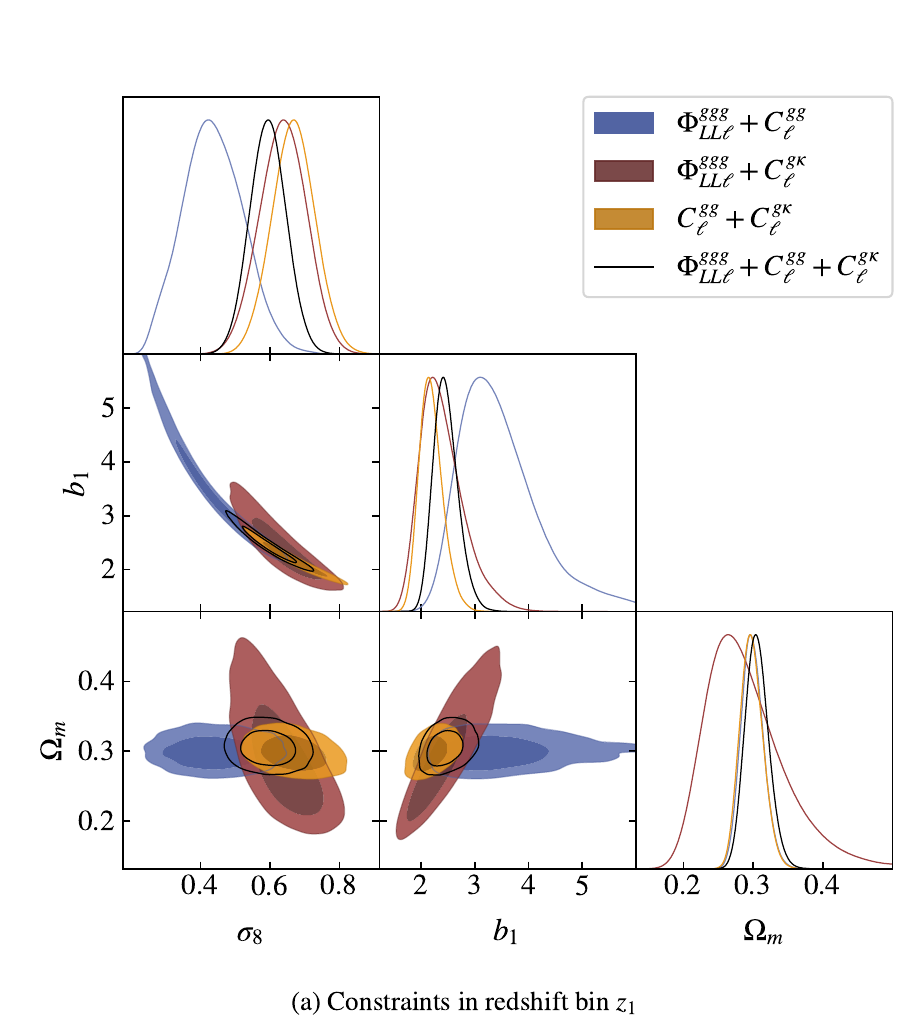}
        \includegraphics[width=0.4\linewidth]{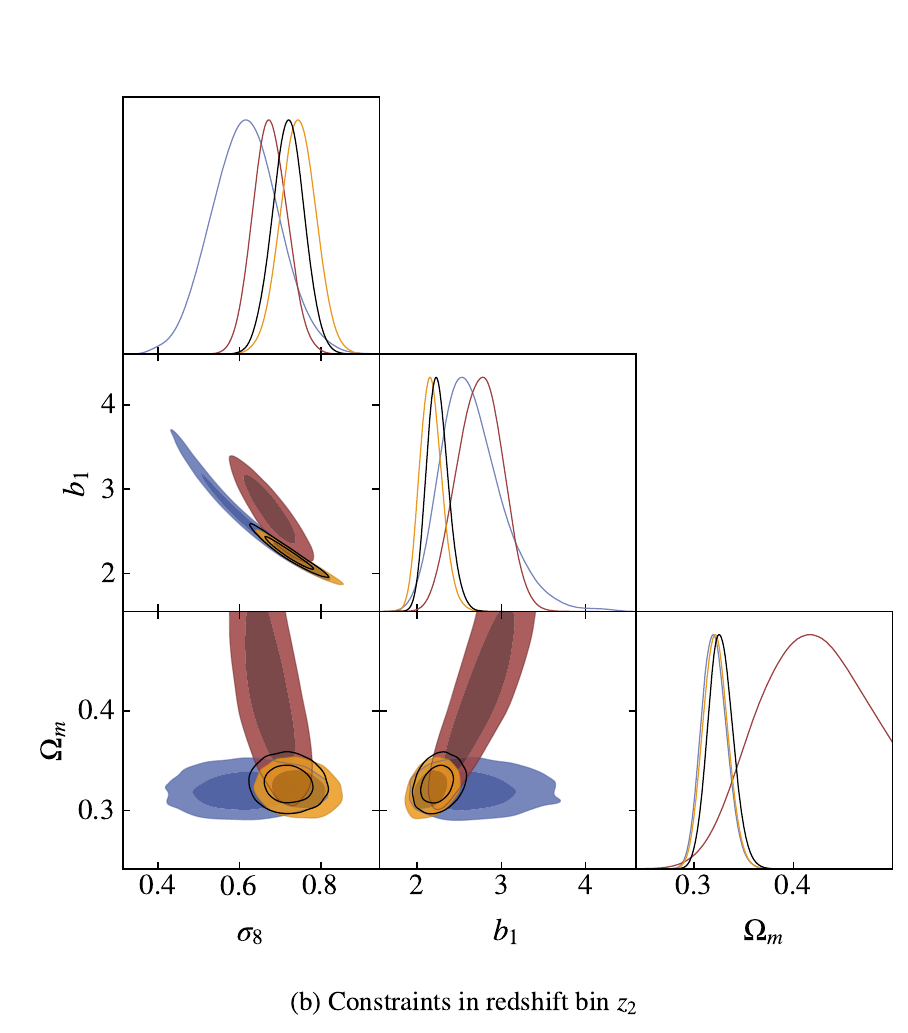}
        \includegraphics[width=0.4\linewidth]{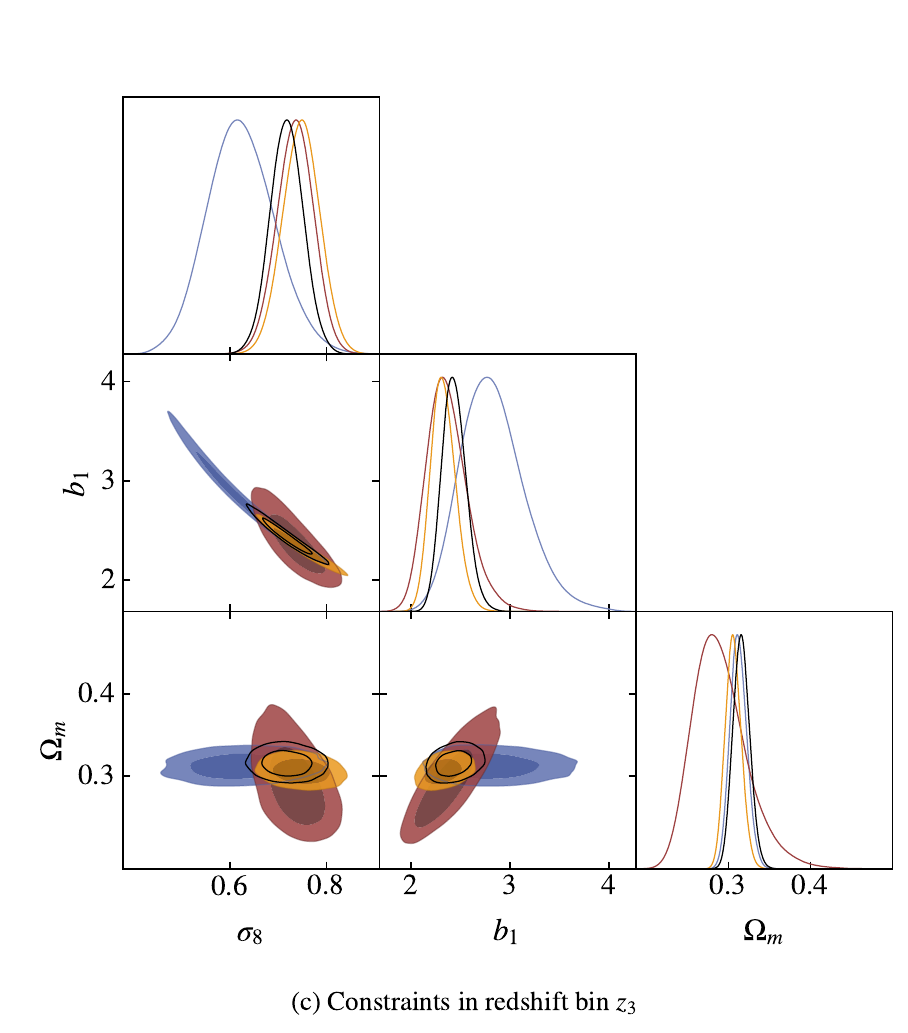}
        \includegraphics[width=0.4\linewidth]{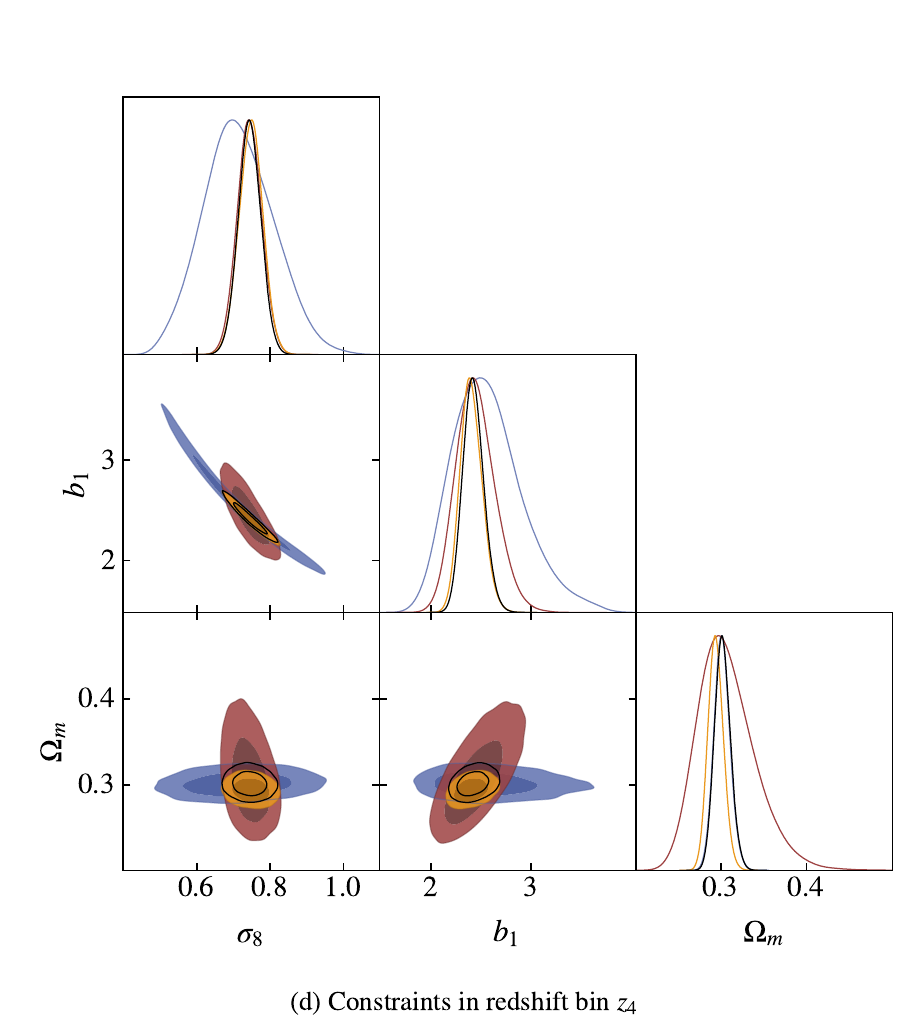}
        \caption{Constraints on $\sigma_8$, $\Omega_m$ and $b_1$ found in the four different redshift bins. The yellow contours show constraints derived from the standard \txt combination, while the red and blue contours show two combinations of FSBs and power spectra which lead to additional constraints. The black contours show constraints obtained by combining all three probes ($C_\ell^{gg}+C_\ell^{g\kappa}+\fsb{\ell}{LL}{ggg}$). We only show the area of parameter space allowed by the priors.} 
        \label{fig:4constraints}
      \end{figure*}

       In general, for the FSB measurements to be sufficiently complementary to $C_\ell^{gg}$ and $C_\ell^{g\kappa}$ in breaking parameter degeneracies, these scaling indices should be sufficiently different from $\alpha=4$ and $\beta=2$. Although our measurement of the $gg\kappa$ FSB is not sufficiently sensitive, we can exploit the $ggg$ FSB to break degeneracies in the $(\sigma_8, b_1)$ parameter plane using this rationale. Fig. \ref{fig:intersections} illustrates how, by combining $\fsb{\ell}{LL}{ggg}$ with either $C_\ell^{gg}$ or $C_\ell^{g\kappa}$, we are able to constrain $\sigma_8$. For the third redshift bin, the resulting constraints on $\sigma_8$ are\footnote{We report $68\%$ confidence level intervals throughout.}:
      \begin{align}
        \sigma_8 = 0.619^{+0.066}_{-0.073}, \hspace{12pt} & \left(C_\ell^{gg} + \fsb{\ell}{LL}{ggg} \right), \\
        \sigma_8 = 0.724\pm 0.034, \hspace{12pt} & \left(C_\ell^{g\kappa}+\fsb{\ell}{LL}{ggg} \right),\\
        \sigma_8 = 0.745\pm 0.038, \hspace{12pt} & \left(C_\ell^{gg} + C_\ell^{g\kappa}\right).
      \end{align}
      In the top panel of Fig. \ref{fig:intersections} we can see that the degeneracy directions for $C_\ell^{gg}$ and $\fsb{\ell}{LL}{ggg}$ are similar, but not exactly parallel. As we just saw, this is because $\alpha\simeq4.5$ is close to, but different from the value $\alpha=4$ that would make both datasets equivalent in terms of their $\sigma_8$-$b_1$ scaling. This allows us to obtain a measurement of $\sigma_8$ using the clustering of galaxies alone, albeit with an uncertainty that is approximately twice as large as that obtained in combination with $C_\ell^{g\kappa}$. Interestingly, the combination of $C_\ell^{g\kappa}+\fsb{\ell}{LL}{ggg}$ is able to constrain $\sigma_8$ with a similar precision to the \txt combination (and in fact with $10\%$ smaller uncertainties, likely due to the steeper scaling of the FSB with $b_1$ than $C_\ell^{gg}$). The constraints obtained with these different data combinations are compatible with one another at the 1-2$\sigma$ level, with the galaxy-only combination ($gg+ggg$) returning the lowest value of $\sigma_8$. We remind the reader that these constraints must be interpreted with care, due to the overly simplistic nature of the model used: not only have we kept $\Omega_m$ constant, we have also assumed a fixed relationship between the non-linear bias parameters and $b_1$ which, as we have just seen, directly affect the scaling of the bispectrum amplitude that allows us to break parameter degeneracies. These constraints are provided here only to illustrate the complementary nature of the galaxy projected bispectrum.

      These two new probe combinations ($C_\ell^{gg}+\fsb{\ell}{LL}{ggg}$ and $C_\ell^{g\kappa}+\fsb{\ell}{LL}{ggg}$) have exciting advantages in terms of robustness to potential systematics. First, an independent measurement of $\sigma_8$ can be obtained in the absence of any lensing information, by combining $C_\ell^{gg}+ \fsb{\ell}{LL}{ggg}$ (blue contours), thus circumventing lensing systematics altogether, and using only data that is sensitive to structure in a relatively narrow range of redshifts. The $C_\ell^{g\kappa}+ \fsb{\ell}{LL}{ggg}$ combination (red contours) is also useful, allowing us to test for the potential impact of systematics in the $C_\ell^{gg}$ measurements on the final constraints (assuming that systematic sky contamination in the galaxy overdensity maps would affect the 2- and 3-point correlators in different ways).

    \subsubsection{Constraints on $\sigma_8$ and $\Omega_m$}\label{sssec:results.constraints.results}

\begingroup 
    \setlength{\tabcolsep}{8pt} 
    \renewcommand{\arraystretch}{1.5} 
    \setlength\extrarowheight{2pt}
    
    \begin{table*}
        \centering
            \begin{tabular}{cccccccccc}
            \toprule \multirow{2}{*}{Bin} & \multirow{2}{*}{Probe combination} & \multicolumn{2}{c}{$b_1$} & \multicolumn{2}{c}{$\sigma_8$} & \multicolumn{2}{c}{$\Omega_m$} & \multirow{2}{*}{BF $\chi^2 / N_\mathrm{dof}$} & \multirow{2}{*}{BF PTE} \\ 
             &  & BF & Mean & BF & Mean & BF & Mean & & \\ \midrule
            
            \multirow{4}{*}{$z_1$} & \textcolor{bleu}{$ggg + gg$} & $3.27$ & $3.46^{+0.44}_{-0.87}$ & $0.442$ & $0.436\pm 0.086$  & $0.296$ & $0.299^{+0.014}_{-0.017}$ & 17.0 / 17 & 0.45 \\ 
            & \textcolor{rouge}{$ggg + g\kappa$} & $2.19$ & $2.40^{+0.27}_{-0.48}$ & $0.665$ & $0.637\pm 0.067$ & $0.259$ & $0.290^{+0.036}_{-0.067}$ & 14.87 / 17 & 0.6 \\ 
            & \textcolor{jaune}{$gg + g\kappa$} & $2.14$ & $2.19^{+0.18}_{-0.25}$ & $0.676$ & $0.668\pm 0.062$ & $0.295$ & $0.298^{+0.015}_{-0.018}$ & 18.83 / 21 & 0.6 \\ 
            & $ggg + gg + g\kappa$ & $2.42$ & $2.46^{+0.19}_{-0.26}$ & $0.599$ & $0.594\pm 0.052$ & $0.303$ & $0.306^{+0.015}_{-0.018}$ & 29.13 / 29 & 0.46 \\ \midrule 
            
            \multirow{4}{*}{$z_2$} & \textcolor{bleu}{$ggg + gg$} & $2.58$ & $2.67^{+0.26}_{-0.43}$ & $0.623$ & $0.615\pm 0.083$ & $0.318$ & $0.320^{+0.012}_{-0.014}$ & 12.63 / 24 & 0.97 \\ 
            & \textcolor{rouge}{$ggg + g\kappa$} & $2.72$ & $2.75\pm 0.26$ & $0.679$ & $0.676\pm 0.042$ & $0.406$ & $0.414^{+0.057}_{-0.044}$ & 16.02 / 24 & 0.89 \\ 
            & \textcolor{jaune}{$gg + g\kappa$} & $2.15$ & $2.17^{+0.12}_{-0.15}$ & $0.748$ & $0.745\pm 0.044$ & $0.320$ & $0.322^{+0.012}_{-0.013}$ & 22.68 / 29 & 0.79 \\ 
            & $ggg + gg + g\kappa$ & $2.23$ & $2.24^{+0.11}_{-0.14}$ & $0.723$ & $0.720\pm 0.040$ & $0.325$ & $0.327^{+0.012}_{-0.013}$ & 28.47 / 40 & 0.91 \\ \midrule

            \multirow{4}{*}{$z_3$} & \textcolor{bleu}{$ggg + gg$} & $2.80$ & $2.83^{+0.26}_{-0.36}$ & $0.622$ & $0.623^{+0.065}_{-0.073}$ & $0.3107$ & $0.3120^{+0.0095}_{-0.011}$ & 35.36 / 31 & 0.27 \\ 
            & \textcolor{rouge}{$ggg + g\kappa$} & $2.31$ & $2.37^{+0.17}_{-0.23}$ & $0.742$ & $0.735\pm 0.038$ & $0.278$ & $0.290^{+0.025}_{-0.038}$ & 25.87 / 31 & 0.73 \\ 
            & \textcolor{jaune}{$gg + g\kappa$} & $2.32$ & $2.33^{+0.11}_{-0.14}$ & $0.750$ & $0.747\pm 0.039$ & $0.3050$ & $0.3057^{+0.0091}_{-0.010}$ & 26.88 / 37 & 0.89 \\ 
            & $ggg + gg + g\kappa$ & $2.43$ & $2.44^{+0.11}_{-0.13}$ & $0.719$ & $0.718\pm 0.034$ & $0.3146$ & $0.3156^{+0.0095}_{-0.011}$  & 47.63 / 51 & 0.61 \\ \midrule

            \multirow{4}{*}{$z_4$} & \textcolor{bleu}{$ggg + gg$} & $2.54$ & $2.57^{+0.27}_{-0.41}$ & $0.707$ & $0.714^{+0.091}_{-0.10}$ & $0.3003$ & $0.3015^{+0.0086}_{-0.0097}$  & 48.57 / 36 & 0.08 \\ 
            & \textcolor{rouge}{$ggg + g\kappa$} & $2.41$ & $2.45^{+0.17}_{-0.21}$ & $0.746$ & $0.744\pm 0.033$ & $0.296$ & $0.307^{+0.026}_{-0.038}$  & 31.86 / 36 & 0.67 \\ 
            & \textcolor{jaune}{$gg + g\kappa$} & $2.390$ & $2.403^{+0.099}_{-0.12}$ & $0.750$ & $0.748\pm 0.033$ & $0.2934$ & $0.2942^{+0.0085}_{-0.0094}$  & 37.37 / 43 & 0.71 \\
            & $ggg + gg + g\kappa$ & $2.411$ & $2.423^{+0.096}_{-0.11}$ & $0.746$ & $0.744\pm 0.031$ & $0.3012$ & $0.3020^{+0.0086}_{-0.0097}$  & 58.6 / 59 & 0.49 \\
            \bottomrule
        \end{tabular}
        \caption{Parameter best-fit (BF) values and marginalised constraints for each of the probe combinations in all 4 redshift bins. The errors quoted correspond to $68\%$ confidence level intervals; both upper and lower bounds are specified when the posterior distribution is non-Gaussian. We also quote the chi-square statistics $\chi^2$ (along with the corresponding number of degrees of freedom $N_\text{dof}$ for comparison), as well as the associated probability-to-exceed (PTE) of the theoretical prediction corresponding to the best-fit values. } 
        \label{tab:cosmo}
    \end{table*}
 \endgroup


\begin{figure*}
    \centering
    \includegraphics[width=0.9\textwidth]{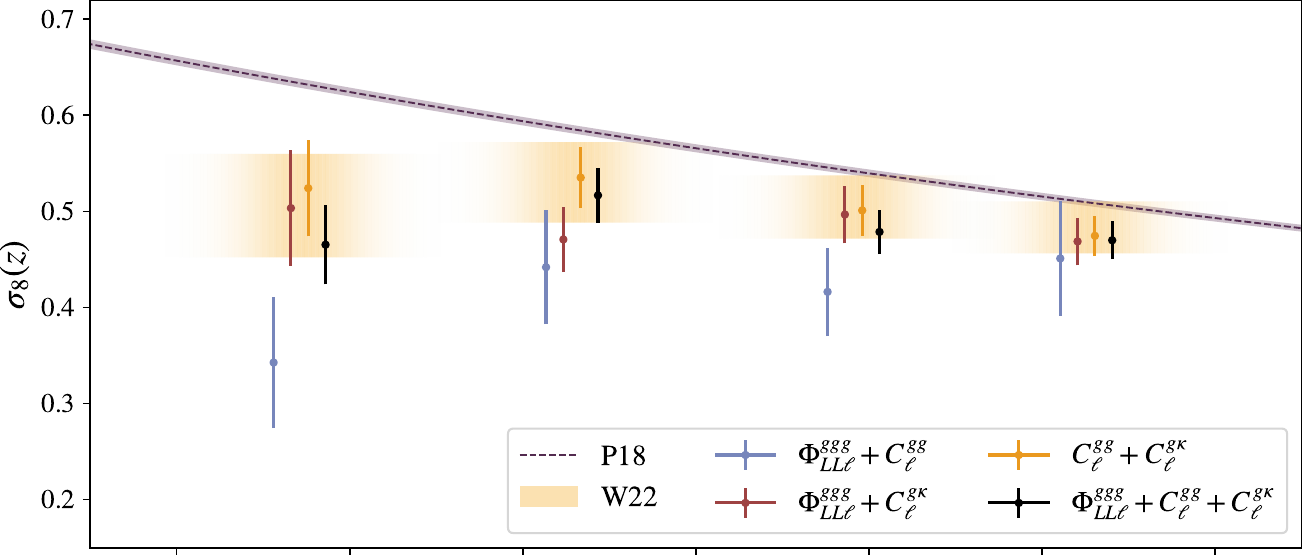} 
    \includegraphics[width=0.9\textwidth]{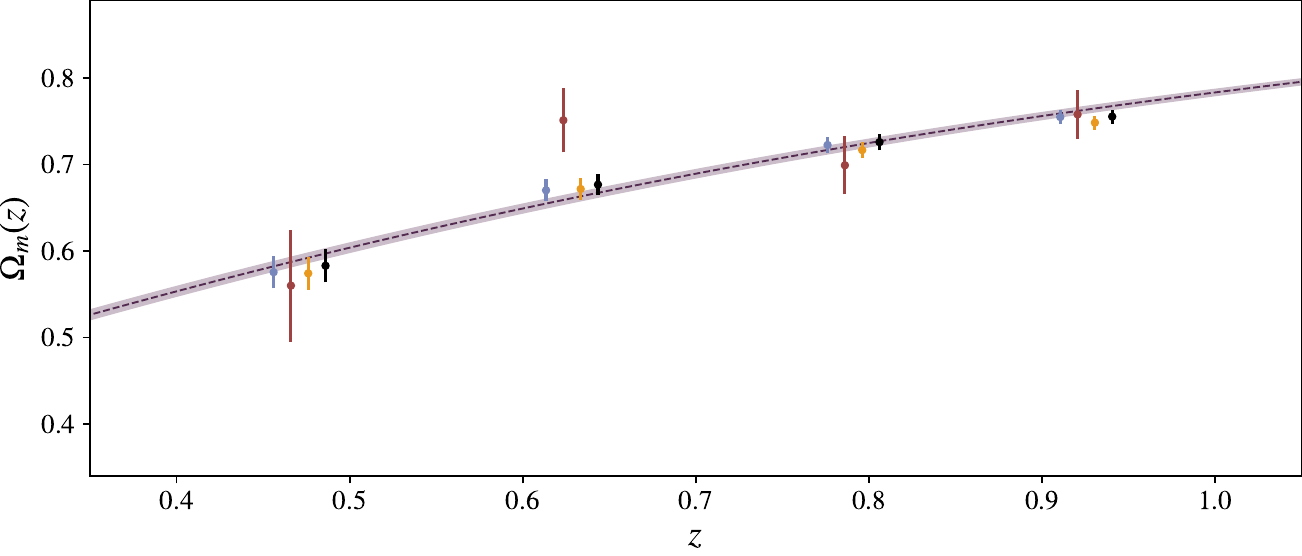}
    \caption{
    Top panel: $\sigma_8$ constraints as a function of redshift, constructed from Eq. \ref{eq:sig8ofz}. We show constraints from all four probe combinations shown in Fig. \ref{fig:4constraints}. For comparison, we also include early-universe constraints on $\sigma_8$ presented in \cite{collaborationPlanck2018Results2020b} -- P18 on the figure -- as well as CMB lensing tomography constraints from \cite{whiteCosmologicalConstraintsTomographic2022} -- shortened to W22 above -- which were derived from the traditional power spectrum combination $C_\ell^{gg} + C_\ell^{g\kappa}$. The W22 constraints were obtained from identical data as used in this work, but with slightly different modelling choices and scale cuts. 
    Bottom panel: $\Omega_m$ constraints as a function of redshift, constructed from Eq. \ref{eq:omofz}. Similarly, this figure shows the constraints on $\Omega_m$ for each probe combination using the same colour scheme. We also show \planck constraints as the dashed purple line.}\label{fig:sig8ofz}
\end{figure*}

      Using the likelihood described in Section \ref{ssec:methods.like}, we now derive constraints on $\sigma_8$ and $\Omega_m$ from different combinations of the measured power spectra and FSBs. We will present constraints found by analysing the data from each redshift bin independently. On the one hand, this simplifies the analysis, reducing the dimensionality of the parameter space to explore, and it allows us to study potential trends with redshift in the recovered cosmological parameters, and in the complementarity of the projected bispectrum. On the other hand, we do this to avoid reporting potentially unrealistic constraints, given the various simplifying assumptions used in this preliminary analysis. This will be addressed in future work. 
      
      The resulting constraints in the $(\sigma_8,\Omega_m,b_1)$ space found for each redshift bin are shown in Fig. \ref{fig:4constraints}. The blue, red, and yellow contours show the constraints found from the combinations $\fsb{\ell}{LL}{ggg}+C_\ell^{gg}$, $\fsb{\ell}{LL}{ggg} + C_\ell^{g\kappa}$, and $C^{gg}_\ell+C_\ell^{g\kappa}$, respectively. The numerical $68\%$ confidence-level constraints on these parameters are listed in Table \ref{tab:cosmo}, which also lists the best-fit $\chi^2$ value, and the associated probability-to-exceed (PTE) assuming a Gaussian likelihood with $N_{\rm dof}=N_{\rm data}-N_{\rm params}$ degrees of freedom, where $N_{\rm data}$ is the size of the data vector, and $N_{\rm params}=3$ is the number of free parameters in our model. We find that the model provides a good description of the data in all cases, with PTE values ranging from 0.08 to 0.97. We again remind readers that these constraints should be interpreted in the context of the relative simplicity of the model used here, in which the power spectrum (and bispectrum) are only described at the tree level, and the non-linear bias parameters are deterministically related to the  linear bias $b_1$. Nevertheless, some interesting insights can be gained from these results.
      
      First, focusing on the $\Omega_m$ constraints, we can see that its derived value is determined by the shape of the galaxy auto-correlation, as its value and statistical uncertainties are consistent across all data combinations including $C_\ell^{gg}$. The combination $\fsb{\ell}{LL}{ggg} + C_\ell^{g\kappa}$, in turn, yields uncertainties that are approximately $3$ times larger, although the recovered constraints on $\Omega_m$ are consistent, well within $1\sigma$, with those found by other data combinations across most redshift bins. The exception in this case is $z_2$, where this data combination recovers a value of $\Omega_m$ that is in $\sim2\sigma$ tension with that preferred by the datasets including $C_\ell^{gg}$, as well as with the value measured by \planck \citep{collaborationPlanck2018Results2020b}. This tension is also apparent in the two-dimensional contours of Fig. \ref{fig:4constraints}. Interestingly, this is the only redshift bin in which one of the model parameters ($\Omega_m$) hits the prior, showing that this data combination is particularly insensitive to this parameter. The preference for a higher $\Omega_m$ in this case may be a statistical fluctuation, or it may be caused by some of our simplifying assumptions. We will explore this further in future work.

      Secondly, we find that, in most cases, all probe combinations are able to obtain constraints on $\sigma_8$ that are consistent within $\sim1\sigma$. Interestingly, we find that all combinations that include the galaxy-$\kappa$ cross-correlation obtain significantly tighter constraints, in remarkable agreement with each other. $C_\ell^{g\kappa}$ is thus vital when measuring $\sigma_8$, and the agreement between different data combinations is a direct consequence of the compatibility between the galaxy two-point and three-point correlators within the model used here. As we anticipated in the previous section, although the constraints from the $C_\ell^{gg}+\fsb{\ell}{LL}{ggg}$ are less precise than other combinations, it is reassuring to see that we are able to obtain constraints on $\sigma_8$ in the absence of lensing information that are largely in agreement with the other probes. Furthermore, the constraints found in the $(\Omega_m,\sigma_8)$ plane are almost orthogonal under this data combination. It is interesting to note that the $gg+ggg$ combination seems to favour consistently lower values of $\sigma_8$ across all redshift bins. This is particularly evident in the first redshift bin, displaying a $\sim2\sigma$ tension between this data combination and the \txt dataset and, to a lesser extent in the third redshift bin, which recovers a $\sim1.5\sigma$ downwards shift. As we discussed in Section \ref{sssec:results.constraints.intersections}, and further explored in Appendix \ref{app:coevol}, the constraints on $\sigma_8$ found from this data combination are critically dependent on our assumptions regarding the relationship between the different bias parameters. Thus, the preference for a lower $\sigma_8$ could be a result of the coevolution relations we have adopted here. In addition to enabling these valuable consistency tests, the inclusion of the galaxy bispectrum in addition to the traditional \txt data combination leads to a $\sim10\%$ improvement in the constraint on $\sigma_8$. We note that, regardless of which data combination is used, the values of $\sigma_8$ we recover are systematically low with respect to that preferred by \planck.

      This fact is evident in the upper panel of Fig. \ref{fig:sig8ofz}. The figure shows the constraints on the amplitude of matter fluctuations $\sigma_8(z)$ at the mean redshift of each of the samples studied here. The time-dependent amplitude is defined as
      \begin{equation} \label{eq:sig8ofz}
          \sigma_8(z)\equiv \sigma_8\,D(z|\Omega_m),
      \end{equation}
      where $D(z|\Omega_m)$ is the linear growth factor at redshift $z$, implicitly dependent on $\Omega_m$, and normalised so that $D(0|\Omega_m)=1$. The figure shows the constraints found from the same data combinations discussed above, in addition to the best-fit prediction from \planck (purple dashed line). We find values of $\sigma_8(z)$ that are consistently lower than the \planck prediction. Qualitatively, this result was also reported in \cite{whiteCosmologicalConstraintsTomographic2022}. The constraints on $\sigma_8(z)$ found there, shown in Fig. \ref{fig:sig8ofz} as yellow bands are largely in agreement with our \txt constraints, in spite of the significant differences in the modelling choices used here (e.g. tree-level bias parametrisation in Eulerian space, different scale cuts, shot noise marginalisation). Our results are also in rough agreement with those found by \cite{2407.04607}. The evidence for a lower $\sigma_8$ is stronger in the first bin, as was also found in these works. We also provide an equivalent figure for the evolution of $\Omega_m$ with redshift (lower panel of Fig. \ref{eq:sig8ofz}). We generate the time-dependent non-relativistic matter fraction using 
      \begin{equation} \label{eq:omofz}
          \Omega_m (z) = \frac{\Omega_m (1+z)^3}{\Omega_m (1+z)^3 + (1 - \Omega_m)}\,.
      \end{equation}
      One data point stands out -- the $ggg+g\kappa$ constraint in $z_2$, as mentioned earlier -- but all other constraints points are in very good agreement with early-universe constraints from \planck, shown by the purple dashed line. It is worth noting that, as discussed above, these constraints are driven by the shape of the large-scale matter power spectrum, which is most tightly constrained by the $C_\ell^{gg}$ measurements.
      Given the relative simplicity of the model used here, we refrain from quantifying or interpreting the level of tension between these results and the CMB predictions. This will be addressed in upcoming work employing a more robust modelling to describe the 2-point and 3-point clustering of galaxies \citep{2510.17796}. Nevertheless, we find that all data combinations recover compatible constraints, which indicates that any potential tension is likely not dominated by systematics in any individual measurement.

\section{Discussion}\label{sec:concl}
This paper presents an attempt at using the projected bispectrum in tandem with the traditional angular power spectrum to conduct a joint cosmological analysis of CMB lensing and galaxy data. To do so, we measure the ``filtered-squared bispectrum'' (FSB) estimator as a proxy for the projected bispectrum. We develop in this paper a self-consistent framework for the generalisation of the FSB, first introduced in H25, to cross-correlations. The FSB estimator is robust to incomplete sky observations and comes with a fully analytical, model-independent covariance which can be efficiently estimated from the data. Theoretical predictions for the FSB estimator can also be built in a straightforward manner from perturbation theory, making the FSB a truly attractive higher-order statistic.

We use data from four photometric DESI Luminous Red Galaxy samples, described in \cite{zhouDESILuminousRed2023} and \cite{whiteCosmologicalConstraintsTomographic2022}, and the CMB lensing convergence map obtained from \planck \citep{collaborationPlanck2018Results2020b} to measure auto- and cross-FSBs. One of our primary results is the detection of both the galaxy bispectrum ($ggg$) at very high significance ($>30\sigma$) in all redshift bins, and the galaxy-galaxy-convergence bispectrum ($gg\kappa$) at lower significance ($>5\sigma$) in three out of the four galaxy samples used in this work. The latter being detected at a low signal-to-noise means that we are not able to use $gg\kappa$ in the second part of the paper, where we conduct a rudimentary cosmological analysis. 

In this first cosmological analysis including the projected bispectrum, we simply focus on three parameters -- the clustering amplitude $\sigma_8$, the non-relativistic matter fraction $\Omega_m$, and the linear bias $b_1$ -- while keeping other parameters fixed.
We identify two new probe combinations that break degeneracies between $\sigma_8$ and $b_1$: combining the $ggg$ FSB measurements with any of the power spectra ($gg$ or $g\kappa$) allows us to derive complementary constraints on $\sigma_8$ and $\Omega_m$, and to test the robustness of the constraints obtained from the standard \txt combination against observational and modelling systematics. The new $ggg + gg$ combination is especially exciting, potentially yielding a fully local measurement of $\sigma_8$ without resorting to using CMB lensing data to break the $\sigma_8$-bias degeneracy. 

The addition of the $ggg$ bispectrum to both power spectra ($ggg+gg+g\kappa$ combination) leads to a modest improvement of $\sim 10\%$ on the $\sigma_8$ constraints. Further improvement may be achievable with more sensitive datasets, allowing us to use the $gg\kappa$ bispectrum, or pushing to scales smaller than the conservative cuts assumed here. We also find that the $g\kappa$ cross-spectrum is indispensable to obtain tight constraints on $\sigma_8$ (and similarly with the $gg$ auto-spectrum for $\Omega_m$). All combinations studied yield consistent results: $gg+g\kappa$ and $ggg+g\kappa$ are consistent between themselves at the $1\sigma$ level in $\sigma_8$, and at the $2\sigma$ level with $ggg + gg$. All combinations give $\sigma_8$ values consistently lower than inferred from \planck measurements, in agreement with previous CMB lensing tomography studies conducted on the same data \citep[][using the standard $gg+g\kappa$ combination with a different modelling approach]{whiteCosmologicalConstraintsTomographic2022}. In particular, we find that the $ggg + gg$ combination favours even lower $\sigma_8$ values than other probe combinations at low redshift. We argue, however, that this result is highly dependent on some of our analysis choices, and thus refrain from fully interpreting this result in terms of a cosmological tension.

Indeed, our modelling of the bispectrum comes with various caveats: in this first approach, we only model the bispectrum at the tree-level (a decision which we must compensate for by implementing conservative scale cuts) and, critically, choose to use empirical coevolution relations to express higher-order bias parameters ($b_2$ and $b_s$). These coevolution relations play an important role in our final results, as they essentially dictate the effective scaling of the bispectrum with $b_1$  -- and thus, how efficiently it can break parameter degeneracies when combined with power spectrum measurements. Upcoming work \citep{2510.17796} will focus on providing a better model for the bispectrum, in combination with a modelling of the power spectrum at the 1-loop level. This will allow us to push to smaller scales and to constrain the higher-order bias parameters directly from the data, without having to resort to using coevolution relations. In this scenario, the bispectrum is a critical tool to improve the constraints on $b_2$ and $b_s$, potentially leading to a relatively larger improvement on the final cosmological constraints obtained from power spectra than we report here. Eventually, the use of hybrid approaches to model galaxy bias \citep[e.g. ][]{modiSimulationsSymmetries2020,hadzhiyskaHeftyEnhancementCosmological2021} will be required to extend the range of scales over which the bispectrum may be used for cosmological inference.

In summary, we have included for the first time the projected bispectrum in a tomographic cosmological analysis of galaxy clustering and CMB lensing. Despite our simplistic bispectrum model and conservative scale cuts, we have shown that the projected bispectrum is able to provide consistent and improved cosmological constraints when combined with power spectrum measurements. This work illustrates the potential of the projected bispectrum as a cosmological probe, and provides a first tentative blueprint for its implementation into inference pipelines using the FSB estimator.

\section*{Acknowledgments}
We would like to thank Sergi Novell-Masot, Emiliano Sefusatti, Francesco Verdiani, and Matteo Zennaro for useful comments and discussions. LH is supported by a Hintze studentship, which is funded through the Hintze Family Charitable Foundation. DA acknowledges support from STFC and the Beecroft Trust. We made extensive use of computational resources at the University of Oxford Department of Physics, funded by the John Fell Oxford University Press Research Fund. AN acknowledges support from the European Research Council (ERC) under the European Union’s Horizon 2020 research and innovation program with Grant agreement No. 101163128.

\bibliographystyle{mnras}
\bibliography{main}

\begin{appendix}

\section{Effect of using coevolution relations}
\label{app:coevol}

\begin{figure}[]
\centering
\includegraphics[width=.31\linewidth]{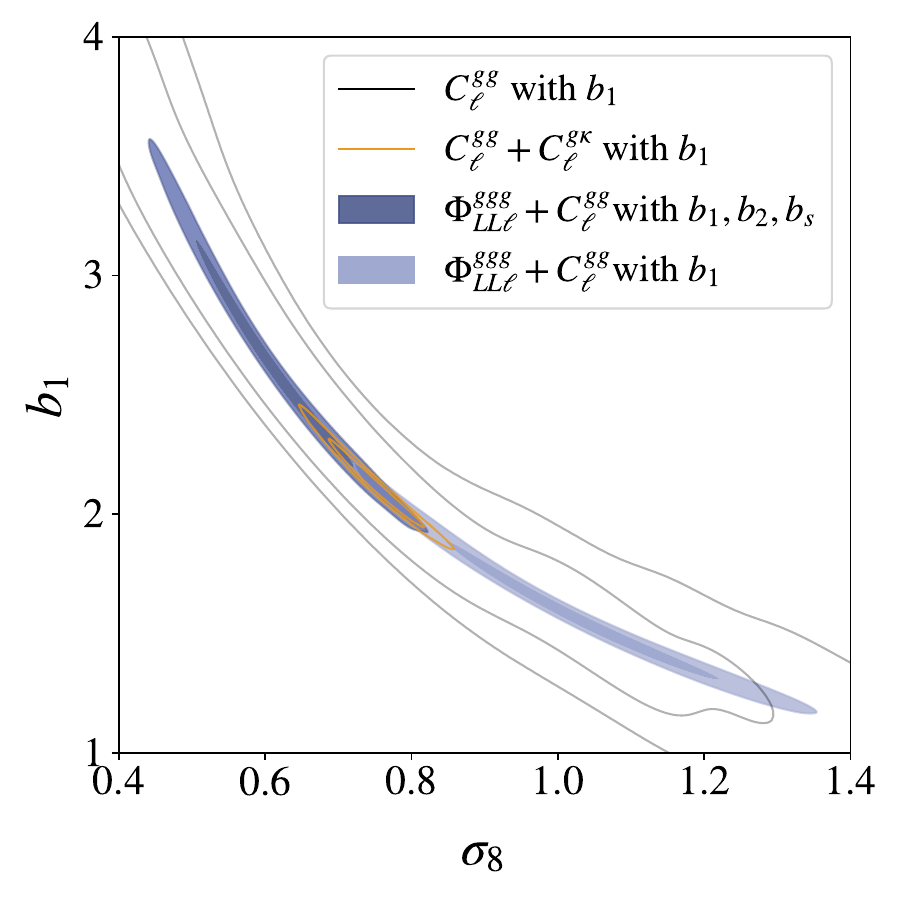}
\includegraphics[width=.31\linewidth]{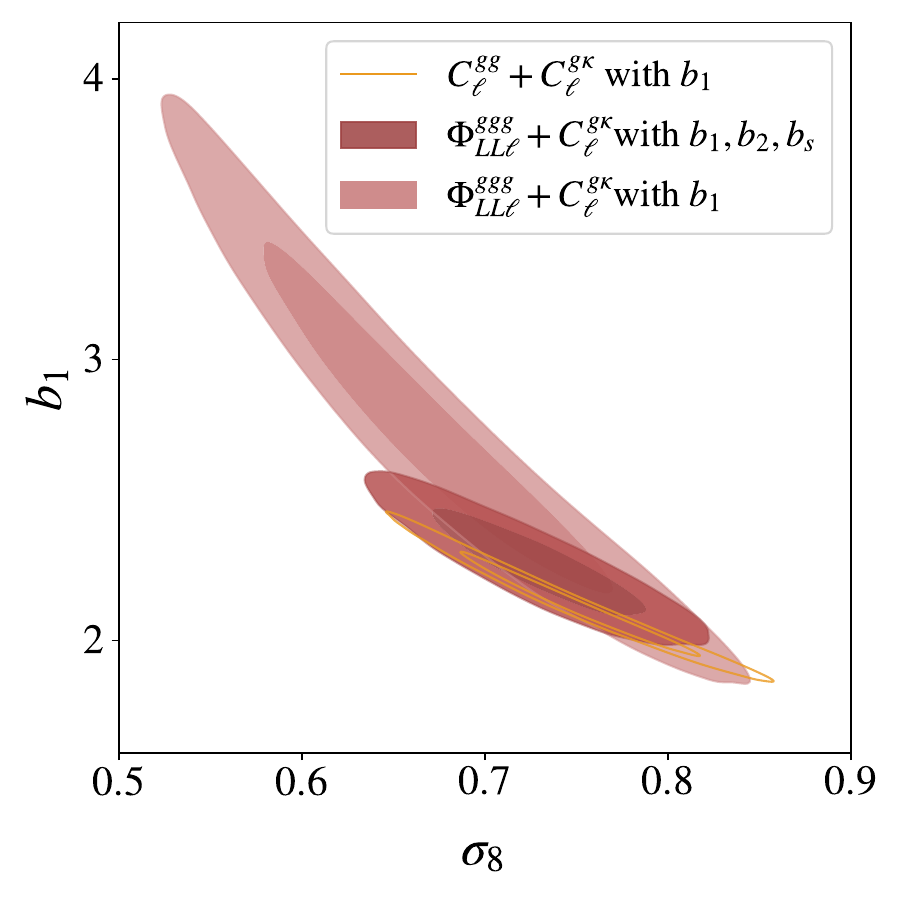}
\includegraphics[width=.31\linewidth]{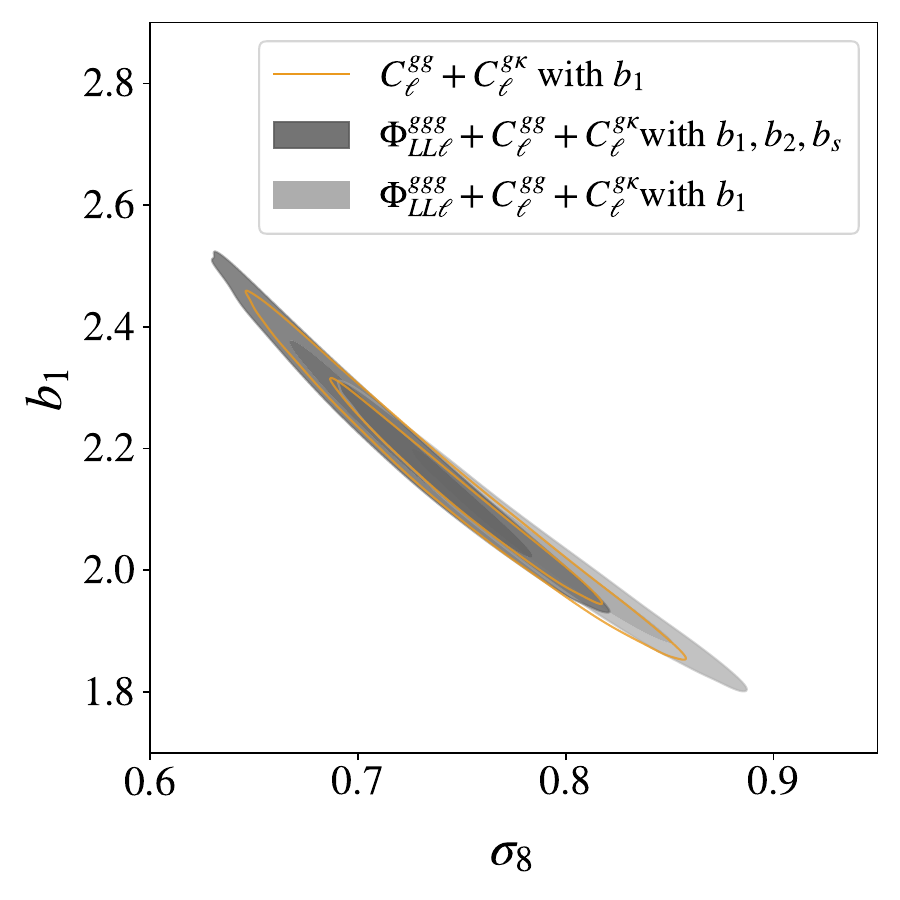}
\caption{Plots showing the effect of using $b_1$, $b_2$ and $b_s$ while fixing the latter two using coevolution relations (dark contours), and that of using $b_1$ only in the bispectrum analysis (i.e. setting $b_2=b_s=0$, shown by the lighter contours). The yellow contours show for comparison the power-spectra-only constrains, which are unaffected by this test as they only depend on the linear bias in this analysis (thus the yellow contours remain the same in all three panels). These contours were generated for the galaxy data in the second redshift bin, $z_2$.}
\label{fig:nob2bs}
\end{figure}

Throughout this paper, we use coevolution relations of bias parameters to simplify our inference pipeline and redirect the constraining power of the FSB probes towards parameters of cosmological interest. For completeness, we repeat their expressions here \citep{lazeyrasPrecisionMeasurementLocal2016, lazeyrasLIMDBiasMeasurement2018}:
\begin{align}
    b_2(b_1) & = 0.412 - 2.143 b_1 + 0.929 b_1^2 + 0.008 b_1^3 \,, \label{eq:b2ofb1} \\
    b_s(b_1) & = -\frac27 (b_1 -1) \,. \label{eq:bsofb1}
\end{align}

In this appendix we explore how these coevolution relations affect our analysis. We show this by comparing different approaches to modelling the bispectrum. Note that here, for simplicity and ease of interpretation,  we will fix $\Omega_m$ to its fiducial value (see Table \ref{tab:fiducialcosmo}), and only vary $\sigma_8$ and the bias parameters.

\begin{figure}
    \centering
    \includegraphics[width=0.8\linewidth]{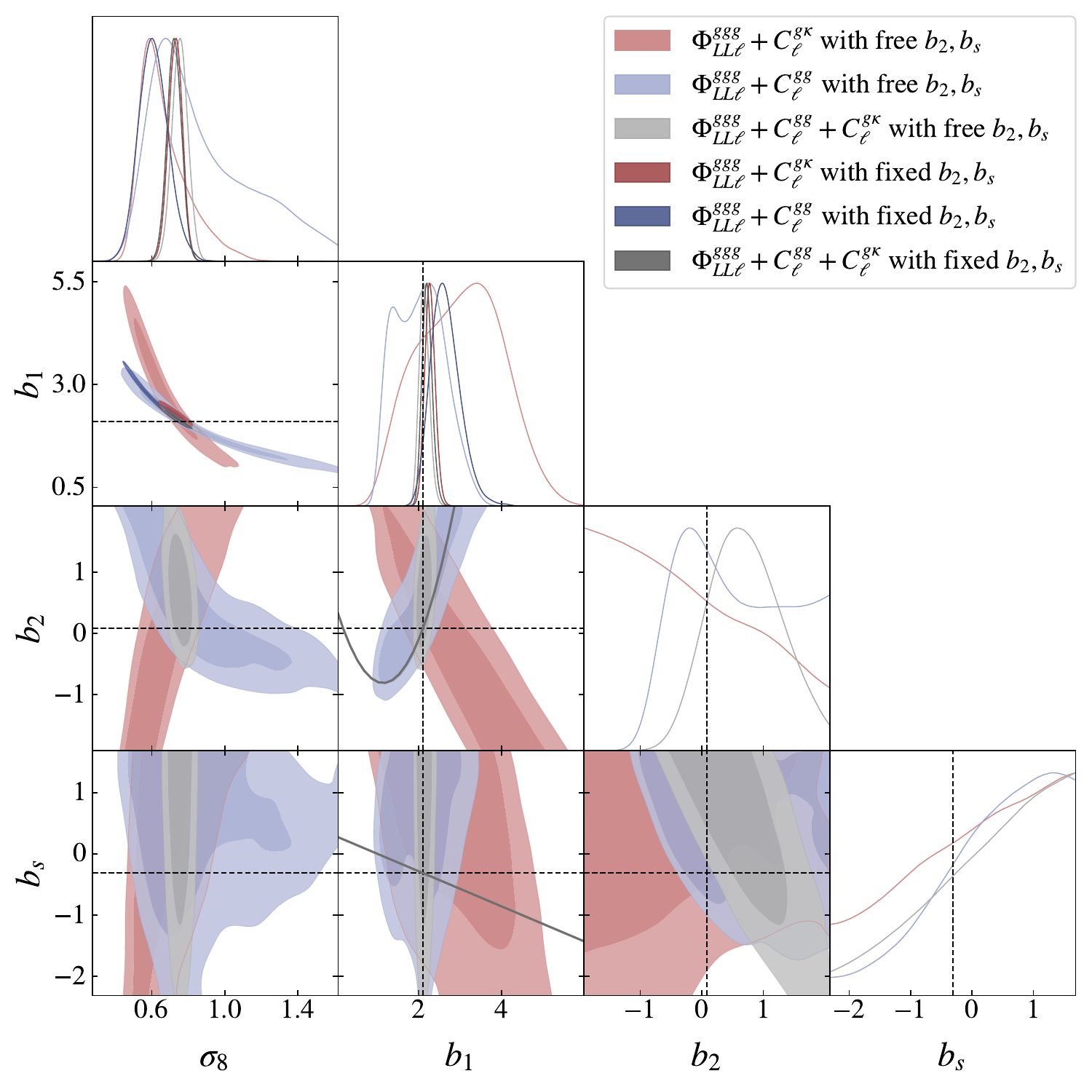}
    \caption{Plot showing constraints obtained when freeing the $b_2$ and $b_s$ parameters (light contours) as opposed to fixing them using coevolution relations (darker contours in the $b_1 - \sigma_8$ block). We show the coevolution relations (Eqs. \ref{eq:b2ofb1} and \ref{eq:bsofb1}) as dark grey lines in the $b_2 - b_1$ and $b_s - b_1$ blocks. The black dashed lines along the $b_2$ and $b_s$ rows and columns show the values of those parameters obtained via coevolution, for the published value of the linear bias for this redshift bin ($z_2$) -- we use those as central values for the uniform priors placed on $b_2$ and $b_s$. We only show the area of parameter space allowed by the priors.} 
    \label{fig:freeb2bs}
\end{figure}

The first approach was to simply set both of the $b_2$ and $b_s$ parameters to zero, effectively making the corresponding terms in the tree-level bispectrum (Eqs. \ref{eq:bggg} and \ref{eq:bggk}) vanish. Fig. \ref{fig:nob2bs} shows the difference in the parameter contours resulting from this choice with respect to the results obtained assuming coevolution relations to determine $b_2$ and $b_s$. The effect is particularly significant in the case of the $\Phi_{LL\ell}^{ggg} + C_\ell^{gg}$ probe combination, for which setting $b_2=b_s=0$ leads to a significant shift in the contours along the $C_\ell^{gg}$ degeneracy direction (shown as the grey contours of the first panel in Fig. \ref{fig:nob2bs}). Using coevolution relations pushes the constraints towards lower values of $\sigma_8$ (and higher $b_1$ values, as discussed in Section \ref{sssec:results.constraints.intersections}). The combination containing all probes similarly gets pushed towards higher $\sigma_8$ values, although the effect is less extreme (as the constraints are dominated by the power spectra, which are insensitive to $b_2$ and $b_s$ at the tree level). The addition of the quadratic and tidal biases via coevolution significantly shrinks the contours for the $\Phi_{LL\ell}^{ggg} + C_\ell^{g\kappa}$ probe combination, pushing it closer to the power-spectra contour (this time towards slightly higher $\sigma_8$ values, and much lower bias). 

We also explore the impact of allowing $b_2$ and $b_s$ to vary freely, instead of fixing them to $b_1$ via coevolution relations. The priors defined for these two parameters still make use of the coevolution relations, but allow a wider range of bias values: we use $\mathcal U [b_2(b_1)-2, b_2(b_1)+2]$ and $\mathcal U [b_s(b_1)-2, b_s(b_1)+2]$ as priors for $b_2$ and $b_s$ respectively, where $b_1$ here is the published linear bias value in the redshift bin (see Table \ref{tab:lrgs}), taken from \cite{whiteCosmologicalConstraintsTomographic2022}. Interestingly, as can be seen by comparing the light blue and light red contours in both Fig. \ref{fig:nob2bs} and Fig. \ref{fig:freeb2bs}, the constraints obtained in this approach are similar to those obtained when setting the non-linear biases to zero. In general, the $ggg$ FSB does not have enough constraining power on its own to fully constrain $b_1$, $b_2$ and $b_s$ simultaneously, as shown by the light blue, red or gray contours for the non-linear bias parameters in Fig. \ref{fig:freeb2bs}, which are largely unconstrained. In order to self-consistently constrain higher-order bias parameters from the power- and bispectrum, it will thus be crucial to push the analysis to smaller scales by modelling the power spectrum up to 1-loop and potentially moving beyond tree-level bispectra. We leave this analysis for future work.

The coevolution relations thus play an important role in the final $\sigma_8$ constraints recovered from the galaxy-only probes, and for the final $b_1$ constraints obtained from the $\Phi_{LL\ell}^{ggg} + C_\ell^{g\kappa}$ combination. The object of future work will be to test thoroughly these coevolution relations over a large range of galaxy samples.

\section{Stochastic contributions}
\label{app:stoch}

In this appendix we derive the expressions for the noise contributions to the $gg\kappa$ bispectrum, Eq. \ref{eq:noiseggk}. In Appendix D of  H25, we derived the equivalent term for the $ggg$ bispectrum (Eq. \ref{eq:noiseggg}), in which case all fields measured receive a shot-noise contribution due to the discreteness of the tracer. Here, that is not the case for the third field (the CMB convergence $\kappa$). Let us see what this implies for the additional shot noise terms contributing to the $gg\kappa$ bispectrum.

Observationally, we estimate the FSB from pixelised maps of the galaxy overdensity $\delta_p = \frac{N_p}{\bar{N}}-1$, where $N_p$ denotes the number of galaxies in pixel $p$, and $\bar{N}$ is the mean number of galaxies per pixel. Assuming galaxies to be Poisson tracers of the underlying density field, the first two moments of the galaxy density field are given by
  \begin{align}
    \langle N_p\rangle_{\mathrm{Pois}}&=\bar{N}_p=\bar{N}(1+\delta_p),\nonumber\\
    \langle N_p N_{q}\rangle_{\mathrm{Pois}}&=\bar{N}_p\bar{N}_{q}+\delta_{pq}^K\bar{N}_p,\label{eq:stochasticity:Poisson}
  \end{align}
  where $\delta_{pq}^{K}$ denotes the Kronecker delta and we have only performed the Poisson ensemble average  \citep{1980Peebles}. 

Similarly, we can denote the value of the convergence field at pixel $p$ as $K_p$. The first moment of the convergence field, is simply $\langle K_p \rangle_\mathrm{Pois} = K_p$ as the convergence is not a discretely-sampled field.

Now let us look at cross-moments, which we will need for the upcoming calculations: we have
\begin{align}
    \langle N_p K_q \rangle_\mathrm{Pois} & = \bar N_p K_q \\
    \langle N_p N_q K_r \rangle_\mathrm{Pois} & = (1 - \delta^{\cal K}_{pq}) \bar N_p \bar N_q K_r + \delta^{\cal K}_{pq} (\bar N_p^2 + \bar N_p) K_r \nonumber \\
    & = \bar N^2 (1+\delta_p + \delta_q + \delta_p \delta_q ) K_r + \delta^{\cal K}_{pq} \bar N(1+\delta_p) K_r 
\end{align}

With these tools in hand, we can now try to derive the 3-point correlation functions of fields $g$, $g$ and $\kappa$:
\begin{align} \nonumber
\langle gg\kappa \rangle_\mathrm{Pois} |_{\mathrm{pix} = p,q,r} & \equiv \bigg\langle \left(\frac{N_p - \bar N}{\bar N} \right) \left(\frac{N_q - \bar N}{\bar N} \right) K_r \bigg\rangle \\ \nonumber
& = \frac{1}{\bar N ^2} \left[ \langle N_p N_q K_r \rangle - \bar N \langle N_q \rangle K_r - \bar N \langle N_p \rangle K_r + \bar N^2 K_r \right] \\ \nonumber
& = \frac{1}{\bar N ^2} \left[ \bar N^2 (1+\delta_p + \delta_q + \delta_p \delta_q ) K_r + \delta^{\cal K}_{pq} \bar N(1+\delta_p) K_r - \bar N^2 (1 + \delta_q + \delta_p ) K_r \right] \\
 & = \frac{1}{\bar N ^2} \left[ \bar N^2 \delta_p \delta_q  K_r + \delta^{\cal K}_{pq} \bar N(1+\delta_p) K_r \right] 
\end{align}
Now taking the expectation value over fields $g$, $\kappa$, and assuming $\langle g \rangle$ and $\langle \kappa \rangle$ to be $0$ over field realisations, we obtain: 
\begin{equation} \label{eq:ggkensembleaverage}
\langle gg\kappa \rangle = \frac{1}{\bar N ^2} \big\langle ( N_p - \bar N) ( N_q - \bar N ) K_r \big\rangle = \langle \delta_p \delta_q \kappa_r \rangle + \delta^{\cal K}_{pq} \frac 1{\bar N} \langle \delta_p \kappa_r \rangle 
\end{equation}

We now can express Eq. \ref{eq:ggkensembleaverage} in harmonic space, thus obtaining the corresponding projected bispectrum: 
\begin{align} \nonumber
    \left\langle g_{\ell m}g_{\ell' m'}\kappa_{\ell'' m''}\right\rangle & = \sum_{p, q, r}\Omega_\mathrm{pix}^3 \left( \langle \delta_p \delta_q \kappa_r \rangle + \delta^{\cal K}_{pq} \frac 1{\bar N} \langle \delta_p \kappa_r \rangle \right) Y_{\ell m p}^*Y_{\ell' m' q}^*Y_{\ell'' m'' r}^* \\
    & = \Omega_\mathrm{pix}^3 \sum_{p, q, r} \langle \delta_p\delta_{q}\kappa_{r}\rangle Y_{\ell m p}^*Y_{\ell' m' q}^*Y_{\ell'' m'' r}^* + \frac{\Omega_\mathrm{pix}^3}{\bar{N}} \sum_{p,r} \langle \delta_p \kappa_{r} \rangle   Y_{\ell m p}^*Y_{\ell' m' p}^*Y_{\ell'' m'' r}^* \label{eq:twoterms}
\end{align}
where we have used the shorthand $Y_{\ell m p} \equiv Y_{\ell m}(\nv_p)$, and where $\Omega_\mathrm{pix}$ denotes the average pixel area. 

Noticing that
\begin{equation}
    \langle \delta_i \delta_j \delta_k \rangle = \sum_{(\ell m)_{123}} Y_{\ell_1 m_{1} i} Y_{\ell_2 m_{2} j} Y_{\ell_3 m_{3} k} \mathcal G^{\ell_1\ell_2\ell_3}_{m_1m_2m_3} b_{\ell_1\ell_2\ell_3} \quad\quad \mathrm{and} \quad\quad \Omega_\mathrm{pix} \sum_i Y^{}_{\ell m i}Y^*_{\ell' m' i} = \delta^\mathcal{K}_{\ell\ell'} \delta^\mathcal{K}_{mm'} \, , 
\end{equation}
we can rewrite the first term in Eq. \ref{eq:twoterms} as 
\begin{align}
(1) \equiv \Omega_\mathrm{pix}^3 \sum_{p, q, r} \left( \sum_{(\ell m)_{123}} Y_{\ell_1 m_1 p} Y_{\ell_2 m_2 q} Y_{\ell_3 m_3 r} \mathcal G^{\ell_1 \ell_2 \ell_3}_{m_1 m_2 m_3} b_{\ell_1\ell_2\ell_3}^{gg\kappa} \right) Y_{\ell m p}^*Y_{\ell' m' q}^*Y_{\ell'' m'' r}^* = \mathcal G^{\ell\ell'\ell''}_{mm'm''} b_{\ell\ell'\ell''}^{gg\kappa} \, ,
\end{align} 
which simply corresponds to the main projected bispectrum term in the absence of shot noise. 

The second term however is introduced due to the stochastic nature of the galaxy density field; if we write the two-point correlation function in the second term of Eq. \ref{eq:twoterms} as
\begin{equation}
    \langle \delta_i \delta_j \rangle = \sum_{(\ell m)_{12}} Y_{\ell_{1} m_{1} i } Y_{\ell_2 m_{2} j} \delta^{\cal K}_{\ell_1 \ell_2} \delta^{\cal K}_{m_1 m_2} C_{\ell_1}^{\delta \delta} = \sum_{(\ell m)} Y_{\ell m i} Y_{\ell m j} C_{\ell}^{\delta \delta}\, ,
\end{equation}
we can simplify the term to: 
\begin{align} \nonumber
(2) & \equiv \frac{\Omega_\mathrm{pix}^2}{\bar{N}} \sum_{p,q} Y_{\ell m p}^*Y_{\ell' m' p}^*Y_{\ell'' m'' q}^* \sum_{(\ell_1 m_1)} Y_{\ell_1 m_1 p} Y_{\ell_1 m_1 q} C_{\ell_1}^{g \kappa} \\
& = \frac{\Omega_\mathrm{pix}}{\bar{N}} C_{\ell''}^{g \kappa} \sum_{p} Y_{\ell m p}^*Y_{\ell' m' p}^* Y_{\ell'' m'' p}
\end{align}
We use the property of spherical harmonics' complex conjugates, $Y_{\ell m}^* = (-1)^m Y_{\ell -m}$, to rename our indices: $m\rightarrow -m$ and $m'\rightarrow -m'$. This allows us to recover a term that resembles (discrete) Gaunt coefficients: 
\begin{equation}
    (2) = \frac{1}{\bar{N}} C_{\ell''}^{g \kappa} \mathcal G_{mm'm''}^{\ell\ell'\ell''} \, .
\end{equation}

Therefore the Poisson contribution for an FSB which is measured on two discretely-sampled fields (here the galaxy overdensity field $\delta_g$) and a continuous one (like the CMB convergence $\kappa$), is of the form 
\begin{align} 
    \Phi^{gg\kappa}_{LLb, \, \mathrm{stoch}} = & \sum_{(\ell)_{123}} h^2_{\ell_1\ell_2\ell_3} K^{LLb}_{\ell_1\ell_2\ell_3}\, \left[ \frac{1}{\bar{n}} C^{g\kappa}_{\ell_3} \right] 
\end{align}
As this term depends on the cross power spectrum of $\delta_g$ and $\kappa$, it will scale with bias accordingly. This additional bias dependence could potentially further help break degeneracies between bias parameters and parameters of cosmological interest, like $\sigma_8$.

\section{Multifield Covariance}
\label{app:multicov}

\begin{figure}
    \centering    
    \includegraphics[width=\linewidth]{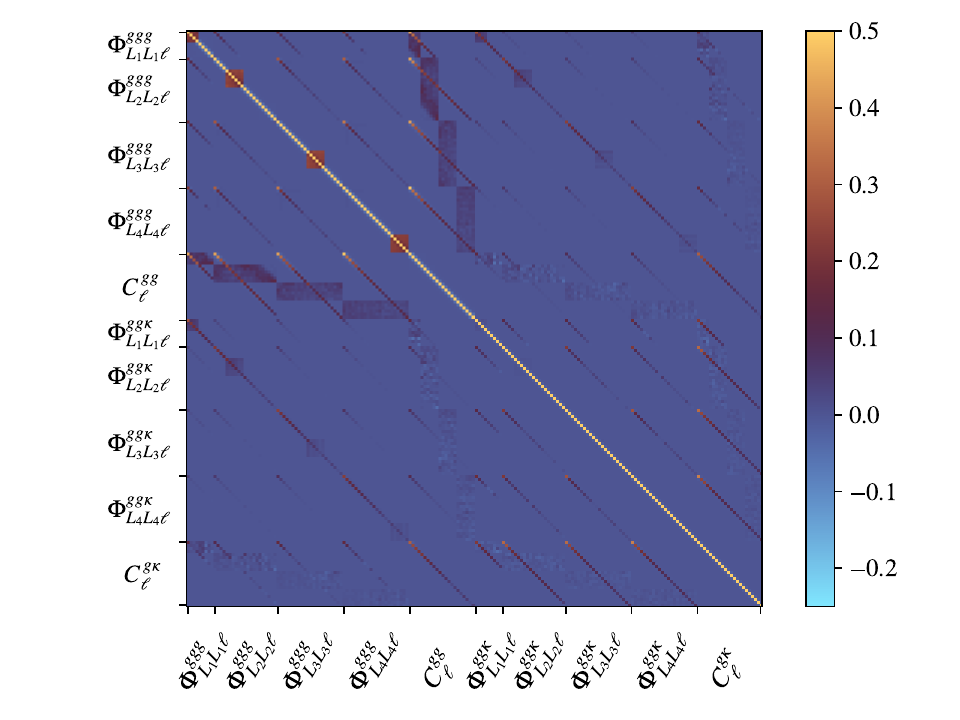}
    \caption{Application of the multi-field covariance derivations presented in this Appendix to the galaxy and convergence data. The example data vector here contains $ggg$ FSB measurements, the $gg$ power spectrum measurement, $gg\kappa$ FSBs, and the $g\kappa$ power spectrum. We plot the correlation matrix (a normalised covariance matrix, where the diagonal is by construction unity) to better highlight off-diagonal contributions -- note that we set the maximum value of the colour map to $0.5$, as essentially only the main diagonal lies above this threshold. The $N_{222}$ contributions can be seen as small brown or purple squares along the main diagonals in the four large FSB-FSB blocks (although it is barely visible in the $gg\kappa$-$gg\kappa$ blocks). The $N_{32}$ contributions show up as dark purple bands in the FSB-$C_\ell$ blocks. Due to the weak signal in the generalised $gg\kappa$ and $g\kappa\kappa$ FSBs, this contribution may sometimes appear as a negative contribution (light blue pixels within the $N_{32}$ bands in the $gg\kappa$-$gg$, $ggg$-$g\kappa$ and $gg\kappa$-$g\kappa$ blocks).}
    \label{fig:bigcov}
\end{figure}

The aim of this appendix is to provide a comprehensive derivation of the terms that contribute to the full covariance used in this analysis. 
Our data vector is made up of four different types of probes: the galaxy density FSBs $\Phi_{LLb}^{ggg}$, the galaxy density PCLs $C_b^{gg}$, the galaxy-convergence FSBs $\Phi_{LLb}^{gg\kappa}$, and the galaxy-convergence PCLs $C_b^{g\kappa}$. This new data vector offers new potential combination of probes for which we derive, following the logic of the single-field case, the leading off-diagonal terms -- denoted by $N_{222}$ and $N_{32}$ in H25. These two terms contribute respectively to the FSB auto-covariance and to the FSB-PCL cross-covariance, and only within the scales filtered in the FSB estimator. 

Their contributions to the corresponding correlation matrix (shown in Fig. \ref{fig:bigcov}) amount to $\lesssim 30\%$ in the case of $N_{222}$, and $\lesssim 10\%$ for $N_{32}$. These are non-negligible contributions which must be accounted for, as they can lead to significant changes in the distribution of residuals (which follow a near-Gaussian distribution once $N_{222}$ and $N_{32}$ are included, as seen in H25) and in the final parameter constraints. These effects are particularly important in our analysis in Section \ref{ssec:results.constraints}: Fig. \ref{fig:datavectors} shows that most, if not all in $z_1$, FSB data points correspond to the filtered scales, where $N_{222}$ increases the error on measurement. In fact, we find that including $N_{222}$ to the simple Gaussian estimate can, in addition to widening the constraints, lead to a $\sim 1.5\sigma$ shift in $\sigma_8$ in the most extreme cases (the number quoted corresponds to the $C_\ell^{gg} + \fsb{\ell}{LL}{ggg}$ in redshift bin $z_4$, but the effect is generally not as egregious for the full data vector, $C_\ell^{gg} + C_\ell^{g\kappa} + \fsb{\ell}{LL}{ggg}$). Adding $N_{32}$ to the covariance which already includes $N_{222}$ only leads to minute shifts in the $\sigma_8$ contours (a few percent of $\sigma$).

We present below the $N_{222}$ and $N_{32}$ derivations for the 
$\Phi_{LLb}^{ggg} - \Phi_{LLb}^{gg\kappa}$ and 
$\Phi_{LLb}^{gg\kappa} - \Phi_{LLb}^{gg\kappa}$ FSB covariances, as well as for the 
$\Phi_{LLb}^{ggg} - C_b^{g\kappa}$, 
$\Phi_{LLb}^{gg\kappa} - C_b^{gg}$, 
and $\Phi_{LLb}^{gg\kappa} - C_b^{g\kappa}$ FSB-PCL cross covariances.

\subsection{FSB covariances}

\subsubsection{$gg\kappa \times gg\kappa$ covariance}

We remind the reader we use the following definition of the FSB in this section:
\begin{equation}
    \Phi^{g g \kappa}_{LL\ell} = \frac{1}{(2 \ell + 1)} \sum_{m, (\ell m)_{12}} \mathcal G^{\ell_1 \ell_2\ell}_{m_1 m_2 m} W_{\ell_1}^L W_{\ell_2}^L \; \langle g_{\ell_1 m_1} g_{\ell_2 m_2} \kappa_{\ell m} \rangle\,,
\end{equation}
and that the covariance of random variables $X$ and $Y$ is given by
\begin{equation} \label{eq:covdef}
    \mathrm{Cov} (X, Y) \equiv \langle XY \rangle - \langle X \rangle \langle Y \rangle
\end{equation}
where $\langle XY \rangle$ is the main term and $\langle X \rangle \langle Y \rangle$ the counter-term. From this definition we obtain a 6 point correlation function as the main term, which we will break down into smaller contributions in order to generate analytical predictions for leading-order terms: 
\begin{align} \nonumber
\langle \hat{\Phi}^{gg\kappa}_{LL\ell} \hat{\Phi}^{gg\kappa}_{L'L'\ell'} \rangle = \frac{1}{(2\ell +1)(2\ell' +1)} & \sum_{ \ell_{1, 2} , \ell'_{1, 2}}  W_{\ell_1}^L W_{\ell_2}^L W_{\ell_1'}^{L'} W_{\ell_2'}^{L'} \\ \times & \,\sum_{m, m', m_{1, 2}, m'_{1, 2}} \mathcal G^{\ell, \ell_1, \ell_2}_{-m, m_1, m_2} \mathcal G^{\ell', \ell_1', \ell_2'}_{-m', m_1', m_2'} \; \langle g_{\ell_1 m_1} g_{\ell_2 m_2}\,\kappa_{\ell -m} g_{\ell_1' m_1'} g_{\ell_2' m_2'}\,\kappa_{\ell' -m'} \rangle \label{eq:6pcf_cov}
\end{align}

We take a look at the ``$2+2+2$'' decomposition since we know from the single-field case (see Appendix C.1 of H25) that it should give rise to additional contributions along the main diagonal. To keep track of individual combinations of the 6 fields involved in this covariance, we write $$ \langle g_{\ell_1 m_1} g_{\ell_2 m_2} \kappa_{\ell m} g_{\ell_1' m_1'} g_{\ell_2' m_2'} \kappa_{\ell' m'} \rangle \equiv \langle abcdef \rangle .$$
There are 15 combinations of these 6 fields into pairs:
\begin{align}
\nonumber \langle abcdef \rangle_{222} =  & \cancel{\langle ab \rangle \langle cd \rangle \langle ef \rangle} + \cancel{\langle ab \rangle \langle ce \rangle \langle df \rangle} + \cancel{\langle ab \rangle \langle cf \rangle \langle de \rangle} + \\ \nonumber 
& \cancel{\langle ac \rangle \langle bd \rangle \langle ef \rangle} + \cancel{\langle ac \rangle \langle be \rangle \langle df \rangle} + \cancel{\langle ac \rangle \langle bf \rangle \langle de \rangle} + \\ \nonumber 
& \cancel{\langle ad \rangle \langle bc \rangle \langle ef \rangle} + \langle ad \rangle \langle be \rangle \langle cf \rangle + \langle ad \rangle \langle bf \rangle \langle ce \rangle + \\ \nonumber 
& \cancel{\langle ae \rangle \langle bc \rangle \langle df \rangle} + \langle ae \rangle \langle bd \rangle \langle cf \rangle + \langle ae \rangle \langle bf \rangle \langle cd \rangle + \\ 
& \cancel{\langle af \rangle \langle bc \rangle \langle de \rangle} + \langle af \rangle \langle bd \rangle \langle ce \rangle + \langle af \rangle \langle be \rangle \langle cd \rangle  
\end{align}

Some vanish due to the closed triangle condition enforced by the bispectrum definition. The remaining terms can be split into 2 groups, noticing that swapping $a\leftrightarrow b$ or $d \leftrightarrow e$ (corresponding to galaxy density fields filtered on scales $L$ and $L'$ respectively) or $c \leftrightarrow f$ (corresponding to convergence fields) does not change the expression. 
Because the two FSBs might have two different filters however, $a, b \leftrightarrow d, e$ is not equivalent (yet).
This gives us 2 groups: 
\begin{align} \nonumber D_{222} & = \langle ad \rangle \langle be \rangle \langle cf \rangle + \langle ae \rangle \langle bd \rangle \langle cf \rangle \\ & = 2 \; \left( \delta^\mathcal{K}_{\ell_1 \ell_1'} W_{\ell_1}^L W_{\ell_1'}^{L'} C_{\ell_1}^{g} \times \delta^\mathcal{K}_{\ell_2 \ell_2'} W_{\ell_2}^L W_{\ell_2'}^{L'} C_{\ell_2}^{g} \times \delta^\mathcal{K}_{\ell \ell'} C_{\ell}^{\kappa} \right) \end{align}
and
\begin{align} \nonumber N_{222} & = \langle ad \rangle \langle bf \rangle \langle ce \rangle + \langle af \rangle \langle bd \rangle \langle ce \rangle + \langle ae \rangle \langle bf \rangle \langle cd \rangle + \langle af \rangle \langle be \rangle \langle cd \rangle \\ 
& \equiv 4 \left( \delta^\mathcal{K}_{\ell_1 \ell_1'} W_{\ell_1}^L W_{\ell_1'}^{L'} C_{\ell_1}^{g} \times \delta^\mathcal{K}_{\ell_2 \ell'} W_{\ell_2}^L C_{\ell'}^{g \kappa} \times \delta^\mathcal{K}_{\ell_2' \ell} W_{\ell_2'}^{L'} C_{\ell}^{g \kappa} \right) 
\end{align}
where we have already noticed that both of these two terms will only contribute when filters are identical due to the products $\delta^\mathcal{K}_{\ell_i \ell_i'} W_{\ell_i}^L W_{\ell_i'}^{L'}$, which means all sub-contributions are equivalent ($a, b \leftrightarrow d, e$ is now valid since the FSBs must be the same). 

Inserting these expressions back in the original formula for the FSB covariance, we find that the first group contributes in the diagonal only, while the second contributes within the filtered ranges along the main diagonal. These correspond respectively to $D_{222}$ and $N_{222}$ (already introduced in H25) in the 2-field case.

Their full expressions are:
\begin{align}
\text{Cov}_{D_{222}} \left( \hat{\Phi}^{gg\kappa}_{LL\ell} , \hat{\Phi}^{gg\kappa}_{L'L'\ell'} \right) & = \frac{2 C_{\ell}^{\kappa\kappa}}{(2\ell +1)} \delta^\mathcal{K}_{\ell \ell'}  \sum_{\ell_{12}} \frac{(2\ell_1+1)(2\ell_2+1)}{4\pi} \begin{pmatrix} \ell & \ell_1 & \ell_2 \\ 0&0&0 \end{pmatrix}^2 W_{\ell_1}^{L,L'} W_{\ell_2}^{L,L'} \, C_{\ell_1}^{gg} C_{\ell_2}^{gg} 
\end{align}

\begin{align} \label{eq:n222}
\text{Cov}_{N_{222}} \left(  \hat{\Phi}^{L(gg\kappa)}_{\ell} , \hat{\Phi}^{L'(gg\kappa)}_{\ell'} \right) & = \frac{ W_{\ell'}^L W_{\ell}^{L'}  C_{\ell}^{g \kappa} C_{\ell'}^{g \kappa}}{\pi} \sum_{ \ell_{1}} (2\ell_1 +1) \begin{pmatrix} \ell & \ell_1 & \ell' \\ 0&0&0 \end{pmatrix}^2  W_{\ell_1}^L W_{\ell_1}^{L'} \, C_{\ell_1}^{gg}
\end{align}
We have used the following property of the Gaunt coefficients to further simplify the two expressions: 
\begin{equation}
    \sum_{m_1m_2m_3}\left({\cal G}^{\ell_1\ell_2\ell_3}_{m_1m_2m_3} \right)^2=\frac{(2\ell_1+1)(2\ell_2+1)(2\ell_3+1)}{4\pi}\wtj{\ell_1}{\ell_2}{\ell_3}{0}{0}{0}^2.
\end{equation}
There have been slight changes compared to the 1-field case: the spectra outside the sum are cross-spectra but the one inside is the auto spectrum of field $g$; but it still retains its original behaviour -- the term vanishes when the two filters are not identical. 

Note that once again, we can write the $N_{222}$ using the mode-coupling matrix $\mathcal M$ \citep[MCM; Eq. 14 in ][]{alonsoUnifiedPseudoC_2019}: 
\begin{align}
 \mathrm{\ref{eq:n222}} & = \frac{4 W_{\ell'}^L W_{\ell}^{L'}  C_{\ell}^{g \kappa} C_{\ell'}^{g \kappa}}{(2\ell' +1)} \sum_{ \ell_1} \frac{(2\ell' +1) (2\ell_1 +1)}{4\pi} \begin{pmatrix} \ell & \ell' & \ell_1 \\ 0&0&0 \end{pmatrix}^2  \left[ W_{\ell_1}^L W_{\ell_1}^{L'} \, C_{\ell_1}^{gg} \right] = \frac{4 W_{\ell'}^L W_{\ell}^{L'}  C_{\ell}^{g \kappa} C_{\ell'}^{g \kappa}}{(2\ell' +1)} \mathcal M \left[ W_{\ell_1}^L W_{\ell_1}^{L'} \, C_{\ell_1}^{gg} \right] 
\end{align}

\subsubsection{$gg\kappa \times ggg$ covariance}

It is rather straightforward to derive this cross-covariance, by simply swapping one $\kappa$ field for a $g$ field in the expressions derived in the previous section. This gives 
\begin{align}
\text{Cov}_{D_{222}} \left( \hat{\Phi}^{gg\kappa}_{LL\ell} , \hat{\Phi}^{ggg}_{L'L'\ell'} \right) & = \frac{2 C_{\ell}^{g\kappa}}{(2\ell +1)} \delta^\mathcal{K}_{\ell \ell'}  \sum_{\ell_{12}} \frac{(2\ell_1+1)(2\ell_2+1)}{4\pi} \begin{pmatrix} \ell & \ell_1 & \ell_2 \\ 0&0&0 \end{pmatrix}^2 W_{\ell_1}^{L,L'} W_{\ell_2}^{L,L'} \, C_{\ell_1}^{gg} C_{\ell_2}^{gg} \,, \\
\text{Cov}_{N_{222}} \left( \hat{\Phi}^{gg\kappa}_{LL\ell} , \hat{\Phi}^{ggg}_{L'L'\ell'} \right) & = \frac{ W_{\ell'}^L W_{\ell}^{L'}  C_{\ell}^{g g} C_{\ell'}^{g \kappa}}{\pi} \sum_{ \ell_{1, 2}} (2\ell_1 +1) \begin{pmatrix} \ell & \ell_1 & \ell' \\ 0&0&0 \end{pmatrix}^2  W_{\ell_1}^L W_{\ell_1}^{L'} \, C_{\ell_1}^{gg} \,. \label{eq:n222cross}
\end{align}

\vspace{3mm}
\subsection{FSB-PCL covariances}

\subsubsection{$gg\kappa \times g\kappa$ covariance}

The main term in the cross-covariance of an FSB and a power spectrum yields this time a 5 point correlation function:
\begin{align} \label{eq:5pcf_cov}
\left\langle \hat{\Phi}^{gg\kappa}_{LL\ell} \; \hat C_{\ell'}^{g\kappa} \right\rangle & = \frac{1}{(2\ell +1)(2\ell' +1)} \; \sum_{\ell_{1} \ell_{2}} W_{\ell_1}^L W_{\ell_2}^L \; \sum_{ m, m', m_1, m_2} \mathcal G_{m m_1 m_2}^{\ell \ell_1 \ell_2} \, \langle g_{\ell_1 m_1} g_{\ell_2 m_2} \,\kappa_{\ell m} \,g^{}_{\ell' m'}\,\kappa_{\ell' m'} \rangle
\end{align}
As before, we write $$ \langle g_{\ell_1 m_1} g_{\ell_2 m_2} \kappa_{\ell m} g_{\ell' m'} \kappa_{\ell' m'} \rangle \equiv \langle abcde \rangle $$for simplicity, and now look at the ``$3+2$'' decomposition which gives rise to noticeable off-diagonal contributions in the 1-field case. This gives us the 10 following combinations: 
\begin{align} 
\langle abcde \rangle_{32} = &  \quad \; \cancel{\langle cde \rangle \langle ab \rangle} + \cancel{\langle bde \rangle \langle ac \rangle} + \langle bce \rangle \langle ad \rangle + \langle bcd \rangle \langle ae \rangle \nonumber \\ 
& + \cancel{\langle ade \rangle \langle bc \rangle} + \langle ace \rangle \langle bd \rangle + \langle acd \rangle \langle be \rangle \nonumber \\ 
& + \langle abd \rangle \langle ce \rangle + \langle abe \rangle \langle cd \rangle \nonumber \\ & + \langle abc \rangle \langle de \rangle \nonumber
\end{align}
where we cancelled the terms that vanish due to the closed triangle condition. The last contribution cancels out with the counter-term $\langle X\rangle \langle Y\rangle$ of the covariance (Eq. \ref{eq:covdef}), and as such we will ignore it for now. In the 1-field case, there were only 2 independent groups among the remaining 6 terms -- the only discriminating factor was the number of filtered fields in each 2-point and 3-point functions respectively, but now one also needs to keep track of the number of $\kappa$ fields in each. 

This gives us 4 different groups; there are two diagonal terms (let's label them $D_{32}^a$ and $D_{32}^b$) and two off-diagonal terms ($N_{32}^a$ and $N_{32}^b$). They are the following:
\begin{align}
    N_{32}^a & = \langle bce \rangle \langle ad \rangle + \langle ace \rangle \langle bd \rangle = 2 \langle g_{\ell_2 m_2} \kappa_{\ell m} \kappa_{\ell' m'}  \rangle \langle g_{\ell_1 m_1} g_{\ell' m'} \rangle = 2 \,W_{\ell'}^L W_{\ell_2}^L\, C_{\ell'}^g \, \delta^\mathcal{K}_{\ell' \ell_1} \, \mathcal G_{m m' m_2}^{\ell \ell' \ell_2} b_{\ell \ell' \ell_2}^{\kappa \kappa g} \\
    D_{32}^a & = \langle abd \rangle \langle ce \rangle \equiv \langle g_{\ell_1 m_1} g_{\ell_2 m_2} g_{\ell' m'}  \rangle \langle \kappa_{\ell m} \kappa_{\ell' m'} \rangle = W_{\ell_1}^L W_{\ell_2}^L\, C_{\ell}^\kappa \, \delta^\mathcal{K}_{\ell \ell'} \, \mathcal G_{m m_1 m_2}^{ \ell \ell_1 \ell_2} b_{\ell \ell_1 \ell_2}^{ggg} \\
    D_{32}^b & = \langle abe \rangle \langle cd \rangle + \langle abc \rangle \langle de \rangle \equiv 2 \langle g_{\ell_1 m_1} g_{\ell_2 m_2} \kappa_{\ell' m'}  \rangle \langle \kappa_{\ell m} g_{\ell' m'} \rangle = 2 W_{\ell_1}^L W_{\ell_2}^L\, C_{\ell}^{g \kappa} \, \delta^\mathcal{K}_{\ell \ell'} \, \mathcal G_{m m_1 m_2}^{ \ell \ell_1 \ell_2} b_{\ell \ell_1 \ell_2}^{\kappa gg} \\ 
    N_{32}^b & = \langle bcd \rangle \langle ae \rangle + \langle acd \rangle \langle be \rangle \equiv 2 \langle g_{\ell_2 m_2} \kappa_{\ell m} g_{\ell' m'}  \rangle \langle g_{\ell_1 m_1} \kappa_{\ell' m'} \rangle = 2 \,W_{\ell'}^L W_{\ell_2}^L\, C_{\ell'}^{g \kappa} \, \delta^\mathcal{K}_{\ell' \ell_1} \, \mathcal G_{m m' m_2}^{\ell \ell' \ell_2} b_{\ell \ell' \ell_2}^{\kappa g g}
\end{align}

Inserting each into Eq. \ref{eq:5pcf_cov} gives
\begin{align}
    \text{Cov}_{D_{32}} \left( \hat{\Phi}^{gg\kappa}_{LL\ell} , \hat{C}_{\ell'}^{g\kappa} \right) & = \text{Cov}_{D_{32}^a} \left( \hat{\Phi}^{gg\kappa}_{LL\ell} , \hat{C}_{\ell'}^{g\kappa} \right) + \text{Cov}_{D_{32}^b} \left( \hat{\Phi}^{gg\kappa}_{LL\ell} , \hat{C}_{\ell'}^{g\kappa} \right) = \frac{\delta^\mathcal{K}_{\ell \ell'} }{(2\ell +1)} \left(C_{\ell}^{\kappa\kappa} \Phi^{ggg}_{LL\ell} + \, C_{\ell}^{g \kappa} \Phi^{gg\kappa}_{LL\ell} \right) \\
    \text{Cov}_{N_{32}} \left( \hat{\Phi}^{gg\kappa}_{LL\ell} , \hat{C}_{\ell'}^{g\kappa} \right) & = \text{Cov}_{N_{32}^a} \left( \hat{\Phi}^{gg\kappa}_{LL\ell} , \hat{C}_{\ell'}^{g\kappa} \right) + \text{Cov}_{N_{32}^b} \left( \hat{\Phi}^{gg\kappa}_{LL\ell} , \hat{C}_{\ell'}^{g\kappa} \right) = \frac{2 \, W_{\ell'}^L }{(2\ell' +1)} \; \left( C_{\ell'}^{gg} \, \Phi_{L\ell'\ell}^{g \kappa \kappa} + C_{\ell'}^{g \kappa}\, \Phi_{L\ell'\ell}^{g g \kappa} \right)
\end{align}

\subsubsection{$gg\kappa \times gg$ and $ggg \times g\kappa$ covariances}

Similarly to our approach for the $gg\kappa \times ggg$ covariance, we swap one $\kappa$ field from the section above for a $g$ field. The corresponding expressions now read 
\begin{align}
    \text{Cov}_{D_{32}} \left( \hat{\Phi}^{gg\kappa}_{LL\ell} , \hat{C}^{gg}_{\ell'} \right) & = \frac{2\, \delta^\mathcal{K}_{\ell \ell'} }{(2\ell +1)} C_{\ell}^{g\kappa} \Phi^{ggg}_{LL \ell} \\
    \text{Cov}_{N_{32}} \left( \hat{\Phi}^{gg\kappa}_{LL\ell} , \hat{C}^{gg}_{\ell'} \right) & = \frac{4 \, W_{\ell'}^L }{(2\ell' +1)} \; C_{\ell'}^{g g}\, \Phi_{L\ell'\ell}^{g g \kappa}
\end{align}
for the $gg\kappa \times gg$ covariance, and
\begin{align}
    \text{Cov}_{D_{32}} \left( \hat{\Phi}^{ggg}_{\ell} \hat{C}^{g\kappa}_{\ell'} \right) & = \frac{\delta^\mathcal{K}_{\ell \ell'} }{(2\ell +1)} \left(C_{\ell}^{g\kappa} \Phi^{ggg}_{LL\ell} + \, C_{\ell}^{g g} \Phi^{gg\kappa}_{LL\ell} \right) \\
    \text{Cov}_{N_{32}} \left( \hat{\Phi}^{ggg}_{LL\ell} \hat{C}^{g\kappa}_{\ell'} \right) & = \frac{2 \, W_{\ell'}^L }{(2\ell' +1)} \; \left( C_{\ell'}^{gg} \, \Phi_{L\ell'\ell}^{g g \kappa} + C_{\ell'}^{g \kappa}\, \Phi_{L\ell'\ell}^{g g g} \right)
\end{align}
for the $ggg \times g\kappa$ case.

\vspace{7mm}

\subsection{Calibrating the $f_\mathrm{sky}$ correction}
\label{app:multicov.fsky}

\begin{figure}[t]
    \centering
    \includegraphics[width=0.45\linewidth]{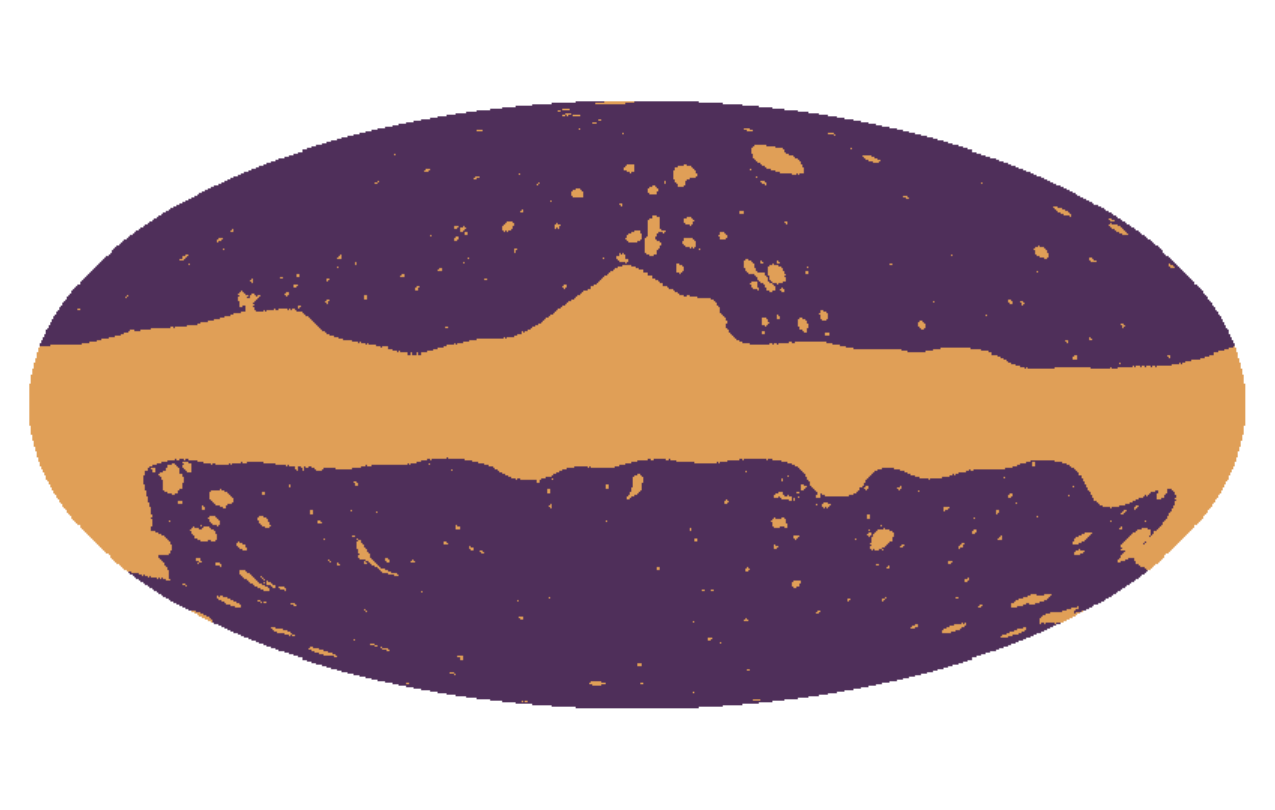}
    \includegraphics[width=0.45\linewidth]{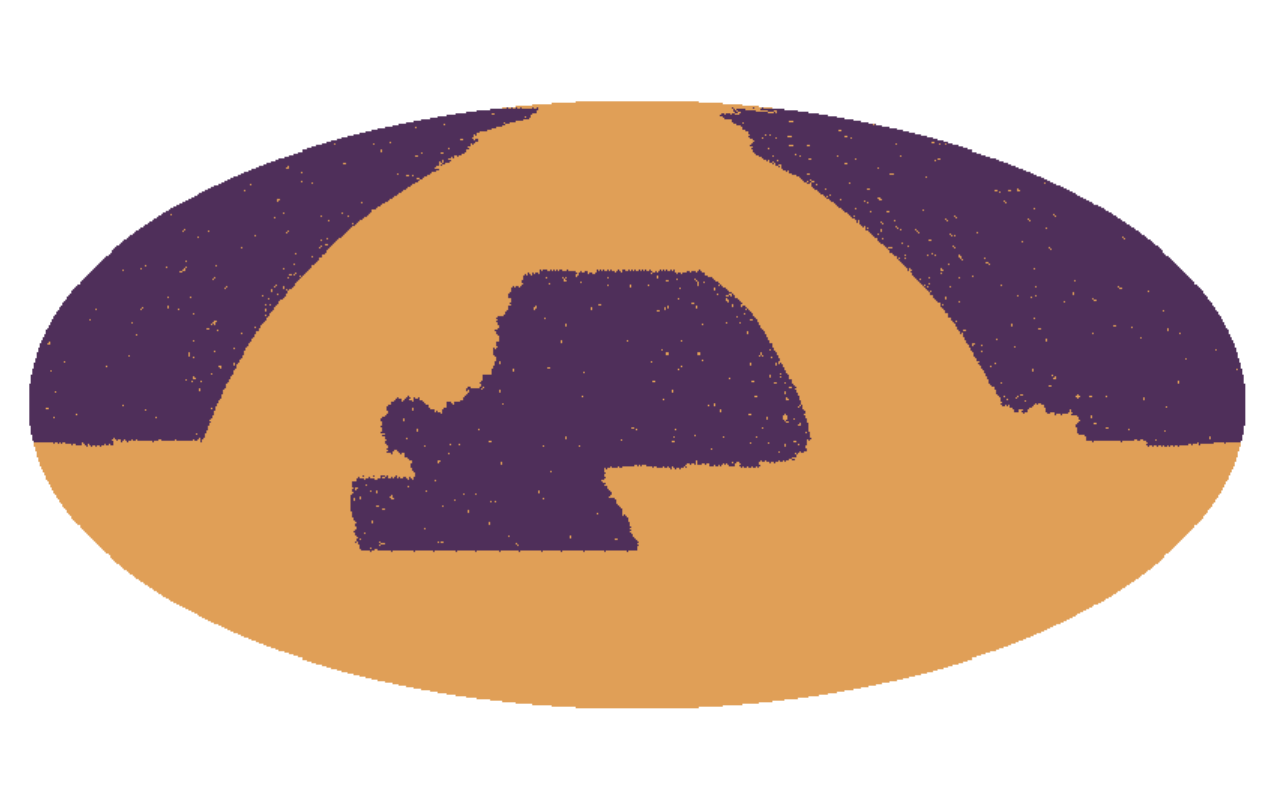}
    \centering
    \caption{Binary masks used to check the accuracy of the $f_\text{sky}$ scalings. The first mask (left) corresponds to the lensing convergence data used in this paper, while the second is typical of galaxy surveys like DESI. Note that we use both Galactic (left) and Equatorial (right). Their overlap thus leads to an $f_\text{sky}$ value that is quite low compared to their individual $f_\text{sky}$ values, making it easier to identify the correct $f_\text{sky}$ scaling.}
    \label{fig:masks}
\end{figure}

As detailed in H25, we account for mode loss due to incomplete observations on the sky by dividing the off-diagonal contributions ($N_{222}$ and $N_{32}$) by the effective sky fraction observed: since the sky fraction, defined as the mean of the square of the survey mask, is less than unity (signifying incomplete observations), the rescaling operation leads to an increase in the covariance. This empirical rescaling in the case of the FSB agrees well with the covariance obtained from simulations. 

This poses no issues in the 1-field case, where the original field and its filtered-squared version share the same footprint on the sky; the choice of rescaling factor becomes less obvious when the two fields each have their respective masks. Take Eq. \ref{eq:n222} for instance; the expression includes $C^{gg}_\ell$ and $C^{g\kappa}_\ell$, which were measured on different effective sky areas -- in this case, should the $N_{222}$ contribution be rescaled using $f_\mathrm{sky} = \langle w_g w_g \rangle$ or $f_\mathrm{sky} = \langle w_g w_\kappa \rangle$ ? The most conservative approach would be to always use the smallest effective sky fraction, as to not underestimate measurement errors, but this might be suboptimal.

Let us look at this problem from a theoretical perspective. Let us partition the mask associated with a field $f$ into smaller, equal-area masks; those smaller masks make up a set which we label $\mathcal S$. For this we do not require the mask to be binary, as the following should also hold for non-binary masks as well. The power spectrum of $f$ can be approximated on small scales by an average over the power spectra $x_i$ of individual set elements $i \in \mathcal S$: 
\begin{equation}
    X = \frac{1}{N_{\mathcal S}} \sum_{i\in \mathcal S} x_i
\end{equation}
where $N_{\mathcal S}$ is the number of elements in the set (which we can liken to $f_\mathrm{sky} = N_{\mathcal S} / N_\mathrm{sphere}$). The covariance of such an estimator $X$ and another $\tilde X$ defined on a different mask, assuming $\langle x_i \rangle = \langle \tilde x_i \rangle = 0$, is given by 
\begin{equation}
    \mathrm{Cov} \left( X, \tilde X \right) = \frac{1}{N_{\mathcal S} N_{\tilde{\mathcal S}}} \sum_{i\in \mathcal S} \sum_{j\in \tilde{\mathcal S}} \langle x_i \tilde x_j \rangle. 
\end{equation}
As we do not expect different patches on the sky to be strongly correlated, we assume that $\langle x_i \tilde x_j \rangle$ only contributes when $i=j$ (a contribution which we label $\sigma$ in that case). Of course, this approximation breaks in case the masks have no overlap.
This allows us to further simplify the expression to 
\begin{equation} \label{eq:fskyscaling}
    \mathrm{Cov} \left( X, \tilde X \right) = \frac{1}{N_{\mathcal S} N_{\tilde{\mathcal S}}} \sum_{i\in \mathcal S \cap \tilde{\mathcal S}} \sigma = \frac{\sigma}{N_{\mathcal S} N_{\tilde{\mathcal S}}} N_{\mathcal S \cap \tilde{\mathcal S}} \, ,
\end{equation}
where $N_{\mathcal S \cap \tilde{\mathcal S}}$ is the number of patches belonging to both sets $\mathcal S$ and $\tilde{\mathcal S}$. We can now interpret this result for our case study of the galaxy and convergence field:
\begin{itemize}
    \item For the covariance of $\hat{\Phi}^{gg\kappa}_{LL\ell}$ and $\hat{\Phi}^{gg\kappa}_{L'L'\ell'}$, both quantities are defined over the same effective sky fraction, given by the intersection of the $g$ and $\kappa$ masks. In this case, $N_{\mathcal S}$, $N_{\tilde{\mathcal S}}$ and $N_{\mathcal S \cap \tilde{\mathcal S}}$ are all identical, and consequently the covariance should be scaled by $f_\mathrm{sky} = \langle w_g w_\kappa \rangle$.
    \item For the covariance of $\hat{\Phi}^{gg\kappa}_{LL\ell}$ and $\hat{\Phi}^{ggg}_{L'L'\ell'}$, $N_{\tilde{\mathcal S}}$ is a subset of $N_{\mathcal S}$ by definition, since one power spectrum ($ggg$ FSB) is defined over the $g$ mask, while the other ($gg\kappa$) has for effective sky fraction the intersection of the $g$ and $\kappa$ masks. $N_{\tilde{\mathcal S}}$ and $N_{\mathcal S \cap \tilde{\mathcal S}}$ are therefore the same and cancel out, such that the expression must scale with $f_\mathrm{sky} = \langle w_g w_g \rangle$. 
\end{itemize}

We check these results by calibrating this rescaling factor using Gaussian simulations: $N_{222}$ being a fully disconnected term, it should appear in the covariance even if the fields are fully Gaussian. Comparing the possible rescalings of the analytical $N_{222}$ with the covariance obtained from measuring FSBs on Gaussian fields with different masks, should give us an idea of which $f_\mathrm{sky}$ correction is most accurate. To measure $N_{222}$ on simulations, one must make sure that fields $g$ and $\kappa$ are correlated, as $N_{222}$ depends on the cross-power spectrum of the two fields. To enhance the $N_{222}$ signal, we instead use the same field for both $g$ and $\kappa$, but use different masks for each -- we shall now call both fields $g$ and $g'$, and their respective masks $w_g$ and $w_{g'}$. We chose the masks so that possible combinations of effective sky fractions lead to values that are distinguishable from one another. The two masks, shown in Fig. \ref{fig:masks}, have little overlap, and give the following $f_\mathrm{sky}$ values: $\langle w_g w_g \rangle = 0.68$, $\langle w_{g'} w_{g'} \rangle = 0.38$, $\langle w_g w_{g'} \rangle = 0.22$.

\begin{figure*}
    \centering
    \includegraphics[width=\textwidth]{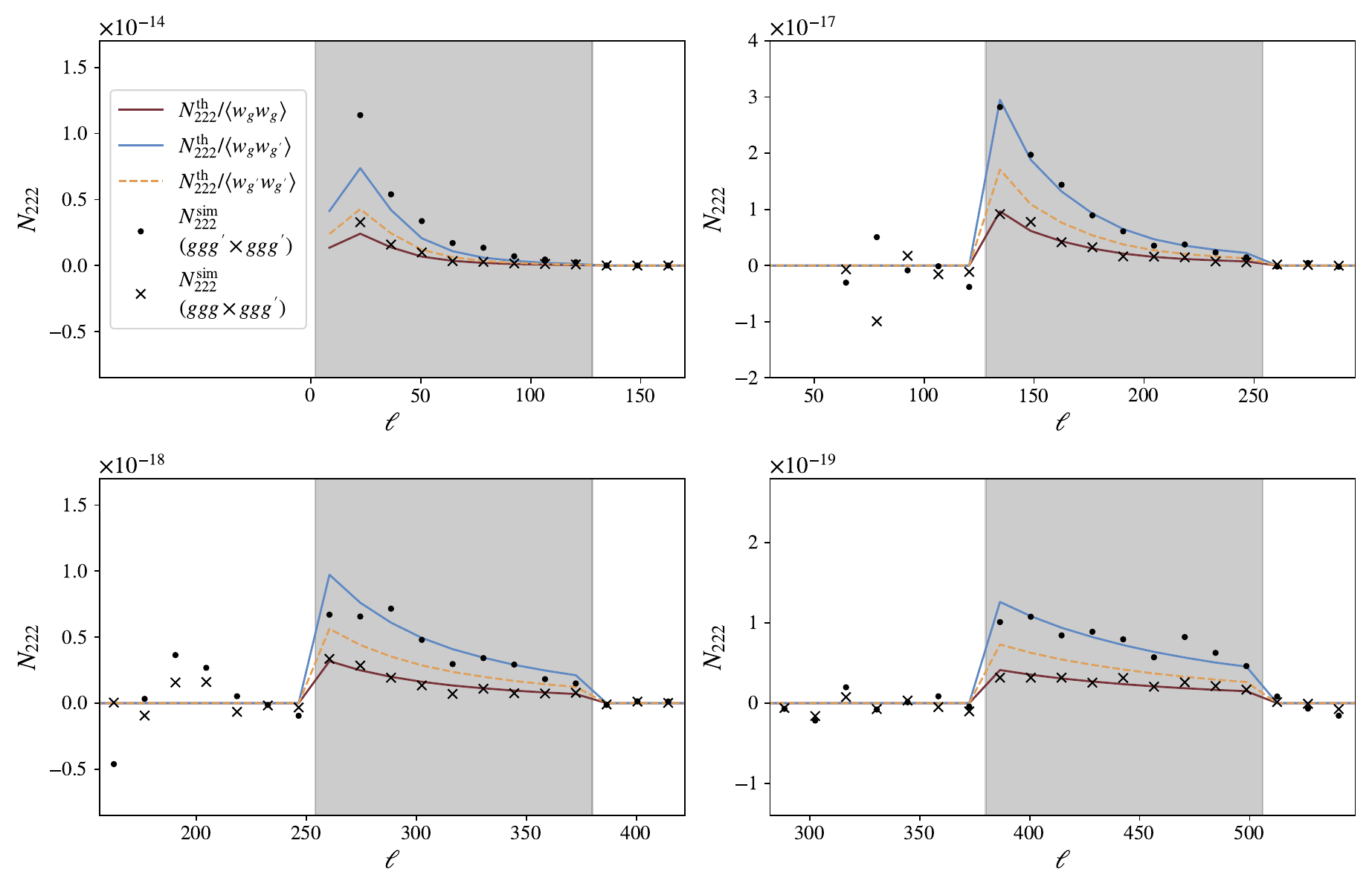} 
    \caption{Scalings of Gaussian covariance term $N_{222}$ with the effective sky fraction. The $N_{222}$ term only contributes within the filtered range, which we have represented as the vertical grey band -- each sub-figure therefore corresponds to a different FSB. 
    The simulation-based $ggg \times ggg'$ cross-covariance (black dots) and $ggg' \times ggg'$ auto-covariance (black crosses) $N_{222}^\text{sim}$ terms follow different scalings, as is shown by the coloured lines which represent the theoretical value of $N_{222}$ in the full-sky scenario, scaled by different values of $f_\mathrm{sky}$. The three values chosen here correspond to $\langle w_g w_g \rangle$ (solid brown line), $\langle w_g w_{g'} \rangle$ (solid blue line), and $\langle w_{g'} w_{g'} \rangle$ (dashed yellow line). \vspace{3mm}}
    \label{fig:n222cali}
\end{figure*}

We computed $N_{222}$ in the case of the auto-covariance $\mathrm{Cov} ( \Phi_{LL\ell}^{ggg'}, \Phi_{L'L'\ell'}^{ggg'})$  from Eq. \ref{eq:n222} and the cross-covariance $\mathrm{Cov} ( \Phi_{LL\ell}^{ggg}, \Phi_{L'L'\ell'}^{ggg'} )$ from Eq. \ref{eq:n222cross} over a total of 2400 Gaussian simulations. The simulation-based, isolated $N_{222}$ terms were computed as the difference between the simulation-based covariance and the Gaussian covariance computed from assuming $g_L^2$ and $g$ are Gaussian fields (see Section \ref{sssec:methods.review.cov}). Taking this difference should successfully isolate $N_{222}$, as the subdominant connected terms we do not account for in our covariance model vanish for Gaussian fields. 

The $N_{222}$ expressions for $\mathrm{Cov} ( \Phi_{LL\ell}^{ggg'}, \Phi_{L'L'\ell'}^{ggg'} )$ and $\mathrm{Cov} ( \Phi_{LL\ell}^{ggg}, \Phi_{L'L'\ell'}^{ggg'} )$ depend on auto- and cross-power spectra of $g$ and $g'$, but Fig. \ref{fig:n222cali} shows that they have different $f_\mathrm{sky}$ scalings, which seem to agree with our theoretical expectations: the $ggg' \times ggg'$ term does indeed scale with $\langle w_g w_{g'} \rangle$ (the black dots on Fig. \ref{fig:n222cali} closely follow the blue line across all FSBs), and the $ggg' \times ggg'$ term with $\langle w_g w_{g} \rangle$ (the black crosses match the brown line).

Our measurements of $N_{222}$ are therefore easily explained by our simple theoretical considerations (Eq. \ref{eq:fskyscaling}), and can be generalised to the connected $N_{32}$ terms. We repeat below all relevant scalings for completeness: 
\begin{align}
    & \mathrm{Cov} \left( C_\ell^{gg}, C_{\ell'}^{gg} \right) \sim \mathrm{Cov} \left( \Phi_{LL\ell}^{ggg}, C_{\ell'}^{gg} \right) \sim \mathrm{Cov} \left( \Phi_{LL\ell}^{ggg}, \Phi_{\ell'}^{ggg} \right) \propto \langle w_g w_g \rangle^{-1}, \\ 
    & \mathrm{Cov} \left( C_\ell^{g\kappa}, C_{\ell'}^{gg} \right) \sim  \mathrm{Cov} \left( \Phi_{LL\ell}^{gg\kappa}, C_{\ell'}^{gg} \right) \sim \mathrm{Cov} \left( \Phi_{LL\ell}^{ggg}, C_{\ell'}^{g\kappa} \right) \sim \mathrm{Cov} \left( \Phi_{LL\ell}^{gg\kappa}, \Phi_{L'L'\ell'}^{ggg} \right)  \propto \langle w_g w_g \rangle^{-1}, \\ 
    & \mathrm{Cov} \left( C_\ell^{g\kappa}, C_{\ell'}^{g\kappa} \right) \sim  \mathrm{Cov} \left( \Phi_{LL\ell}^{gg\kappa}, C_{\ell'}^{g\kappa} \right) \sim \mathrm{Cov} \left( \Phi_{LL\ell}^{gg\kappa}, \Phi_{L'L'\ell'}^{gg\kappa} \right)  \propto \langle w_g w_\kappa \rangle^{-1}. 
\end{align}

\vspace{5mm}

\section{Interpolator Performance}
\label{app:interp}

We test the performance of the interpolator used in this paper, both in terms of accuracy and runtime. There are technically two stages in the bispectrum interpolator where we make use of interpolation: 
\begin{enumerate}[label=(\roman*)]
    \item First, instead of computing the bispectrum at each $(\ell_1, \ell_2, \ell_3)$, we define a subset of multipoles, then interpolate linearly in 3D to evaluate missing configurations. The multipole subset we use here contains all multipoles between $0$ and $10$, then every $20$ multipole from $10$ to $96$, and $50$ multipoles logarithmically spaced between $100$ and $\ell_\mathrm{max}$. 
    \item We then build a cubic $\Omega_m$-interpolator, as the bispectrum $\Omega_m$ dependence is more complex and cannot be accounted for by a simple linear rescaling of the templates. We compute the FSB templates for $100$ values of $\Omega_m$ linearly spaced in the range $0.05 \leq \Omega_m < 0.5$; for a given value of $\Omega_m$, we can then interpolate the FSB from neighbouring values of $\Omega_m$. 
\end{enumerate}
The final interpolator is then a result of these two interpolations, at the level of the projected bispectrum templates (multipole interpolation), and at the FSB templates level ($\Omega_m$ interpolation). 

\begin{figure*}
    \centering
    \includegraphics[width=0.4\linewidth]{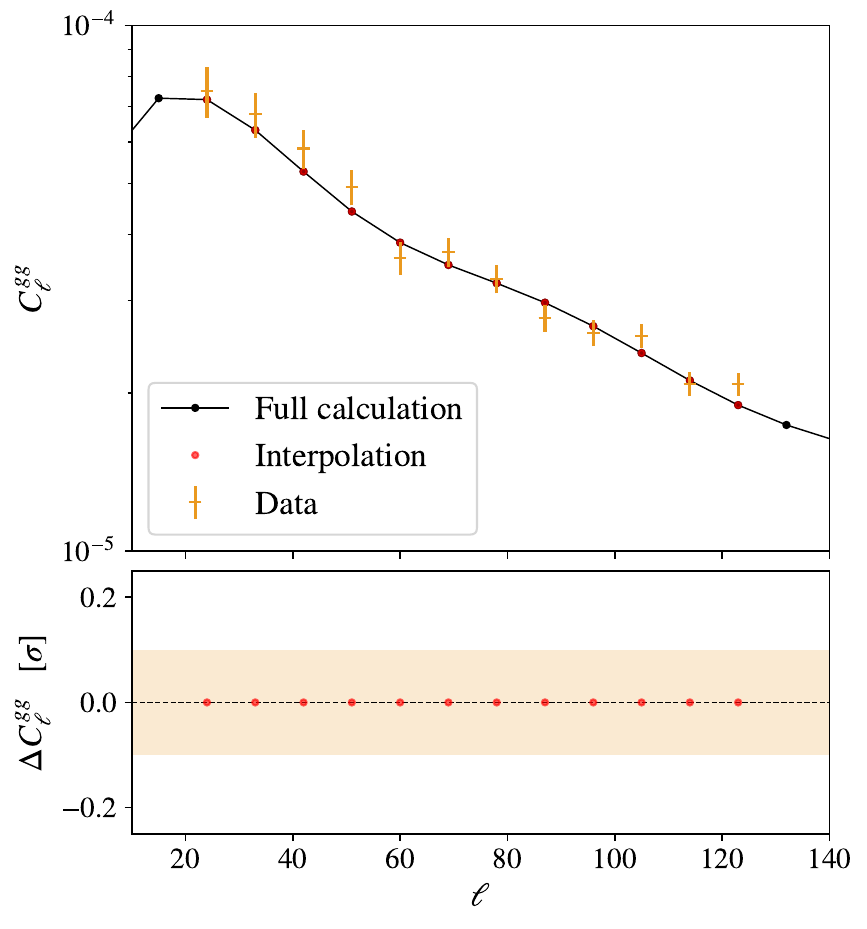}
    \quad\quad\quad
    \includegraphics[width=0.393\linewidth]{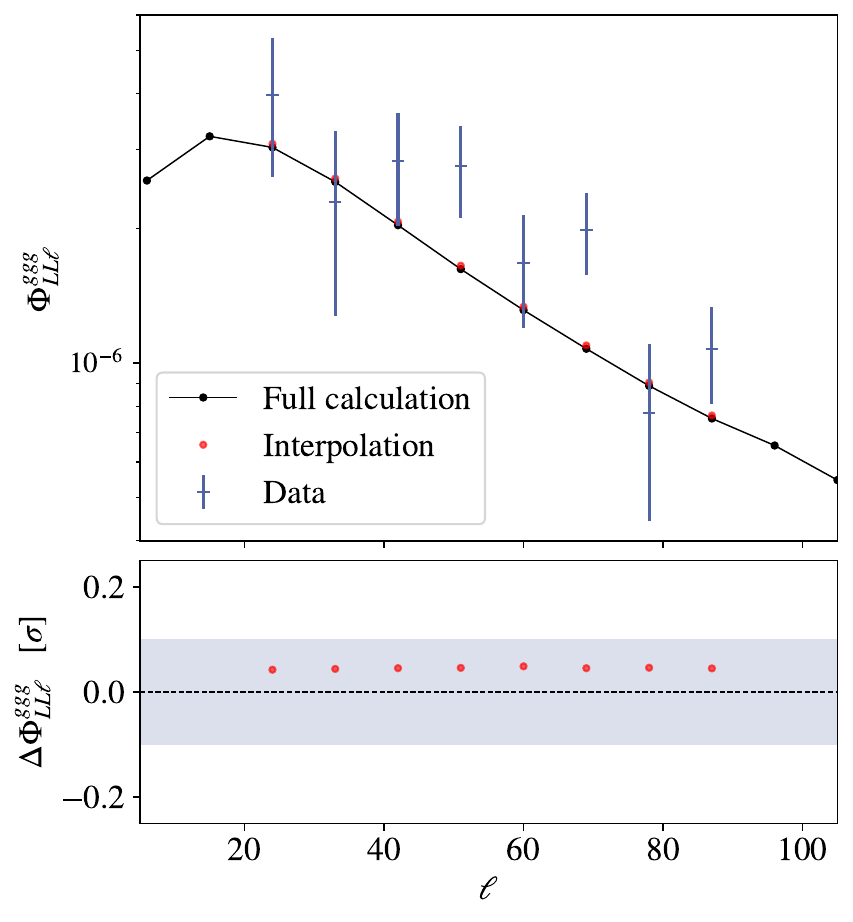}
    \caption{Interpolator accuracy. \textit{Left panel}: We compare the power spectrum best-fit obtained from the interpolation (which comprises an interpolation at the multipole level, and an interpolation between values of $\Omega_m$) and the full calculation of the power spectrum. The difference between the two approaches is less than 0.003\% of the error $\sigma$ associated with the measurement. \textit{Right panel}: We show the same comparison for the $ggg$ FSB best-fit. The difference between the interpolation and full calculation corresponds in that case to about 5\% of the error associated with the measurement. In both lower panels, we show in colour the interval corresponding to $10\%$ of the error bars in the upper panels.}
    \label{fig:interp_accuracy}
\end{figure*}

\begin{figure*}
    \centering
    \includegraphics[width=0.43\linewidth]{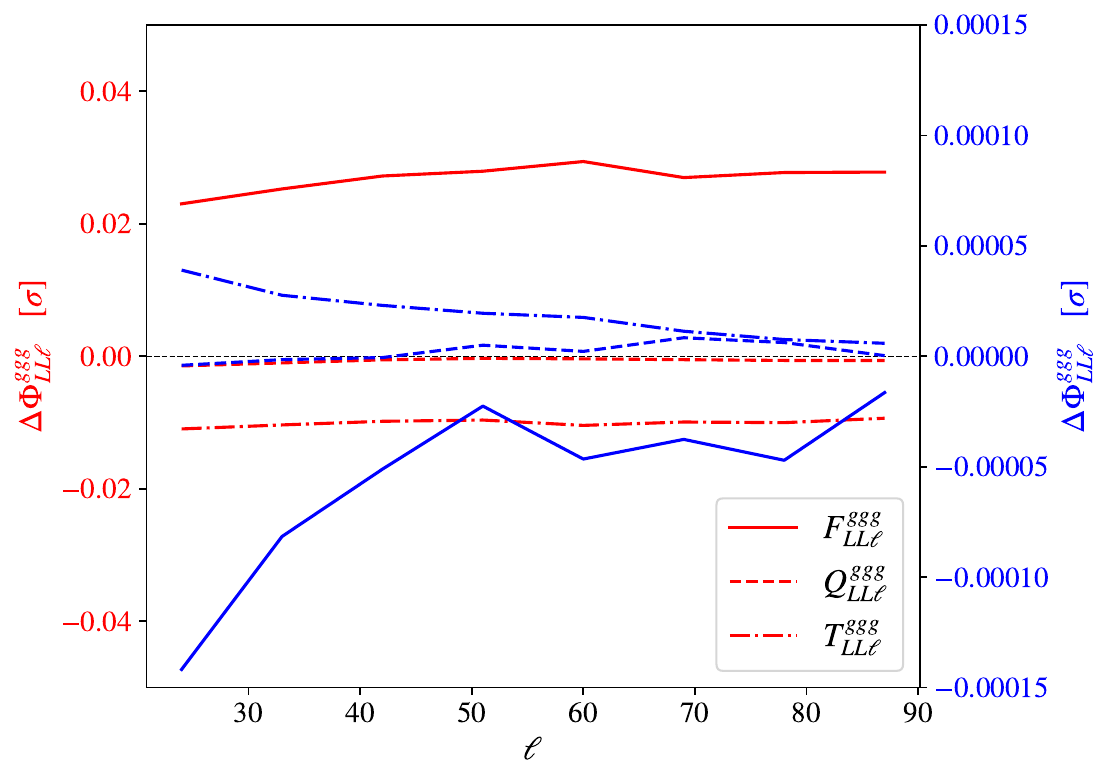}
    \quad\quad\quad
    \includegraphics[width=0.36\linewidth]{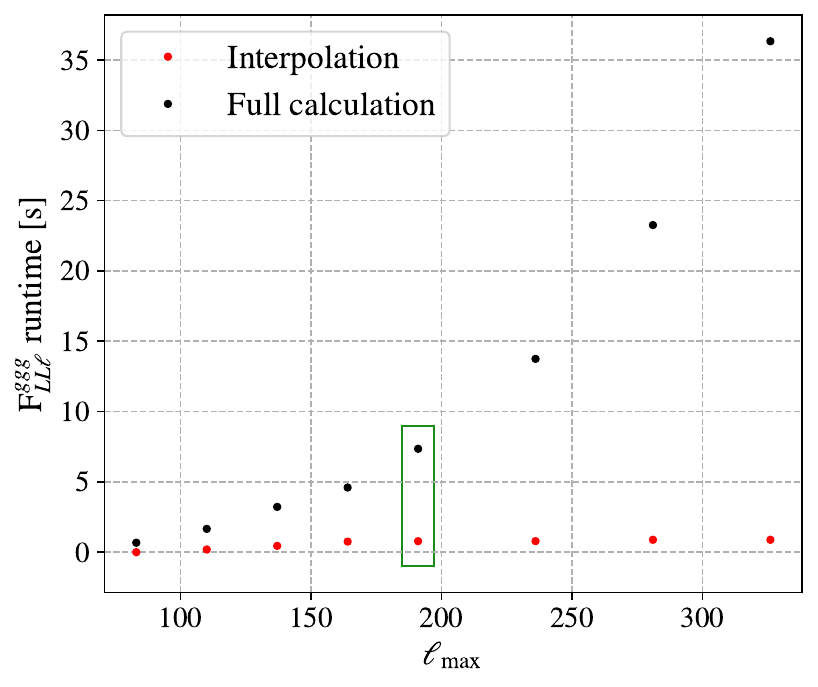}
    \caption{Details by templates. \textit{Left panel}: Interpolated best-fit template errors with respect to the full calculation, for the individual templates of the $ggg$ FSB. We do not show the interpolated noise templates, which depend on power spectra (which are themselves recovered to high precision by the interpolator, as shown in the left panel of Fig. \ref{fig:interp_accuracy}). The red lines were obtained using linear interpolation between multipoles, while the blue lines show the equivalent for a cubic interpolator. Note how the right axis (cubic interpolation) is about $300$ times smaller than the left axis (linear interpolation). \textit{Right panel}: Runtime performance of the interpolator (red dots) compared to computing the full calculation (black dots) for one $F^{ggg}_{LL\ell}$ template, as a function of maximum scale $\ell_\mathrm{max}$. As we increase $\ell_\mathrm{max}$, the number of triangle configurations contributing to the bispectrum dramatically increases, making the full calculation unsuitable for inference purposes. We highlight the data points relevant to this paper in a green box: for a scale cut of $\ell_\mathrm{max} = 191$, interpolating is more than $9$ times faster than computing the full bispectrum.} 
    \label{fig:interp_runtime}
\end{figure*}

We start by verifying that the overall accuracy of the interpolator is acceptable: we show in Fig. \ref{fig:interp_accuracy} that the interpolator works exquisitely well for the power spectra, and is slightly biased in the case of the bispectrum. Indeed, we report a difference between interpolation and full calculation of the FSB corresponding to $\sim 5\%$ of the errors on the $ggg$ measurement, which we deem acceptable. We investigate the source of this bias by computing the difference between the two approaches for each template contributing to the deterministic part of the $ggg$ FSB (see definitions in Eq. \ref{eq:bsp_kernels}); we show this in the left panel of Fig. \ref{fig:interp_runtime}. The largest error on individual projected templates (at fixed $\Omega_m$ and $\sigma_8$, and before accounting for the bias rescaling) is found for the $T_2$ template with a bias of about $\lesssim 15\%$ of the full-calculation template. However, it is not the main contribution to the overall error budget, as the linear contribution $F_2$ accounts for most of the bispectrum signal. In blue, we also show the much smaller errors that could be achieved using a cubic interpolator at the multipole level (instead of a linear one, shown in red). In this case, the interpolation bias is barely noticeable, but the making of such cubic interpolators is very costly in 3D, even at small $\ell_\text{max}$. To generate $100$ FSB templates (i.e. 5 templates for each $\Omega_m$ value to build the interpolator), and allocating the same computational resources in both cases, the linear approach to interpolation takes about 4 minutes, but the cubic approach is closer to 38 hours. This could easily be sped up using parallelisation, and once built, the interpolator is fast to evaluate -- whether it is linear or cubic; however, given our error budget and various approximations taken in this work, it is not yet necessary to refine the bispectrum interpolator to achieve sub-percent accuracy.

In the right panel of Fig. \ref{fig:interp_runtime}, we also show the time gain that can be obtained from using the interpolator. This gain evidently depends on the number of triangle contributions that need to be computed, which increases with the maximum multipole used in the analysis, $\ell_\mathrm{max}$. Since we selected a large-scale filter for the inference part of this work, the signal of the FSB dies off at reasonably low $\ell$, but the time gain is already consequential: to estimate a single $F_2$ template, for identical binning and cosmological parameters, the interpolator is 9 times faster to evaluate than the full bispectrum. However, we could have selected larger filters in the highest redshift bins -- Fig. \ref{fig:datavectors} makes this obvious -- which would have then made us compute the bispectrum up to $\ell_\mathrm{max} \sim 300$. Reading from the plot, this corresponds to a $\sim \times 30$ improvement in runtime when using the interpolator over the full calculation\footnote{We do not claim however that our implementation of the tree-level bispectrum calculation is fully optimized.}. We do not show the same plot for $Q_2$ and $T_2$, since $F_2$ can partly be derived from these two templates -- once you compute $F_2$, you effectively get $Q_2$ and $T_2$ for free if you save the various terms of $F_2$ separately (see Eq. \ref{eq:bsp_kernels}). A full MCMC chain for the $ggg$ data vector with $50,000$ samples takes 0.6 hours to run on 8 cores, therefore we can expect a chain to take more than 5 hours to complete on the same number of cores using the full bispectrum calculation -- which is not unfeasible given appropriate computing resources, but might become so if we push the calculation to smaller scales.


\end{appendix}

\end{document}